\documentclass[12pt]{article}
\usepackage{xr-hyper}
\makeatletter
\usepackage{xr}
\usepackage{hyperref}
\usepackage{enumitem}
\usepackage{longtable}
\usepackage{xcolor}
\usepackage{enumitem}
\usepackage{amssymb}
\usepackage[]{natbib}
\usepackage{amsmath,amsfonts,bm}
\usepackage[parfill]{parskip}
\usepackage{amsthm}
\usepackage[toc,page]{appendix}

\usepackage{setspace}
\usepackage[pagewise]{lineno}

\usepackage{ragged2e}

\usepackage{microtype}
\usepackage{graphicx}
\usepackage{booktabs} 
\usepackage{multirow} 

\usepackage{comment}
\usepackage{caption}
\usepackage{bbm}
\usepackage{subcaption}
\usepackage{enumitem}
\setitemize{noitemsep,topsep=0pt,parsep=0pt,partopsep=0pt}

\usepackage{algorithm}
\usepackage{chngcntr} 
\usepackage[noend]{algpseudocode}
\algrenewcommand\textproc{}
\algdef{SE}[SUBALG]{Indent}{EndIndent}{}{\algorithmicend\ }%
\algtext*{Indent}
\algtext*{EndIndent}

\newcommand{\Mean}{{\mathbb{E}}}
\newcommand{\Var}{{\mbox{Var}}}

\newcommand{\Cov}{{\mbox{Cov}}}

\newtheorem{thm}{Theorem}

\newtheorem{assumption}{Assumption}

\newtheorem{lemma}{Lemma}

\newtheorem{cor}{Corollary}

\usepackage{threeparttable}

\numberwithin{equation}{section}

\usepackage{algorithm}
\usepackage{algcompatible}
\hypersetup{
	unicode=false,
	pdftoolbar=true,
	pdfmenubar=true,
	pdffitwindow=false,     
	pdfstartview={FitH},    
	pdfkeywords={}, 
	pdfnewwindow=true,      
	colorlinks=true,       
	linkcolor=cyan,          
	citecolor=blue,        
	filecolor=cyan,      
	urlcolor=cyan           
}

\usepackage[utf8]{inputenc}

\newcommand{\neighbor}[1]%
{\overline{#1}}

\newcommand{\blind}{1}
\usepackage{colortbl}
\definecolor{lightblue}{rgb}{0.8, 0.9, 1}
\usepackage{mathtools}
\graphicspath{{fig/}}

\begin{document}

\def\spacingset#1{\renewcommand{\baselinestretch}%
{#1}\small\normalsize} \spacingset{1.4}

\if1\blind
{
  \title{ \bf ARMA-Design: Optimal Treatment Allocation Strategies for A/B Testing in Partially Observable Experiments}
  \author{Ke Sun$^{1}$, Linglong Kong$^{1}$, Hongtu Zhu$^{2}$ and Chengchun Shi$^3$ \\ 
\smallskip\\
$^1$Department of Mathematical and Statistical Sciences, University of Alberta \\ $^2$ Department of Biostatistics, University of North Carolina at Chapel Hill \\ $^3$Department of Statistics, London School of Economics and Political Science}
\date{}
  \maketitle
} \fi

\if0\blind
{
  \bigskip
  \bigskip
  \bigskip
  \begin{center}
    {\LARGE\bf Off-policy Evaluation in Doubly Inhomogeneous Environments}
\end{center}
  \medskip
} \fi
	
\bigskip
\vspace{-1em}
\begin{abstract}
Online experiments 
are frequently employed in many technological companies to evaluate the performance of a newly developed policy, product, or treatment relative to a baseline control. In many applications, the experimental units receive a sequence of treatments over time. To handle these time-dependent settings, existing A/B testing solutions typically assume a fully observable experimental environment that satisfies the Markov condition. However, this assumption often does not hold in practice.
   
This paper studies the optimal design for A/B testing in partially observable online experiments.  We introduce a controlled (vector) autoregressive moving average model to capture partial observability. We introduce a small signal asymptotic framework to simplify the calculation of asymptotic mean squared errors of average treatment effect estimators under various designs. We develop two algorithms to estimate the optimal design: one utilizing constrained optimization and the other employing reinforcement learning. We demonstrate the superior performance of our designs using two dispatch simulators that realistically mimic the behaviors of drivers and passengers to create virtual environments, along with two real datasets from a ride-sharing company. A Python implementation of our proposal is available at \url{https://github.com/datake/ARMADesign}. 
\end{abstract}


\spacingset{1.7} 

	\section{Introduction}\label{sec:intro}

	
\noindent \textit{\textbf{Background}}. A growing number of 
companies, particularly 
multi-sided platforms like  Airbnb, DoorDash, Uber, and retail marketplaces such as  Amazon and Zara are increasingly harnessing data-driven approaches to evaluate and refine their policies and products. In particular, A/B testing, which conducts online experiments to compare a standard control policy ``A'' to an alternate version ``B'', plays a crucial role in informing business decisions within these companies and has proven invaluable for their growth and development~\citep{koning2022experimentation}.   For instance, ride-sharing platforms, including Uber, Lyft, and DiDi Chuxing, constantly develop new order dispatching, driver repositioning, pricing policies and assess their improvements through A/B testing \citep{qin2024reinforcement}. Accurate A/B testing enables decision-makers to choose better policies that meet more ride requests, enhance passenger satisfaction, increase driver income, and thus benefit the entire transportation ecosystem \citep{xu2018large}. 
 
\noindent \textit{\textbf{Challenges}}.
In many applications, the experimental units receive treatments sequentially over time. 
A/B testing in these experiments 
poses four major challenges:
\begin{enumerate}[leftmargin=*]
    \item \textbf{Small sample size}.  
    Online experiments are often constrained to a short duration, typically several weeks~\citep{luo2024policy}. This limited timeframe leads to large variances in estimating the difference in expected outcomes between the new and standard policies, referred to as the average treatment effect (ATE). 
    \item \textbf{Small signal}. The ATE is usually quite small \citep{farias2022markovian,athey2023semi,xiong2023data}, posing considerable challenges 
    in distinguishing between the two policies. 
    For instance, in ride-sharing companies, the ATE generally ranges from 0.5$\%$ to 2$\%$~\citep{tang2019deep}. 
    \item \textbf{Carryover effects}. Carryover effects are ubiquitous in online experiments, where the treatment assigned at a given time can influence future outcomes \citep{bojinov2019time,han2022detecting,shi2022dynamic,xiong2023data,chen2024experimenting}. These effects are typical in ride-sharing companies where past policies can alter the distribution of drivers in the city, which in turn affects future outcomes; refer to Figure \ref{fig:carryover} for detailed illustrations. 
    Such phenomena lead to violations of the stable unit treatment value assumption \citep[SUTVA, see][Section 1.6]{imbens2015causal}, rendering many existing A/B testing solutions \citep[see, e.g.,][]{johari2017peeking,azevedo2020b,wang2023b,larsen2024statistical,quin2024b,waudby2024anytime} and causal inference methods \citep[see, e.g.,][]{imai2013estimating,belloni2017program,chernozhukov2017double,armstrong2021finite,athey2021matrix,viviano2023synthetic,ding2024first} ineffective.

    \begin{figure*}[t!]
		\centering
		\includegraphics[width=13.5cm]{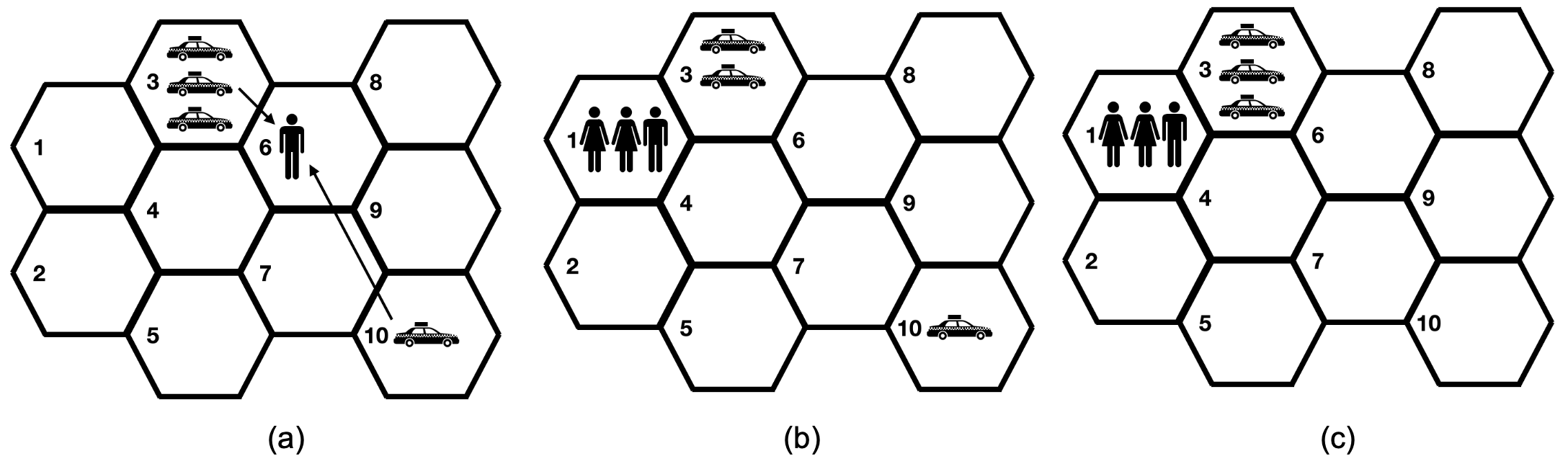}
		\caption{\small Illustration of the carryover effect in ride-sharing, taken from \citet{li2024evaluating}. (a) A city is divided into ten regions, and a passenger from Region 6 orders a ride. Two actions are available: assigning a driver from Region 3 or Region 10. These actions will lead to different future outcomes, as illustrated in (b) and (c). (b) Assigning a driver from Region 3 might result in an unmatched future request in Region 1 due to the driver in Region 10 being too far from Region 1. (c) Assigning the driver in Region 10 preserves all three drivers in Region 3, allowing all future ride requests to be easily matched.} 
		\label{fig:carryover}
\end{figure*}

	\begin{figure*}[t!]
		\centering
		\begin{subfigure}[t]{0.75\textwidth}
			\centering
			\includegraphics[width=\textwidth,trim=0 0 0 0,clip]{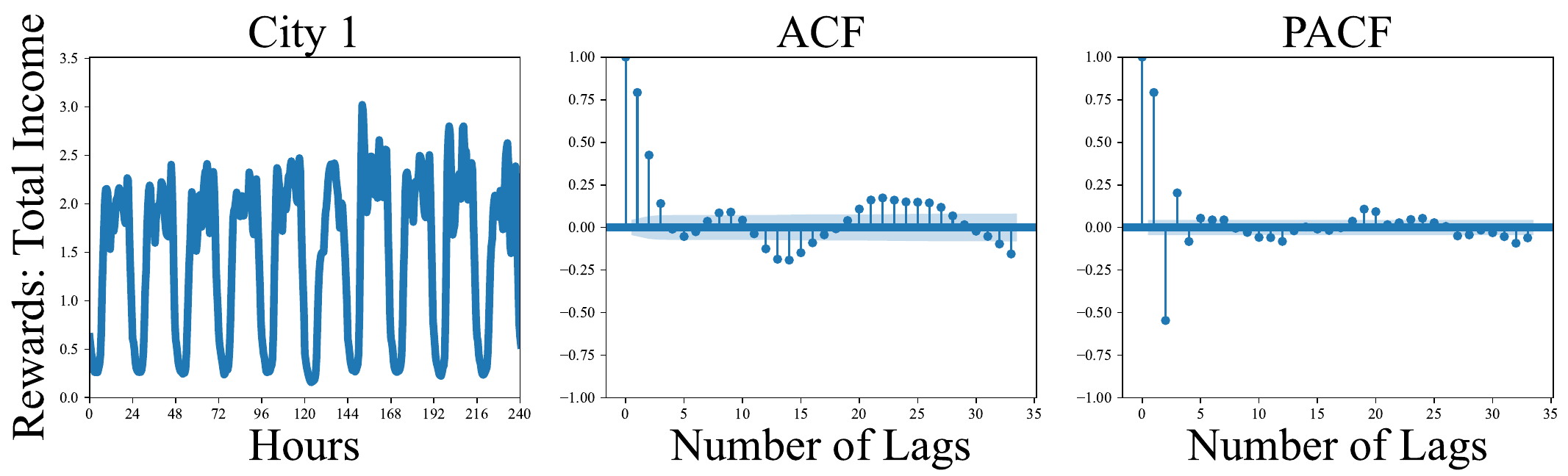}	
  \end{subfigure}
		\begin{subfigure}[t]{0.75\textwidth}
			\centering
			\includegraphics[width=\textwidth,trim=0 0 0 0,clip]{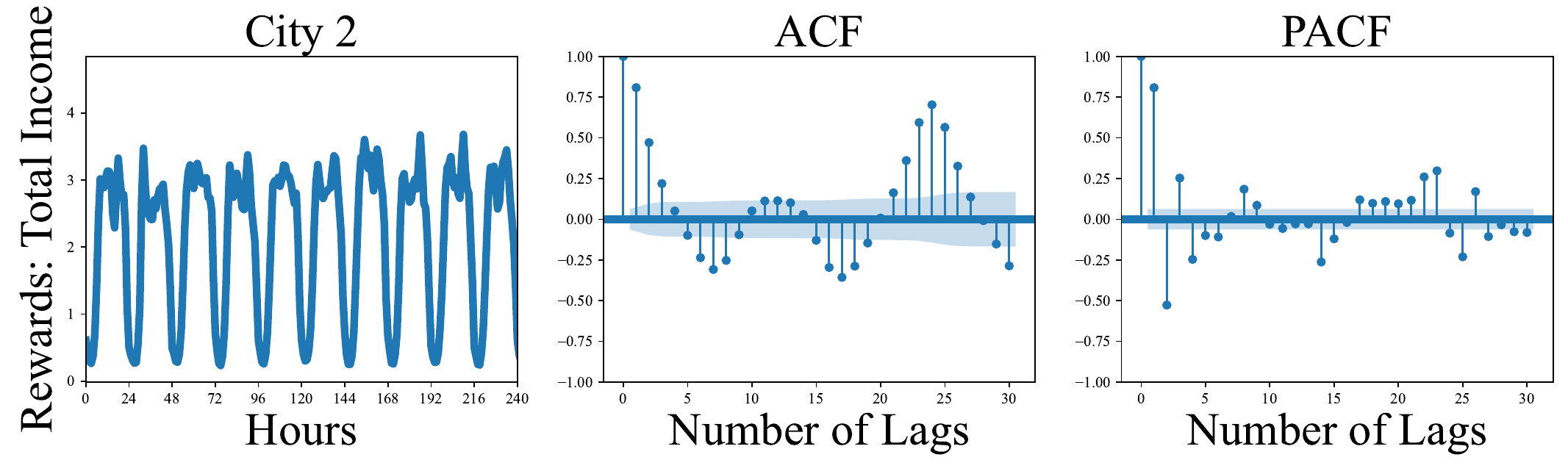}
  \end{subfigure}
		\caption{\small Visualizations of two sequences of driver income collected from a ride-sharing platform in two cities. Each row plots the data from one of the cities. Left panels: The trend in all drivers' total income over time. Middle panels: The ACF of the residuals of these income sequences (after filtering the seasonal effects within each day and regressing on other relevant market features). Right panels: The PACF of the residuals of these income sequences.}
		\label{fig:motivation}
	\end{figure*}
    
    \item \textbf{Partial observability}. Partial observability frequently occurs in online experiments. Assuming the underlying time series follows a Markov chain or Markov decision process \citep[MDP,][]{puterman2014markov}, full observability requires its state to be completely recorded. 
    In contrast, partial observability means only part of the state is observable, leading to the violation of the Markov property \citep{krishnamurthy2016partially}. It is often the rule rather than the exception in real applications, where recording all relevant features to ensure the ``memoryless" property proves impractical. To elaborate, consider our motivating ride-sharing example. The left panels of Figure \ref{fig:motivation} visualize two sequences of driver income from two cities, both exhibiting strong daily patterns. The middle and right panels display the auto-correlation function (ACF) and partial ACF (PACF) of the residuals of these income sequences after filtering the seasonal effects within each day and regressing on other relevant market features. Notably, the PACF exhibits significant higher-order lags, which demonstrates the non-Markovian nature of the data.
\end{enumerate}

\noindent \textit{\textbf{Contributions}}. Our primary objective is to develop a statistical framework for A/B testing that addresses the above challenges. Our contributions include the development of optimal designs, efficiency indicators, statistical modeling, estimation methods, and theoretical frameworks. We detail them as follows.
\begin{enumerate}[leftmargin=*]
    \item To tackle the first two challenges, we focus on carefully designing the experiment to optimize the data generation process from the online experiments, so as to minimize the mean squared error (MSE) of the resulting average treatment effect (ATE) estimator. In particular, we propose two innovative algorithms to learn the optimal design: one based on constrained optimization and the other via reinforcement learning (RL). Our empirical studies, which leverage a synthetic dispatch simulator and a city-level real-data-based simulator -- both constructed using physical models to realistically simulate driver and passenger behaviors -- along with two real datasets from a ride-sharing company, demonstrate that our proposed designs consistently outperform existing state-of-the-art. 
    \item Additionally, we derive two efficiency indicators 
    to compare the statistical efficiencies of three frequently employed designs in estimating the ATE: the alternating-day design, the uniform random design, and the alternating-time design. 
    \item To address the last two challenges, we introduce a controlled (vector) autoregressive moving average ((V)ARMA) model for fitting experimental data. The proposed model is a variant of classical (V)ARMA models \citep[Chapters 3]{brockwell2002introduction} and represents a rich sub-class of partially observable MDP models \citep[POMDP, see, e.g.,][]{monahan1982state}. It employs the autoregressive component to accommodate carryover effects and incorporates the moving average error structure to allow for partial observability. 
    \item We devise the parameter estimation procedures for the controlled (V)ARMA model and introduce a novel small signal asymptotic framework to substantially simplify the computation of asymptotic MSEs of ATE estimators under various designs. 
\end{enumerate}  
To summarize, our proposal integrates cutting-edge machine learning algorithms, such as RL, with asymptotic theories derived from classical time series models in econometrics to offer guidance for policy deployment in real-world applications. 

\noindent \textbf{\textit{Outline}}. We discuss the related literature in Section \ref{sec:related}. In Section~\ref{sec:controlledVARMA}, we introduce the controlled ARMA model, elaborate its connection with POMDPs, and develop the associated estimating procedure for ATE. We further derive the asymptotic MSEs of different designs under the small signal assumption and propose two efficiency indicators to assess their effectiveness. In Section~\ref{sec:optimaldesign}, we present the proposed algorithms to estimate the optimal design. In Section~\ref{sec:experiments}, we demonstrate the efficacy of the proposed designs and efficiency indicators 
through 
two dispatch simulators and two real datasets from a ride-sharing platform. Finally, we conclude our paper in Section \ref{sec:discussion}. 
 
\section{Related Literature}\label{sec:related}
Our proposal intersects with a wide range of research fields, including econometrics, statistics, management science, operational research, and machine learning. It particularly engages with three main research branches: experimental designs, POMDPs, ARMA and state space models. 

\noindent \textit{\textbf{Experimental Designs}}. The design of experiments, also known as experimental design, is a classical problem in statistics, driven by diverse applications in biology, psychology, agriculture, and engineering \citep{fisher1966design}. Within the statistics literature, our proposal specifically relates to works that focus on identifying treatment allocation strategies tailored for clinical trials~\citep[see, e.g.,][]{robbins1952some,pocock1975sequential,begg1980treatment,atkinson2007optimum,jones2009d,rosenblum2020optimal,liu2022balancing,ma2024new}. These studies typically focus on non-dynamic settings (referred to as contextual bandit settings in the machine learning literature) where observations are assumed to be independent, excluding any carryover effects. In contrast, our research accommodates carryover effects and addresses the more complex challenge of temporal dynamics.   While traditional crossover designs \citep{laird1992analysis,jones2003design} can deal with long-lasting carryover effects, they often require extended washout periods, making them less practical for modern A/B testing with short durations. 

More recently, there has been a growing body of literature in management science, econometrics, and machine learning that explores experimental designs for A/B testing in technological companies. Our work differs from them in several aspects:
(i) Many papers consider settings without carryover effects over time \citep{bajari2021multiple,wan2022safe,viviano2023causal,wang2023b,basse2024randomization}. (ii) Some existing works adopt an RL framework to model the experimental data 
\citep{glynn2020adaptive,li2023optimal, wen2024analysis}, where the data follows a fully observable MDP. In contrast, our framework is more general and accommodates partial observability, which is a more typical scenario in real applications. (iii)  Several recent studies focus on switchback designs where policies alternate at specified intervals under various optimality conditions~\citep{hu2022switchback,bojinov2023design,xiong2023data,wen2024analysis}.  In contrast, our approach considers a broader class of designs that allow each treatment assignment to be influenced by the entire treatment history (see Section \ref{sec:optimaldesign}).

In the RL literature, the design of the experiment is also referred to as the behavior policy search problem, 
in which \cite{mukherjee2022revar, hanna2017data} explored the optimal behavior policy by minimizing the MSE of the policy value estimator in MDPs. Meanwhile, \citet[Section 3.3]{agarwal2022reinforcement} employed the D-optimal design for policy learning in MDPs. In contrast to these works, we focus on the evaluation of ATE -- the difference between two policy value estimators — and allow partial observability, offering a more realistic scenario in practice.

Finally, recent works have approached the design problem from an optimization perspective \citep{zhao2024experimental}. In particular, works in the machine learning literature have proposed the use of deep learning or RL to numerically compute a Bayesian version of the optimal design \citep{foster2021deep,blau2022optimizing}. In contrast to these methods, we employ a frequentist approach and focus specifically on the evaluation of ATE.  

\noindent \textit{\textbf{POMDPs}}. 
Partial observability often arises in real applications, including autonomous driving~\citep{levinson2011towards}, resource allocation~\citep{bower2005resource}, recommendation~\citep{li2010contextual}, and medical management systems~\citep{hauskrecht2000planning}. POMDP is the most commonly used model to characterize the partial observability of a stochastic dynamics system. Learning the optimal policy in general POMDPs requires the agent to infer the latent belief state~\citep{krishnamurthy2016partially}, which is both statistically and computationally intractable in general~\citep{papadimitriou1987complexity,vlassis2012computational}. Despite these challenges, it is possible to focus on a sub-class of POMDPs to make the estimation tractable~\citep{kwon2021rl, liu2022partially}. Our proposal follows this principle by introducing a controlled (V)ARMA model under a weak signal condition to streamline estimation and design. Different from existing works that proposed partial history importance weighting \citep{hu2023off} or value-function-based methods \citep{uehara2022future} to construct policy value estimators, we focus on the experimental design, aiming to optimize the data collection process to enhance policy evaluation.

\noindent \textit{\textbf{ARMA and State Space Models.}} The ARMA model, a cornerstone in time series analysis, has been widely employed in various domains, particularly in econometrics~\citep{brockwell1991time,hendry1995dynamic,fan2003nonlinear,box2015time,hamilton2020time}. 
Additionally, it is closely related to state space models, which plays a vital role in analyzing continuous dynamic systems~\citep{harvey1990forecasting, durbin2012time, aoki2013state, kim2017state,komunjer2020likelihood}. The ARMA and state space models are also related to POMDPs, which can be seen as controlled state space models with an added dimension of the action or treatment space, allowing state transitions to be influenced by treatments \citep{krishnamurthy2016partially}; see Section \ref{subsec:ARMAmodel} for detailed discussions about their connections. 

In the causal inference literature, \cite{menchetti2021estimating} proposed causal versions of ARIMA models. 
More recently, \cite{liang2023randomization} proposed linear state space models for causal inference. Despite the similarity in the models, these works differ from ours primarily in their focus: they focused on estimating and inferring causal effects, whereas we concentrate on the experimental design to ``optimize'' the estimated causal effect. Consequently, these works did not utilize the small signal framework we propose to derive closed-form solutions for the asymptotic MSEs under different designs, which we use as criteria to optimize the design. Nor did they explore RL approaches to finding the optimal design. This difference in objectives also influences the choice of models. For instance, the causal ARIMA model is designed for settings with a single persistent treatment over time \citep[Assumption 1]{menchetti2021estimating}, making it unsuitable for studying general designs. 

{\singlespacing	
\section{The Controlled ARMA Model and Its Applications in A/B Testing}\label{sec:controlledVARMA}
}
This section presents the proposed controlled (V)ARMA model and demonstrates its usefulness in estimating the ATE and comparing different treatment allocation strategies. We first describe the data collected from time series experiments, define the ATE for A/B testing, and introduce three commonly used designs in Section \ref{subsec:data}. We further introduce the proposed controlled ARMA model, discuss its connections to POMDPs, and present the estimation procedure for ATE in Section \ref{subsec:ARMAmodel}. Next, we propose the small signal asymptotic framework,  establish the asymptotic MSE of the estimated ATE, and then derive two efficiency indicators to compare the estimation efficiency under the three designs in Section \ref{subsec:smallsignal}. Finally, we generalize these results to accommodate multivariate observations and exogenous variables based on the proposed controlled VARMA model in Section \ref{subsec:VARMA}. 

\subsection{Data, ATE, and Designs}\label{subsec:data}	
\noindent \textit{\textbf{Data}}. 
We divide the experimental period into a series of non-overlapping time intervals, and during each of the time intervals, a specific policy or treatment is implemented. In our collaboration with a ride-sharing company, time intervals are typically set to 30 minutes or 1 hour. The data gathered from the online experiments can be summarized as a sequence of observation-treatment pairs, denoted by $\{(\mathbf{Y}_t, U_t): 1\le t\le T\}$, where $T$ represents the termination time of the experiment. Here, the notations are consistent with those used in control engineering \citep{aastrom2012introduction}: $\mathbf{Y}_t$ denotes a potentially multivariate observation collected at time $t$, and $U_t$ represents a scalar treatment applied at time $t$. In detail: 
\begin{itemize}[leftmargin=*]
    \item $Y_{t,1}$, the first element of $\mathbf{Y}_t$,  denotes the outcome of interest, such as total driver income or total number of completed orders at the $t$-th time interval in a ride-sharing platform.
    \item The subsequent elements of $\mathbf{Y}_t$ denote additional relevant market features, which can contain the drivers' online time and the number of call orders at the $t$-th interval on the online platform in the context of ride-sharing. These features represent the supply and demand of the ride-sharing platform and can significantly influence the outcome \citep{zhou2021graph}. Our experiments suggest that jointly modeling both the outcome and market features can substantially improve the estimation of the ATE, achieving a reduction in the MSE by 10 to 100 times compared to approaches that model only the outcome, such as those in \cite{menchetti2021estimating} and \cite{liang2023randomization}.   
    \item $U_t \in \{-1, 1\}$ specifies the policy implemented during the $t$-th interval. By convention, $1$ denotes a new treatment, while $-1$ represents the standard control.
\end{itemize}

\noindent \textbf{\textit{ATE}}. Our ultimate goal lies in estimating the ATE, defined as the difference in the cumulative outcome between the treatment and the control, 
			\begin{equation}\begin{aligned}\label{eqn:ATE}
					\text{ATE} =\lim_{T\to \infty}    \mathbb{E}_{1} \left[\frac{1}{T}\sum_{t=1}^{T} Y_{t,1}\right] - \lim_{T\to \infty}\mathbb{E}_{-1} \left[\frac{1}{T}\sum_{t=1}^{T} Y_{t,1}\right],
			\end{aligned}\end{equation}
  provided the limit exists. Here, $\mathbb{E}_{1}$ and  $\mathbb{E}_{-1}$ denote expectations under which the treatment $U_t$ is consistently set to $1$ and $-1$ at every time $t$, respectively. This objective is a central focus in A/B testing with carryover effects 
  \citep[see, e.g.,][]{hu2022switchback,li2023optimal,xiong2023data,wen2024analysis}. Both terms on the right-hand-side (RHS) of \eqref{eqn:ATE} should be understood as potential outcomes \citep{imbens2015causal}, representing the average outcome that would have been observed if either the new treatment or the control had been assigned at all times. Nonetheless, as we focus on experimental design,  it eliminates concerns about unmeasured confounders. To simplify the presentation, we choose not to use potential outcome notations. 
  Interested readers may refer to   \citet{ertefaie2014constructing}, \citet{luckett2019estimating} and \citet{viviano2023synthetic} for detailed discussions on potential outcomes in dynamic settings.
	
\noindent \textit{\textbf{Design}}. In our context, each design corresponds to a sequence of treatment allocation strategies $\pi=\{\pi_t\}_{t\ge 1}$ where each $\pi_t$ specifies the conditional distribution of $U_t$ given the past data history up to time $t-1$, denoted by $\mathbf{H}_{t-1}=\{\mathbf{Y}_1,U_1,\ldots,\mathbf{Y}_{t-1}, U_{t-1}\}$. Informally speaking, each design determines the probabilities of applying the treatment and control at each time, given the past history. Our focus is on \textit{observation-agnostic} designs, where each $\pi_t$ depends on $\mathbf{H}_{t-1}$ only through $\{U_1,U_2,\ldots, U_{t-1}\}$, independent of past observations. This class covers the following three special examples:

\textbf{Example I: Alternating-time (AT) design}. This design alternates between treatment and control at adjacent time intervals and is frequently employed in many ride-sharing companies, such as Lyft and DiDi Chuxing, to compare different order dispatching policies \citep{Chamandy2016,luo2024policy}. To implement the AT design, the initial treatment $U_1$ is randomly generated with equal probabilities: $\pi_1(1)=\pi_1(-1)=0.5$. For subsequent times, we set $\pi_t(-U_{t-1}|\mathbf{H}_{t-1})=1$ and $\pi_t(U_{t-1}|\mathbf{H}_{t-1})=0$ so that $U_t=-U_{t-1}$ almost surely. 
    
\textbf{Example II: Alternating-day (AD) design}. This design assigns the same treatment throughout each day and switches to the opposite treatment on the following day. Similar to the AT design, the initial treatment $U_1$ in the AD design is also uniformly randomly determined. 
Let $\tau$ represent the number of time intervals per day. The treatment assignment ensures that $U_1 = U_2 = \cdots = U_\tau = -U_{\tau+1} = -U_{\tau+2} = \cdots = -U_{2\tau} = U_{2\tau+1} = \cdots$, maintaining consistency within each day and alternating on a daily basis.

\textbf{Example III: Uniform random (UR) design}. This design independently assigns treatment and control randomly with equal probabilities each time. 
    Specifically, $\pi_t$ remains a constant function with a value of 0.5, regardless of $t$ and $\mathbf{H}_{t-1}$. Despite its simplicity, designs of this type have been widely adopted in clinical trials. 

To conclude this section, we make two remarks here. First, both AT and AD fall under the category of switchback designs, where the duration of each treatment varies from a single time interval to an entire day. Second, while many studies have explored these designs in fully observable Markovian environments, less is known about their efficacy in more realistic, partially observable environments. Addressing this gap is one of our main objectives. 

{\singlespacing    
\subsection{The Controlled ARMA Model, Connection to POMDPs, and Estimation of ATE}\label{subsec:ARMAmodel}
}
\begin{figure}[t!]
    \centering
    \includegraphics[width=8.5cm]{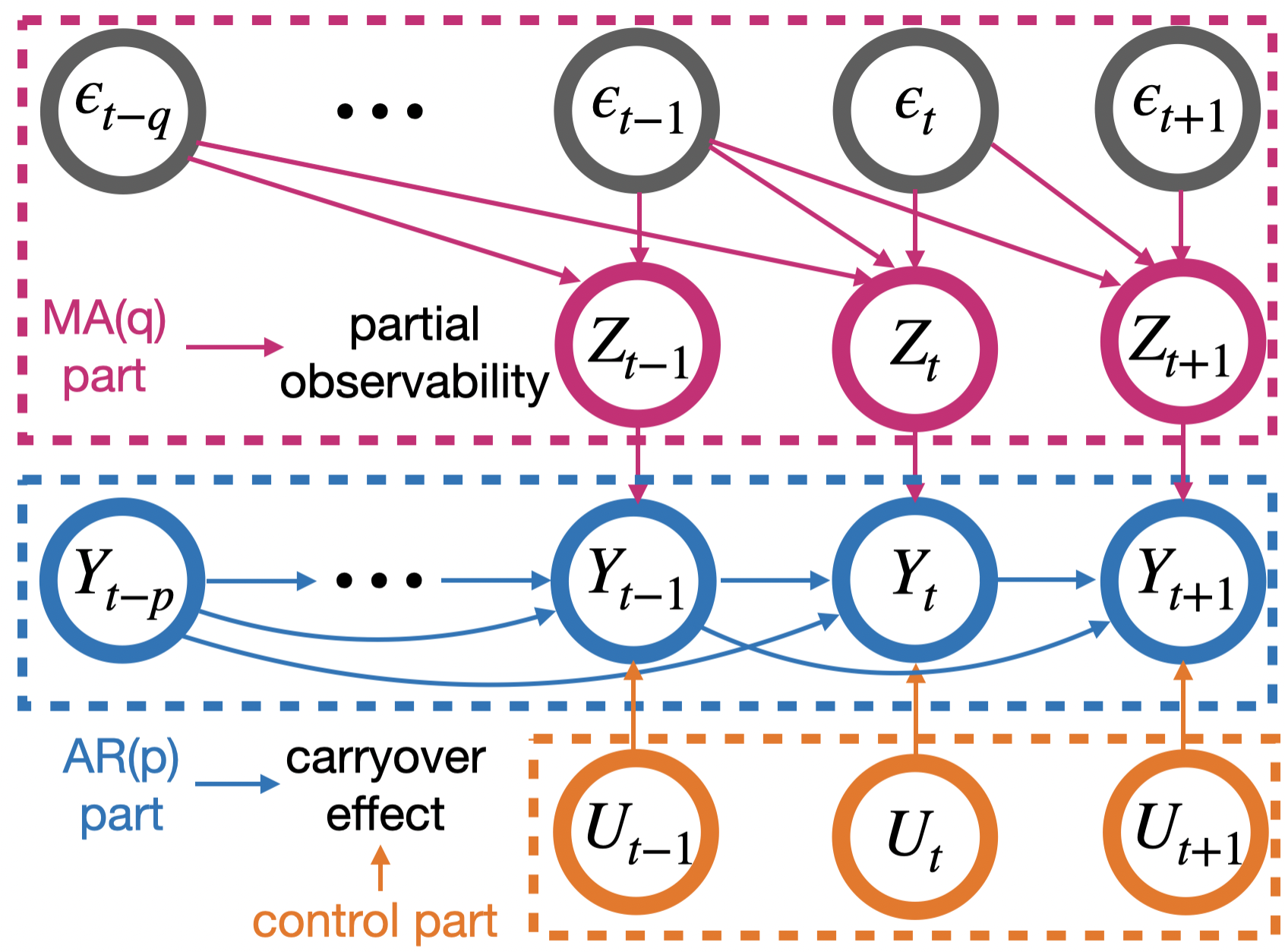}
    \caption{\small Visualization of the proposed controlled (V)ARMA model: $Y_t$ denotes the observation, $U_t$ denotes the treatment, $Z_t$ denotes the residual $\sum_{j=0}^{q} \theta_j \epsilon_{t-j}$,  and $\epsilon_t$ denotes the latent white noise. The model features two key properties: (i) the existence of both the autoregressive and control parts enables the pathway from $U_{t-1}$ to $Y_{t-1}$ and then to $Y_t$ and $Y_{t+1}$, capturing the carryover effects; (ii) the inclusion of the moving average part allows for partial observability, as the pathway $Y_{t-1}\leftarrow Z_{t-1} \leftarrow \epsilon_{t-1} \to Z_{t+1}\to Y_{t+1}$ is unblocked by $Y_t$ and $U_t$, resulting in the conditional dependence between $Y_{t-1} $ and $Y_{t+1}$ given $Y_t$ and $U_t$. 
    }\label{fig:dag}
\end{figure}

\noindent \textit{\textbf{Controlled ARMA($p, q$).}} We first introduce the controlled ARMA model, a sub-class of POMDPs, designed to capture carryover effects and partial observability in online experiments; see Figure \ref{fig:dag} for a graphical visualization. The one-dimensional controlled ARMA($p, q$) model is formulated as:
		\begin{equation}
		\begin{aligned}\label{eq:controlledARMA}
			Y_t =\mu+ {\color{blue}\sum_{j=1}^{p} a_j  Y_{t-j}} + {\color{orange}b U_{t}} + {\color{purple}\sum_{j=0}^{q} \theta_j \epsilon_{t-j}}, 
		\end{aligned}
	\end{equation}
where $\mu$ denotes the intercept, $b,a_1,\ldots,a_p,\theta_1,\ldots,\theta_q$ are parameters, and by convention, $\theta_0=1$. Model (\ref{eq:controlledARMA}) consists of three main components:
\begin{itemize}[leftmargin=*]
    \item{} The first term in blue on the RHS of model \eqref{eq:controlledARMA} represents the autoregressive component with the parameters $\{a_j\}_{j=1}^p$, capturing the influence of past observations $\{Y_{t-j}\}_{j=1}^p$ on its current observation $Y_t$.
    \item{} The second term in orange on the RHS of model \eqref{eq:controlledARMA} incorporates the treatment into the model, affecting the observation $Y_t$ at each time. Its treatment effect is measured by the parameter $b$.
    \item{}  The last term in purple represents the residual, denoted by $Z_t$, which is modeled by a moving average process with the parameters $\{\theta_j\}_{j=1}^q$, i.e., $Z_t=\sum_{j=0}^{q} \theta_j \epsilon_{t-j}$. We assume 
    the white noises $\{\epsilon_{t}\}_{t}$ are i.i.d. with zero mean and variance $\sigma^2$.  
\end{itemize}

We next illustrate how model \eqref{eq:controlledARMA} allows carryover effects and partial observability. First, when $p>0$, the autoregressive structure and the control component allow $U_{t-1}$ to have an indirect effect on subsequent observations (e.g., $Y_t$ and $Y_{t+1}$) through its impact on $Y_{t-1}$, effectively capturing the carryover effects; see the pathway $U_{t-1}\to Y_{t-1}\to Y_t \to Y_{t+1}$ in Figure \ref{fig:dag}. Second, when $q>0$, the inclusion of the moving average process renders the time series non-Markovian. For instance, consider the pathway $Y_{t-1}\leftarrow Z_{t-1} \leftarrow \epsilon_{t-1} \to Z_{t+1}\to Y_{t+1}$ in Figure \ref{fig:dag}. This pathway is not blocked by $Y_t$ and $U_t$, thus violating the Markov assumption and resulting in a partially observable environment. 

Finally, different sets of parameters play different roles in A/B testing: the autoregressive coefficients ($\{a_j\}_{j=1}^p$) and the control parameter ($b$) determine the ATE,  whereas the moving average coefficients ($\{\theta_j\}_{j=1}^q$) influence the residual correlation, which in turn determines the optimal design. Formal statements can be found in Lemma \ref{lemma:ATE} and Theorem \ref{them:asymptotic}. 

\vspace{0.25em}

\noindent \textit{\textbf{Connection to POMDPs.}} We next show that the proposed controlled ARMA($p, q$) model is in essence a sub-class of POMDPs, which have been widely employed to model partially observable environments. Consider the following POMDP with linear state transition and observation emission functions: 
		\begin{equation}
		\begin{aligned}\label{eq:statespace}
			\text{State}:&   \ \boldsymbol{X}_{t+1} =F \boldsymbol{X}_t + B \boldsymbol{U}_t +\boldsymbol{V}_t \\ 
			\text{Observation}:& \quad \boldsymbol{Y}_t \  =H \boldsymbol{X}_t+ C \boldsymbol{U}_t + \boldsymbol{W}_t.
		\end{aligned}
	\end{equation}
 In this model:
(i) Both the observation $\boldsymbol{Y}_t$ and the treatment $\boldsymbol{U}_t$ can be multi-dimensional.
(ii) $\boldsymbol{X}_t$ denotes a vector-valued latent state such that any dependence between the past and future will ``funnel'' through this latent state.
(iii) $\boldsymbol{V}_t$ and $\boldsymbol{W}_t$ are the measurement errors. $F, B, H,$ and $C$ are the parameter matrices, respectively. This model can also be viewed as a variant of the linear state space or dynamic linear model, which incorporates an extra treatment variable $\boldsymbol{U}_t$.

By setting $\boldsymbol{X}_t$ to linear combinations of current and past treatments and observations, the proposed controlled ARMA($p, q$) model \eqref{eq:controlledARMA} can be transformed into a linear POMDP. See Appendix~\ref{appendix:equivalence} of the Supplementary Material for formal proof. The advantage of utilizing the controlled ARMA model over a linear POMDP lies in its ability to provide concise and closed-form expressions for the asymptotic MSE of the ATE estimator (see, e.g., Corollary \ref{cor:efficiency_ARMA}), which is crucial for deriving the optimal design.

According to the Wold decomposition theorem \citep{wold1938study}, any stationary process can be decomposed into two mutually uncorrelated processes: a linear combination of lags of a white noise process (MA($\infty$) process) and a linear combination of its past values (AR($\infty$) process). The stationarity assumption can typically be satisfied in practice by applying periodic filtering to remove seasonal effects, as detailed in Section \ref{subsec:VARMA} and our data analysis in Section \ref{sec:realdata}. This underlying principle in time series theory indicates that our model is broadly applicable and can represent a diverse range of linear POMDPs.
	

\noindent \textit{\textbf{Estimation of ATE.}} We begin by deriving the closed-form expression for the ATE under the proposed controlled ARMA$(p,q)$ model.
\begin{assumption}[No unit root]\label{assump:ergodicity}
     All the roots of the polynomial $1-\sum_{j=1}^p a_j y^j$  lie outside the unit circle.
\end{assumption}
\begin{lemma}\label{lemma:ATE}
    Under Assumption \ref{assump:ergodicity}, ATE equals $2b/(1-a)$,  where $a=a_1 + \ldots+ a_p$. 
\end{lemma}

We make three remarks. First, Assumption \ref{assump:ergodicity} guarantees the ergodicity of the proposed controlled ARMA model, which in turn validates the limits in the definition of the ATE (see \eqref{eqn:ATE}). Second, the ATE can be decomposed into the sum of $2b+2ab/(1-a)$ where the first term corresponds to the direct effect of $U_t$ on $Y_t$ and the second term represents the indirect effect mediated by the past observations $\{Y_{t-j}\}_{j\ge 1}$. 
Third, as commented earlier, the ATE is exclusively determined by the autoregressive coefficients and the control parameter, and it remains independent of the moving average coefficients. This motivates us to apply \textit{the method of moments} \citep[e.g., the Yule-Walker method,][]{yule1927vii,walker1931periodicity} to estimate the ATE.

Notably, directly applying the ordinary least square method to minimize $\sum_t (Y_{t}-\mu -\sum_{j=1}^{p} Y_{t-j}-bU_{t})^2$ will fail to produce consistent estimators. This failure is due to the correlation between the residual $Z_{t}$ and predictors $\{Y_{t-j}\}_{j=1}^{p}$ under partial observability, as illustrated by the causal pathway $Y_{t-1}\leftarrow Z_{t-1} \leftarrow \epsilon_{t-1}\to Z_{t}$ in Figure \ref{fig:dag}. To deal with such exogenous predictors, we employ historical observations $\{Y_{t-q-j}\}_{j=1}^{p}$ as instrumental variables \citep{angrist1996identification}, which are uncorrelated with $Z_{t}$ to construct unbiased estimating equations. Specifically, by multiplying these historical observations on both sides of \eqref{eq:controlledARMA} and taking the expectation, we obtain the following Yule-Walker equations:
\begin{equation}\label{eqn:YW}
\begin{cases}
   \Mean(Y_{t} Y_{t-q-1})=\mu  \Mean (Y_{t-q-1}) + \sum_{j=1}^{p} a_j  \Mean (Y_{t-j} Y_{t-q-1})+b \Mean (U_{t} Y_{t-q-1}),\\
      \Mean (Y_{t} Y_{t-q-2})=\mu \Mean (Y_{t-q-2})+ \sum_{j=1}^{p} a_j  \Mean (Y_{t-j} Y_{t-q-2})+b \Mean (U_{t} Y_{t-q-2}),\\
      \qquad\qquad\vdots\\
    \Mean (Y_{t} Y_{t-q-p})=\mu \Mean (Y_{t-q-p})+ \sum_{j=1}^{p} a_j \Mean (Y_{t-j} Y_{t-q-p})+b \Mean (U_{t} Y_{t-q-p}).
\end{cases}
\end{equation}
It yields $p$ equations, but we have $p+2$ parameters to estimate, including $p$ autoregressive coefficients, a control parameter, and an intercept. In light of our concentration on observation-agnostic designs, under which each treatment is independent of the residual process,  we further multiply $U_{t}$ and $1$ on both sides of model \eqref{eq:controlledARMA} and take the expectation, leading to: 
\begin{eqnarray}\label{eqn:YW2}
   \begin{aligned}
    \Mean (Y_{t} U_{t}) & =\mu \Mean (U_{t}) +\sum_{j=1}^p a_j \Mean(Y_{t-j} U_{t})+b,\\
     \Mean (Y_t)&=\mu +\sum_{j=1}^p a_j \Mean(Y_{t-j})+b  \Mean(U_t).
   \end{aligned}
\end{eqnarray}
We next replace the expectations in \eqref{eqn:YW} and \eqref{eqn:YW2} by their sample moments from $t=p+q+1$ to $T$ and construct $p+2$ estimating equations. Subsequently, we solve these equations to obtain the Yule-Walker estimators $\{\widehat{a}_j\}_j$ and $\widehat{b}$ for $\{a_j\}_j$ and $b$, respectively, by which we construct the following  estimator for ATE:
\begin{equation}\label{eqn:ATEest}
    \widehat{\textrm{ATE}}=2\widehat{b}/(1-\sum_{j=1}^p \widehat{a}_j).
\end{equation}
By definition, the asymptotic property of \eqref{eqn:ATEest} depends on those of $\{\widehat{a}_j\}_{j=1}^p$ and $\widehat{b}$. However, deriving their asymptotic variances is extremely challenging, and no closed-form expressions are available to the best of our knowledge. To establish the ATE estimator's asymptotic MSE, we introduce a small signal asymptotic framework, 
detailed in the next section.  

{\singlespacing
\subsection{Small Signal Asymptotics, MSEs of ATE Estimators, and Efficiency Indicators}\label{subsec:smallsignal} 
}
We propose a small signal asymptotic framework to simplify the theoretical analysis in the ATE estimator with two key conditions:
\begin{itemize}[leftmargin=*]
    \item \textbf{Large sample}. The first condition is the conventional large sample condition, which requires the sample size $T$ to grow to infinity. In our ride-sharing example, most experiments last for two weeks, divided into 30-minute or 1-hour intervals, resulting in $T=672$ or $336$ time units. 
    \item \textbf{Small signal}. The second condition, which we introduce, requires the absolute value of the ATE to diminish to zero. This is consistent with our empirical observations, where improvements from new strategies typically range only from $0.5\%$ to $2\%$.
\end{itemize}

Next, an application of the Delta method \citep{oehlert1992note} to \eqref{eqn:ATEest} leads to
\begin{equation}\label{eqn:ATE_delta}
    \widehat{\textrm{ATE}}-\textrm{ATE}=\frac{2(\widehat{b}-b)}{1-a}+\frac{2b}{(1-a)^2}\sum_{j=1}^p(\widehat{a}_j-a_j)+o_p(T^{-1/2}).
\end{equation}
Under the first large sample condition, the third term in \eqref{eqn:ATE_delta} -- a high-order reminder term -- becomes negligible. As such, the first two terms, which measure the discrepancies between the Yule-Walker estimators and their oracle values, become the leading terms. However, as mentioned earlier, deriving their asymptotic variances remains extremely challenging under partial observability. 

The second small signal condition further simplifies the calculation in two ways: (i) First, it is immediate to see that the second term is proportional to ATE. Under this condition, the second term also becomes negligible as the ATE decays to zero. The first term, therefore, becomes the sole leading term, and it suffices to calculate the asymptotic variance of the estimated control parameter. (ii) Under this condition, the influence of the treatment on the observation becomes marginal. Consequently, the sequence of treatments becomes asymptotically independent of the sequence of observations, facilitating the derivation of the asymptotic variance of $\widehat{b}$. The following theorem summarizes our findings. 
\begin{thm}\label{them:asymptotic}
	Given an \textit{observation-agnostic} design with its treatment allocation strategy $\pi$, let $\xi_{\pi}= \lim_{t\to\infty}\mathbb{E}(U_t)$. Under Assumption \ref{assump:ergodicity} and the small signal asymptotics with  $T\to +\infty$ and ATE $\to 0$, the ATE estimator under $\pi$, denoted by $\widehat{\textrm{ATE}}(\pi)$,  satisfies: 
			\begin{equation*}
			\begin{aligned}
					\lim_{\substack{T \rightarrow +\infty\\ \textrm{ATE}\to 0}} \text{MSE}( \sqrt{T} \widehat{\text{ATE}}(\pi))  = 	\lim_{T \rightarrow +\infty} \frac{4}{(1-a)^2(1-\xi_{\pi}^2)^2 T} \Var\Big[\sum_{t=1}^T (U_{t}-\xi_{\pi})Z_{t}\Big].
			\end{aligned}
		\end{equation*}
\end{thm}
The proof of Theorem \ref{them:asymptotic} is provided in Appendix~\ref{appendix:optimal}. Theorem \ref{them:asymptotic} may initially appear complex. To elaborate, we first narrow our analysis to the class of controlled AR models by setting $q=0$.  In this simplified scenario, the residuals become uncorrelated, and the data follows a $p$-th order Markov process. Such simplification leads to the following corollary.

\begin{cor}\label{cor:efficiencyAR}
Under the assumptions stated in Theorem \ref{them:asymptotic}, when $q=0$, we have:
\begin{equation*}
    \lim_{\substack{T \rightarrow +\infty\\ \textrm{ATE}\to 0}} \text{MSE}(\sqrt{T}\widehat{\text{ATE}}(\pi))=\frac{4\sigma^2}{(1-a)^2 (1- \xi_{\pi}^2)^2},
\end{equation*}
where, recall, $\sigma^2$ denotes the variance of the white noise $\epsilon_t$.
\end{cor}
According to Corollary \ref{cor:efficiencyAR}, the asymptotic MSE of the ATE estimator is determined by three factors: (i) the variance of the white noise; (ii) the autoregressive coefficients; and  (iii) $\xi_{\pi}$, which measures the percentage of time the new treatment is applied. 
Different designs affect the ATE's asymptotic variance only through $\xi_{\pi}$. In other words, designs with the same $\xi_{\pi}$ achieve the same statistical efficiency in estimating the ATE. This uniformity is due to the uncorrelated residuals in the AR model. Additionally, it turns out that any (asymptotically) balanced design with $\xi_{\pi}=0$ is optimal. This principle holds even when $q>0$, as detailed in Theorem \ref{thm:optimal} in Section \ref{sec:optimaldesign}. These observations align with the findings of \citet{xiong2023data}, highlighting the importance of balancing periodicity in switchback designs under a different model setup.

We now turn our attention to the general controlled ARMA model with $q > 0$. We focus on the three particular designs—AT, AD, and UR—introduced in Section \ref{subsec:data}, denoting their treatment assignment strategies as $\pi_{\text{AD}}$, $\pi_{\text{UR}}$, and $\pi_{\text{AT}}$, respectively. We derive the asymptotic MSEs of ATE estimators under these designs in the following corollary. By definition, it is evident that these three designs are balanced with $\xi_{\pi} = 0$. Specifically for the AD design, we additionally require the number of intervals per day $\tau$ to diverge to infinity as $T$ approaches infinity.
\begin{cor}\label{cor:efficiency_ARMA} 
 Under the assumptions stated in Theorem \ref{them:asymptotic}, we have the simplified asymptotic MSEs of the AT, UR, and AD designs as follows:
        \begin{equation}\label{eqn:cor1}
        \begin{aligned}
            \lim_{\substack{T \rightarrow +\infty\\ \tau\to +\infty}} \text{MSE}( \sqrt{T} \widehat{\text{ATE}}(\pi_{\text{AD}})) & =\frac{4\sigma^2}{(1-a)^2}\Big[ \sum_{j=0}^q \theta_j^2+2\sum_{k=1}^q \sum_{j=k}^q \theta_j\theta_{j-k} \Big],\\
            \lim_{\substack{T \rightarrow +\infty\\ \textrm{ATE}\to 0}} \text{MSE}(\sqrt{T}\widehat{\text{ATE}}(\pi_{\text{UR}})) & =\frac{4\sigma^2}{(1-a)^2} \sum_{j=0}^q \theta_j^2,\\
            \lim_{\substack{T \rightarrow +\infty\\ \textrm{ATE}\to 0}} \text{MSE}(\sqrt{T}\widehat{\text{ATE}}(\pi_{\text{AT}})) & =\frac{4\sigma^2}{(1-a)^2}\Big[ \sum_{j=0}^q \theta_j^2+2\sum_{k=1}^q (-1)^k \sum_{j=k}^q \theta_j\theta_{j-k} \Big].
        \end{aligned}
    \end{equation}
\end{cor}
The proof is provided in Appendix~\ref{appendix:efficiency_ARMA}. According to Corollary \ref{cor:efficiency_ARMA}, the statistical efficiency of the three designs is primarily determined by the second term on the RHS of \eqref{eqn:cor1}, which depends solely on the moving average coefficients $\{\theta_j\}_{j=1}^q$. As previously noted, these coefficients directly influence the correlation of residuals, which in turn affects the designs' efficiencies. Specifically: (i) When all $\theta_j$s are non-negative, it results in non-negatively correlated residuals, and thus AT typically outperforms AD. (ii) Conversely, when the majority of residuals are non-positively correlated, AD tends to outperform AT. These observations align with the findings in \citet{xiong2023data} and \citet{wen2024analysis}.  

Finally, Corollary \ref{cor:efficiency_ARMA}  motivates us to define two efficiency indicators $\textrm{EI}_{\textrm{AD}}=\sum_{k=1}^q \sum_{j=k}^q \theta_j\theta_{j-k}$ and $\textrm{EI}_{\textrm{AT}}=\sum_{k=1}^q (-1)^k \sum_{j=k}^q \theta_j\theta_{j-k}$. By \eqref{eqn:cor1}, it is immediate to see that
\begin{itemize}
    \item AD outperforms UR and AT if and only if $\textrm{EI}_{\textrm{AD}}<0$ and $\textrm{EI}_{\textrm{AD}} <\textrm{EI}_{\textrm{AT}}$;
    \item UR outperforms AD and AT if and only if both $\textrm{EI}_{\textrm{AD}}$ and $\textrm{EI}_{\textrm{AT}}$ are positive;
    \item AT outperforms UR and AD if and only if $\textrm{EI}_{\textrm{AT}}<0$ and $\textrm{EI}_{\textrm{AT}}<\textrm{EI}_{\textrm{AD}}$.
\end{itemize} 
These indicators are useful for comparing the three designs. In practice, one can estimate the moving average coefficients from historical or initial experimental data and plug these estimators into the indicators to determine the most effective design among the three. In Section \ref{sec:optimaldesign}, we discuss methodologies to search the optimal design within the broader class of observation-agnostic designs, beyond just these three. 

To conclude this section, we note that although our asymptotic derivation and the subsequently proposed designs rely on the small signal condition, our proposal remains effective even with large treatment effects. This is because in these scenarios, the design problem itself might not be that critical, and any reasonable design should be able to detect these effects. Therefore, our proposal remains a safe option to use, regardless of whether this assumption holds or not.

\subsection{Extensions}\label{subsec:VARMA}
In this section, we extend the univariate controlled ARMA model by accommodating multivariate observations and exogenous variables, 
derive asymptotic MSEs of the estimated ATEs, and propose the resulting efficiency indicators.
 
 
\noindent \textit{\textbf{Controlled VARMA($p, q$)}}. We define the controlled VARMA($p, q$) model with an additional exogenous variable $\bm{E}_t$ as:
    \begin{equation}\label{eqn:modelVARMA}
        \begin{aligned}
            \mathbf{Y}_{t}=\boldsymbol{\mu} + \sum_{j=1}^{p} \mathbf{A}_j \mathbf{Y}_{ t-j}+ \mathbf{b} U_{t} + \bm{C} \bm{E}_t + \mathbf{Z}_{t}\,\,\,\,\hbox{and}\,\,\,\,
            \mathbf{Z}_{t} = 
            \sum_{j=0}^q \mathbf{M}_j \bm{\epsilon}_{t-j},
        \end{aligned}
    \end{equation}
where the bold vectors $\boldsymbol{\mu}, \mathbf{Y}_{t}$, $\mathbf{Z}_{t}$ and $\bm{\epsilon}_{t}$ denote the $d$-dimensional intercept, observation, residual and the white noise, respectively. 
The treatment $U_t$ remains binary, taking values in $\{-1, 1\}$. 
The purpose of introducing the extra exogenous variable $\bm{E}_t$ is to further enhance model flexibility. This variable remains unaffected by the treatments and can be regarded as the ``non-stationary" components of the model, accounting for a broad range of temporal factors, such as the daily seasonal trends (see Section~\ref{sec:realdata} for the construction of $\bm{E}_t$).

Model \eqref{eqn:modelVARMA} contains four sets of parameters: (i) the autoregressive coefficient matrices $\mathbf{A}_1,\ldots,\mathbf{A}_p \in \mathbb{R}^{d \times d}$; (ii) the control coefficient vector $\mathbf{b} \in \mathbb{R}^{d}$; (iii) the moving average coefficient matrices $\mathbf{M}_1,\ldots,\mathbf{M}_q \in \mathbb{R}^{d \times d}$ and $\mathbf{M}_0 = \mathbb{I} \in \mathbb{R}^{d \times d} $ as an identity matrix; 
(iv) the coefficient matrix $\bm{C}$ for the extra exogenous variable.  

We next introduce the no unit root assumption for the VARMA model and derive the closed-form expression for the ATE using different treatment allocation strategies. 
\begin{assumption}[No unit root]\label{assump:ergodicV}
    All the roots of the determinant of the polynomial matrix $\mathbb{I}-\sum_{j=1}^p \mathbf{A}_j y^j$ lie outside the unit circle. 
\end{assumption}
\begin{lemma}\label{lemma:ATEV}
    Under Assumption \ref{assump:ergodicV}, ATE equals $ 2 \bm{e}^\top (\mathbb{I} - \mathbf{A})^{-1} \mathbf{b}$,  where $\mathbf{A}= \sum_{j=1}^{p} \mathbf{A}_j \in \mathbb{R}^{d \times d}$ and $\bm{e}=(1,0,0,\ldots,0)^\top \in \mathbb{R}^{d} $.
\end{lemma}
Motivated by Lemma \ref{lemma:ATEV}, we similarly employ the method of moments to estimate $\{\mathbf{A}_j\}_j$ and $\mathbf{b}$ and plug in these estimators to construct the ATE estimator $\widehat{\textrm{ATE}}$. To save space, we relegate the details to Appendix~\ref{appendix:efficiency_VARMA} in the Supplementary Material.

\noindent \textit{\textbf{Asymptotic MSEs and Efficiency Indicators}}. Next, we analyze the asymptotic MSE of the ATE estimator in controlled VAMRA($p, q$). The following theorem extends Theorem \ref{them:asymptotic} to accommodate multivariate observations.  
\begin{thm}\label{thm:ATEestV}
   Under Assumption \ref{assump:ergodicV} and the small signal asymptotic framework with $T\to +\infty$ and ATE $\to 0$, the ATE estimator under $\pi$,  denoted by $\widehat{\textrm{ATE}}(\pi)$, satisfies: 
			\begin{equation*}
			\begin{aligned}
					\lim_{\substack{T \rightarrow +\infty\\ \textrm{ATE}\to 0}} \text{MSE}( \sqrt{T} \widehat{\text{ATE}}(\pi))  = 	\lim_{T \rightarrow +\infty} \frac{4}{(1-\xi_{\pi}^2)^2 T} \bm{e}^\top (\mathbb{I} - \mathbf{A})^{-1} \Var\Big[\sum_{t=1}^T (U_{t}-\xi_{\pi})\mathbf{Z}_{t}\Big] (\mathbb{I} - \mathbf{A})^{-1} \bm{e}.
			\end{aligned}
		\end{equation*}
\end{thm}
Similar to Corollary \ref{cor:efficiency_ARMA}, we next present the asymptotic MSEs of $\widehat{\textrm{ATE}}$ under AD, AT, and UR designs in the following corollary to elaborate Theorem \ref{thm:ATEestV}. 
	\begin{cor}\label{cor:efficiency_VARMA}
Under the conditions stated in Theorem \ref{thm:ATEestV}, we have:
        \begin{equation}
            \begin{aligned}\nonumber
                \lim_{\substack{T \rightarrow +\infty\\ \tau\to +\infty}} \text{MSE}( \sqrt{T} \widehat{\text{ATE}}(\pi_{\text{AD}})) & = 4 \bm{e}^\top (\mathbb{I} - \mathbf{A})^{-1} \left( \sum_{j_1=0}^{q}\sum_{j_2=0}^{q}\mathbf{M}_{j_1} \bm{\Sigma} \mathbf{M}_{j_2} 
                \right)  (\mathbb{I} - \mathbf{A})^{-1} \bm{e},\\
                \lim_{\substack{T \rightarrow +\infty\\\textrm{ATE}\to 0}} \text{MSE}(\sqrt{T}\widehat{\text{ATE}}(\pi_{\text{UR}})) & = 4 \bm{e}^\top (\mathbb{I} - \mathbf{A})^{-1} \left( \sum_{j=0}^{q} \mathbf{M}_j \bm{\Sigma} \mathbf{M}_j  \right)  (\mathbb{I} - \mathbf{A})^{-1} \bm{e},\\
                \lim_{\substack{T \rightarrow +\infty\\\textrm{ATE}\to 0}} \text{MSE}(\sqrt{T}\widehat{\text{ATE}}(\pi_{\text{AT}})) & = 4 \bm{e}^\top (\mathbb{I} - \mathbf{A})^{-1} \left( \sum_{j_1=0}^{q} \sum_{j_2=0}^{q}(-1)^{|j_2-j_1|}\mathbf{M}_{j_1} \bm{\Sigma} \mathbf{M}_{j_2} \right)  (\mathbb{I} - \mathbf{A})^{-1} \bm{e},
            \end{aligned}
        \end{equation}
       where $\bm{\Sigma}$ denotes the covariance matrix of $\bm{\epsilon}_t$. 
\end{cor}
The proof of Theorem~\ref{thm:ATEestV} and Corollary~\ref{cor:efficiency_VARMA} are provided in Appendix~\ref{appendix:efficiency_VARMA} in the Supplementary Material. Under the multivariate setting, we define the efficiency indicators as 
    \begin{eqnarray*}
\begin{aligned}
     \textrm{EI}_{\text{AD}}&=\bm{e}^\top (\mathbb{I} - \mathbf{A})^{-1}  \sum_{k=1}^{q}\sum_{j=k}^{q} \mathbf{M}_j \bm{\Sigma} \mathbf{M}_{j-k} (\mathbb{I} - \mathbf{A})^{-1}\bm{e}~~~~\mbox{and} \\
    \textrm{EI}_{\text{AT}} &= \bm{e}^\top (\mathbb{I} - \mathbf{A})^{-1}  \sum_{k=1}^{q}\sum_{j=k}^{q} (-1)^k \mathbf{M}_j \bm{\Sigma} \mathbf{M}_{j-k} (\mathbb{I} - \mathbf{A})^{-1} \bm{e}.
\end{aligned}
\end{eqnarray*}
According to Corollary~\ref{cor:efficiency_VARMA}, 
they enable us to compare the statistical efficiency of the three designs in estimating the ATE in the controlled VARMA model. 


\section{Optimal Treatment Allocation Strategies}\label{sec:optimaldesign}

This section focuses on the optimal observation-agnostic design, where the ATE estimator derived from the experimental data achieves the smallest asymptotic MSE. Identifying the optimal design is computationally intractable. To elaborate, each observation-agnostic design is determined by a sequence of treatment allocation strategies $\pi = \{\pi_t\}_{t=1}^T$, where each $\pi_t$ specifies the conditional distribution of $U_t$ given $U_1, \ldots, U_{t-1}$.
Consider the class of deterministic treatment allocation strategies where each $\pi_t$ is a degenerate distribution. Since $U_t$s are binary, there are $2^t$ possible $\pi_t$ at each time point. Optimizing over such an exponentially growing number of strategies makes the problem NP-hard.

To address this challenge, we propose two solutions, detailed in Sections \ref{sec:optimal_markov} and \ref{sec:optimal_history}, respectively. Specifically, 
in Section \ref{sec:optimal_markov}, we restrict our attention to   Markov and stationary treatment allocation strategies and propose a constrained optimization algorithm to learn the resulting in-class optimal strategy. In Section \ref{sec:optimal_history}, we expand the search space to include general history-dependent policies and propose several optimality conditions to characterize the optimal treatment allocation strategy. These conditions significantly reduce the search space, making the computation feasible. We then develop an RL algorithm based on dynamic programming to learn the optimal treatment allocation strategy.

{\singlespacing
\subsection{A Constrained Optimization Approach}\label{sec:optimal_markov}
}
To simplify the computation, we restrict attention to the class of Markov and stationary treatment allocation strategies in our first approach, where each $\pi_t$ is a function of the most recently assigned treatment $U_{t-1}$ only and remains constant with respect to $t$. In A/B testing, this policy class can be parameterized using two parameters $0 \le \alpha, \beta \le 1$, such that: 
\[
\mathbb{P}(U_{t+1} = 1 | U_{t} = 1) = \alpha \quad \text{and} \quad \mathbb{P}(U_{t+1} = 1 | U_{t} = -1) = \beta.
\] 
By definition, both AT and UR are induced by policies within this class. Specifically, setting $\alpha = 0$ and $\beta = 1$ results in the AT design, whereas $\alpha = \beta = 1/2$ yields the UR design. When $\alpha = 1$, $\beta = 0$, and we alternate the initial treatment on a daily basis, it yields the AD design. This indicates the generality of the considered Markov and stationary policy class, which unifies the AD, UR and AT designs.

Additionally, the sequence $\{U_t\}_t$ forms a Markov chain with binary states. With some calculations, it can be shown that $\xi_{\pi} = (\alpha + \beta - 1) / (\beta + 1 - \alpha)$ in general. To obtain a balanced design, we set $\beta = 1 - \alpha$, leading to $\xi_{\pi} = 0$. 
It remains to identify the optimal $\alpha$ to minimize the asymptotic MSE of resulting the ATE estimator, 
which --- under the small signal asymptotic framework --- can be derived  as 
		\begin{equation}\label{eqn:someequation}
			\begin{aligned}
    4\Big[ 
    c_0+2\sum_{k=1}^q  c_k  (2\alpha - 1)^k  
    \Big],
			\end{aligned}
		\end{equation}
where $c_0 = {\sum_{j=0}^q \theta^2_j}/{(1-a)^2}$ and $c_k={\sum_{j=k}^q \theta_j \theta_{j-k} }/{(1-a)^2}$ under the controlled ARMA($p, q$) model, while under the controlled VARMA($p, q$) model we have
	\begin{equation}
		\begin{aligned}
				c_0 & = \bm{e}^\top (\mathbb{I} - \mathbf{A})^{-1} \left( \sum_{j=0}^{q}  \mathbf{M}_j \Sigma \mathbf{M}_{j} \right)  (\mathbb{I} - \mathbf{A})^{-1} \bm{e}, \\
                c_k &= \bm{e}^\top (\mathbb{I} - \mathbf{A})^{-1} \left( \sum_{j=k}^{q}  \mathbf{M}_j \Sigma \mathbf{M}_{j-k} \right)  (\mathbb{I} - \mathbf{A})^{-1} \bm{e}.
			\end{aligned}
		\end{equation}
See Appendix~\ref{appendix:optimal_markov} for more details. This asymptotic MSE formula motivates us to compute $\alpha$ by solving the following constrained $q$-order polynomial optimization:
		\begin{equation}
		\begin{aligned}\label{eq:optimal_markov_optimization}
				\min_{\alpha} \sum_{k=1}^{q} c_k (2\alpha-1)^k, \quad \text{s.t.} \ \alpha \in [0, 1].
			\end{aligned}
		\end{equation}
The above optimization can be efficiently solved using existing convex optimization techniques, such as the limited-memory Broyden-Fletcher-Goldfarb-Shanno~(L-BFGS) algorithm~\citep{liu1989limited}. Notice that $c_k$ and the optimal number of AR and MA lags, 
$p$ and $q$, depend on the true model, which are typically unknown. However, as discussed in Section \ref{subsec:smallsignal}, they can be effectively estimated or evaluated using historical data in practice. For instance, the optimal $p$ and $q$ can be selected based on the Akaike information criterion~\citep[AIC,][]{akaike1974new} or the Bayesian information criterion~\citep[BIC,][]{schwarz1978estimating}. Alternatively, this procedure can be applied sequentially: use current experimental data up to a specific day to learn $\{c_k\}_k$, $p$ and $q$ to estimate the optimal design. Then, this design will be applied on the following day, and the estimating procedure will continue by incorporating data from the subsequent day.

{\singlespacing
\subsection{A Reinforcement Learning Approach}\label{sec:optimal_history}
}
In this section, we consider the more general history-dependent policy class and propose an RL algorithm to identify the {\textit{unrestricted}} optimal treatment allocation strategy $\pi^*$. The primary objective of RL is to learn an optimal policy, a mapping from time-varying environmental features (referred to as state) to decision rules about which treatment to administer (referred to as action), in order to maximize the expected cumulative outcome (where each intermediate outcome is referred to as a reward). Most existing RL algorithms estimate the optimal policy by modeling these state-action-reward triplets over time as an MDP, wherein each reward and future state are independent of the past history given the current state-action pair. 

We begin by providing an optimality condition in Theorem \ref{thm:optimal} below to characterize 
$\pi^*$. 

\begin{thm}\label{thm:optimal}
    Under Assumption 
    \ref{assump:ergodicV} and the small signal asymptotic framework,
    there exists some $\pi^*$ that satisfies the following five conditions, under which the ATE estimator achieves the smallest MSE asymptotically:
    \begin{enumerate}
        \item \textbf{Balanced}: $\xi_{\pi^*}=0$;
        \item \textbf{Deterministic}: $\pi^*$ is deterministic;
        \item \textbf{Stationary}: $\pi_t^*$ is time-homogeneous, which is independent of $t$ for any $t> q$;
        \item \textbf{$q$-dependent}: $\pi_t^*$ depends on the past treatment history only through the most $q$ recent treatments $U_{t-1},\ldots,U_{t-q}$;
        \item \textbf{Optimal}: 
        The treatment sequence $\{U_t\}_{t}$ generated by $\pi^*$ must minimize 
        \begin{eqnarray}\label{eqn:obj}
       \begin{aligned}
             \lim_{T\to \infty}\sum_{k=1}^q c_k\Big[\frac{1}{T-q} \sum_{t=1}^{T-q}\mathbb{E}(U_t U_{t+k}) \Big], 
       \end{aligned}
        \end{eqnarray}
        where $c_k$ is defined in \eqref{eqn:someequation} under the controlled (V)ARMA($p, q$) model. 
    \end{enumerate}
\end{thm}
We defer the proof of Theorem~\ref{thm:optimal} to Appendix~\ref{appendix:optimal} and make a few remarks: (i) 
Corollary \ref{cor:efficiencyAR} in Section \ref{subsec:smallsignal} proves the optimality of balanced designs for AR processes. Theorem \ref{thm:optimal} extends this to (V)ARMA processes, allowing residuals to be correlated over time.
 (ii)  
 The determinism, stationarity, and $q$-dependency conditions significantly reduce the search space from over $2^{T}$ to less than $2^{q+1}$, simplifying the learning of $\pi^*$. These conditions enable us to focus on this restricted class to find $\pi^*$ by minimizing \eqref{eqn:obj}. 
(iii) 
The proof of Theorem \ref{thm:optimal} draws from existing proofs establishing the Markov and stationarity properties of the optimal policy in RL \citep[see, e.g.,][]{puterman2014markov, ljungqvist2018recursive}. A crucial step in our proof is to construct an MDP and establish the equivalence between learning the optimal policy that maximizes the average reward in this MDP and identifying the optimal treatment allocation strategies that minimize \eqref{eqn:obj}.
To elaborate, we introduce the following sequence of state-action-reward triplets $(\mathbf{S}_t, A_t, R_t)_{t > q}$:
\begin{itemize}
    \item \textbf{State}: $\mathbf{S}_t=(U_{t-1},\ldots,U_{t-q})^\top$, representing the most recently assigned $q$ treatments;
    \item \textbf{Action}: $A_t=U_t$, indicating which treatment to assign at each time; 
    \item \textbf{Reward}: $R_t= - \sum_{k=1}^q c_k U_t U_{t-k}$, designed according to  \eqref{eqn:obj}.
\end{itemize}
Both the future state $\mathbf{S}_{t+1}$ and the immediate reward $R_t$ are functions of $\mathbf{S}_t$ and $A_t$ only, satisfying the MDP assumption. The expected average reward in this MDP aligns with the objective function in \eqref{eqn:obj}. Consequently, the optimal treatment allocation strategies satisfying \eqref{eqn:obj} are equivalent to the optimal policies under this MDP. In RL, the optimal policy is a fixed function of the current state-action pair, proving that the optimal treatment allocation strategy is deterministic, $q$-dependent, and stationary over time.

To identify $\pi^*$ that satisfies the conditions in Theorem \ref{thm:optimal}, we utilize RL as a computational tool to optimize \eqref{eqn:obj}. Specifically, we construct the MDP above and apply dynamic programming to derive the optimal treatment allocation strategy. While an exhaustive policy search might be feasible when \(q\) is small, our RL approach is more computationally efficient in settings with a large \(q\). 
\begin{algorithm}[t]\small
		\caption{Value Iteration for Optimal $q$-dependent Treatment Allocation Stategy}
		\begin{algorithmic}[1] 
			\STATE Initialize the value function $V(s): \{-1, 1\}^q \rightarrow \mathbb{R}$ for all $s \in \mathcal{S}=\{-1, 1\}^q$. Set 
            a small tolerance level $\Delta_0>0$, a large $\Delta$, and a large discount factor $\gamma$ that is close to 1.
			\WHILE{$\Delta > \Delta_0$}
			\STATE $\Delta \leftarrow 0$
			\FOR{each $s = (a_1, \ldots, a_q) \in \mathcal{S}$}
			\STATE$v \leftarrow V(s)$
			\STATE $r \leftarrow -\sum_{k=1}^{q}c_k a \cdot a_k$ for each action $a \in \{-1, 1\}$.
			\STATE $s^\prime \leftarrow \{a\} \cup \{ s \setminus \{a_q\}\}$ for each action $a \in \{-1, 1\}$.
			\STATE$V(s) \leftarrow \max _a  \left(r+\gamma V\left(s^{\prime}\right)\right)$
			\STATE $\Delta \leftarrow \max (\Delta,|v-V(s)|)$
			\ENDFOR
			\ENDWHILE   
			\STATE Output the policy $\pi^*$, such that $\pi^*(s)=\arg \max _a \sum_{s^{\prime}, r} \left(r+\gamma V\left(s^{\prime}\right)\right)$.
		\end{algorithmic}
		\label{alg:optimal_history}
	\end{algorithm} 
We apply the value iteration algorithm \citep{sutton2018reinforcement} for policy learning; refer to Algorithm~\ref{alg:optimal_history} for its pseudocode. The main idea is first to learn an optimal value function $V(s)$, which represents the maximum expected return starting from a given state $s$, and then derive the optimal policy as the greedy policy with respect to this value function (see Line 12 of Algorithm~\ref{alg:optimal_history}). Value iteration updates the value function iteratively using the Bellman optimality equation (see Line 8 of Algorithm~\ref{alg:optimal_history}) until the changes in the estimated value function are below a predefined small threshold (see Line 9 of Algorithm~\ref{alg:optimal_history}), indicating convergence.

	\section{Experiments}\label{sec:experiments}

We demonstrate the finite sample performance of our proposed methods using two dispatch simulators~\citep{xu2018large, tang2019deep} and two 
real datasets from a ride-sharing company. Importantly, the two simulators used in Sections~\ref{sec:exp_simulator_small} and \ref{sec:exp_simulator_large} are based on physical models that simulate the behaviors of drivers and passengers. Therefore, the data generated from these environments does not necessarily follow the proposed controlled (V)ARMA model, providing a robust evaluation of our proposal under model misspecification. 

Our objectives are to (i) validate the effectiveness of the proposed efficiency indicators in comparing AD, UR and AT; (ii) conduct comparisons among the following designs:
\begin{itemize}[leftmargin=*]
    \item The proposed optimal designs via constrained optimization (denoted by \textbf{CO}) and \textbf{RL};
    \item The commonly used \textbf{AD}, \textbf{UR}, and \textbf{AT} designs;
    \item The $\epsilon$-greedy design \citep[denoted by \textbf{Greedy}]{sutton2018reinforcement}, which selects the current best treatment by maximizing an estimated Q-function with probability $1-\epsilon$, and switches to a uniform random policy over the two treatments with probability $\epsilon$;
    \item The \textbf{TMDP} and \textbf{NMDP} designs \citep{li2023optimal}, derived under the assumption that the system follows a time-varying MDP and a non-MDP, respectively;
    \item The optimal switchback design \citep[denoted by \textbf{Switch}]{bojinov2023design}.
\end{itemize}

We note that \textbf{Greedy} is commonly used in online RL for regret minimization. \textbf{TMDP} and \textbf{NMDP} are variants of AD designs that are proven to be optimal under their model assumptions. Finally, \textbf{Switch} is a variant of AT design that switches back and forth over a fixed period rather than at every decision point. The optimal duration of each switch is determined by the order of the carryover effect, and we select the best duration from $\{2,5,10\}$ to report.

\subsection{Synthetic Dispatch Simulator}\label{sec:exp_simulator_small}
	
\noindent \textit{\textbf{Environment.}} We simulate a synthetic ride-sharing environment as in~\citet{xu2018large} and \citet{li2023optimal}, where drivers and customers interact within a 9$\times$9 spatial grid over 20 time steps per day:
\begin{itemize}[leftmargin=*]
    \item \textbf{Orders}.
     We generate 50 orders per day. To simulate realistic traffic conditions with morning and evening peaks, we set their starting locations and calling times as i.i.d. drawn from a truncated two-component mixture of Gaussian distributions. This configuration strategically places the starting locations in two main areas -- representing customers' living and working areas -- and aligns the calling times with the morning and evening peak traffic hours. The destinations of these orders are uniformly distributed across all spatial grids. Each order is canceled if it remains unassigned to any driver for a long time, with customer waiting times until cancellation generated from another truncated Gaussian distribution.
    \item \textbf{Drivers}. We simulate 50 drivers, with their initial locations i.i.d. uniformly distributed over the $9\times 9$ grid. At each time, each driver is either dispatched to serve a customer or remains idle in their current location according to a given order dispatching strategy.
    \item \textbf{Policies}. We compare two order dispatching policies: (i) a conventional distance-based policy that matches idle drivers with unassigned orders by minimizing their total distances at each time, and (ii) an MDP-based policy that solves the matching problem by maximizing the long-term benefits of the ride-sharing platform rather than focusing on total distances at each current time \citep{xu2018large}. 
\end{itemize}

\smallskip
\noindent \textit{\textbf{Implementation.}} The outcome of interest is set to the driver's income earned at each time step. In addition to this outcome, we include two other variables in the observation: the number of unassigned orders and the number of idle drivers each time. Implementing both the proposed efficiency indicators and designs requires estimating the AR and MA parameters. To this end, we first generate a historical dataset that lasts for 50 days. 
Next, we apply the VARMA model to fit this dataset to estimate the AR and MA parameters. The optimal AR and MA orders, $p^*$ and $q^*$, are selected using AIC, resulting in $p^*=q^*=2$. Using these estimators, we compute the proposed efficiency indicators and proceed to implement the proposed designs, comparing them against other previously mentioned designs. Specifically, for each design, we generate 50 days of experimental data 
to estimate the ATE. Finally, we repeat the entire procedure 30 times to compute the MSE of the ATE estimator under each design. The oracle ATE is evaluated via the Monte Carlo method, resulting in a value of 2.24, leading to a 6\% improvement.  

\begin{figure*}[t!]
		\centering
		\begin{subfigure}[t]{0.41\textwidth}
			\centering
			\includegraphics[width=\textwidth,trim=10 37 10 10,clip]{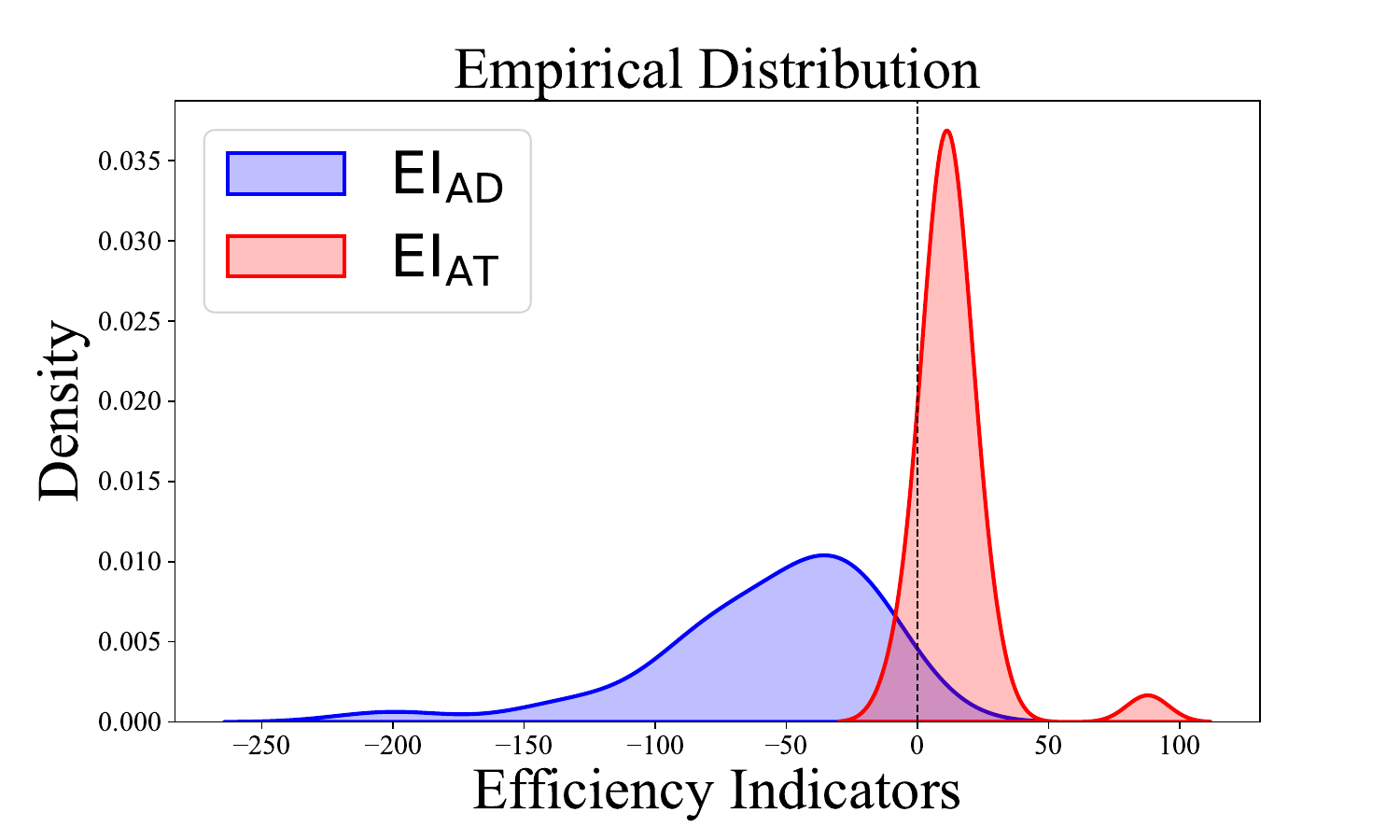}
			\caption{Efficiency indicators}
		\end{subfigure}
		\begin{subfigure}[t]{0.42\textwidth}
			\centering
			\includegraphics[width=\textwidth,trim=15 120 10 12,clip]{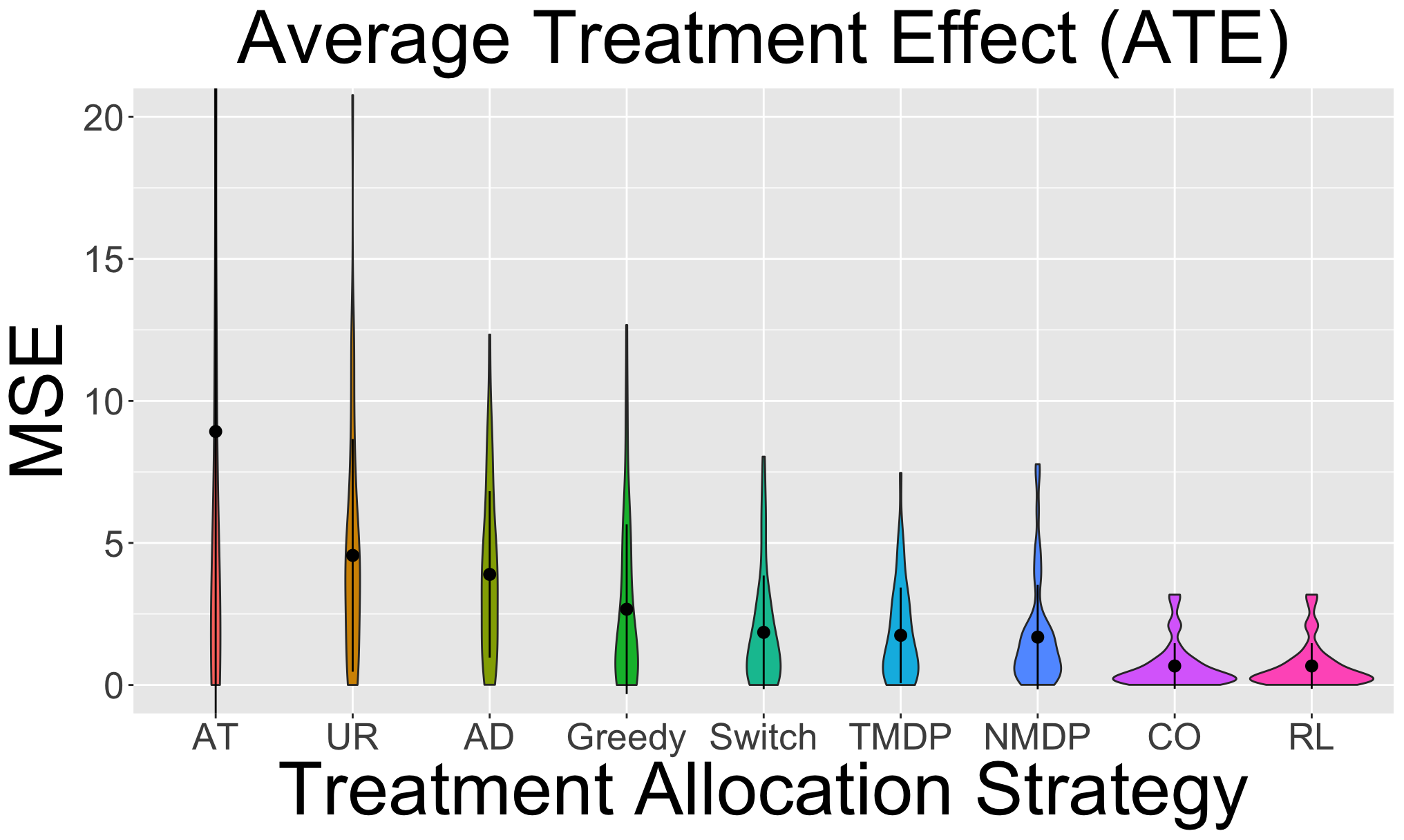}
			\caption{MSE under different designs}
		\end{subfigure}
		\caption{\small (a) The kernel density estimation (KDE) plot for the empirical distribution of the two efficiency indicators $\text{EI}_{\text{AD}}$ and $\text{EI}_{\text{AT}}$.  (b) Violin plots of MSEs of ATE estimators under different designs in environments created by the synthetic simulator. \textbf{Designs are ranked in descending order from left to right according to their average MSEs in estimating the ATE}. The positions of the middle points in each violin denote the average MSEs, while the lengths of violins reflect the variabilities.}\label{fig:experiment1_sim}
\end{figure*}
 
	\begin{table}[b!]
    \centering
    \scalebox{0.9}{
        \begin{tabular}{c|ccccccc|cc}
            \toprule[1pt]
            \textbf{Designs} & AT & UR & AD & Greedy & Switch & TMDP  & NMDP & CO & RL    \\
            \hline
            \textbf{Average MSE} & 8.92 & 4.56 & 3.89 & 2.67 & 1.85 & 1.75 & 1.69 & \bf{0.67} & \bf{0.67}  \\
            \hline
        \end{tabular}
    }
    \caption{\small Average MSEs under different designs in environments created by the synthetic simulator. }
    \label{table:experiment1}
\end{table}

\noindent \textit{\textbf{Results}}. We visualize the efficiency indicators 
and the MSEs of ATE estimators under different designs in Figure \ref{fig:experiment1_sim}. The values of these MSEs are detailed in Table \ref{table:experiment1}. The results are summarized as follows:

\begin{itemize}[leftmargin=*]
    \item \textbf{Efficiency Indicators}. As shown in Figure \ref{fig:experiment1_sim}(a), most of the estimated $\text{EI}_{\text{AD}}$ (colored in blue) are negative across 50 replications, while most estimated $\text{EI}_{\text{AT}}$ (in red) are positive. According to Corollary \ref{cor:efficiency_VARMA}, this pattern suggests that the AD design is likely more efficient than AT and UR in this simulation environment. Figure~\ref{fig:experiment1_sim}(b) and Table \ref{table:experiment1} further verify this finding, showing that both AT and UR result in significantly higher MSEs in estimating the ATE compared to AD. These findings highlight the effectiveness of the proposed efficiency indicators when comparing the three designs.
    \item \textbf{Designs}. As seen in Figure \ref{fig:experiment1_sim}(b) and Table \ref{table:experiment1}, our proposed CO and RL designs lead to the most efficient ATE estimators. 
    Meanwhile, TMDP and NMDP outperform the commonly used AT, UR, AD, Greedy, and Switch but are inferior to our proposed designs. Although Greedy is effective in online experiments for regret minimization by balancing the exploration-exploitation trade-off, it does not necessarily optimize the performance of the resulting ATE estimator. 
\end{itemize}

\subsection{City-level Real-data-based Dispatch Simulator}\label{sec:exp_simulator_large}

\noindent \textit{\textbf{Environment}}. We further conduct A/B testing 
by using a more complicated and realistic city-level order dispatching simulator~\citep{tang2019deep}. To mimic real-world ride-sharing markets, this simulator is trained based on a historical dataset collected from a world-leading ride-sharing company in a particular city. We do not disclose the names of the cities or the company for privacy concerns. 
Compared with the $9\times 9$ synthetic simulator in Section~\ref{sec:exp_simulator_small}, this dispatch simulator is more realistic in the following ways:
\begin{enumerate}[leftmargin=*]
    \item Drivers and customers interact within a real city divided into 85 hexagonal regions, as opposed to a synthetic city with grid-based rectangular regions.
    \item Orders are generated based on historical data rather than being synthetically simulated. The order dispatching policy matches existing unassigned and new orders every 2 seconds, aligning with the company's current practice. 

    \item Drivers are initially distributed according to their empirical distribution in the historical dataset rather than uniformly randomly distributed. Additionally, drivers assigned to orders have the option to reject them, with rejection probabilities computed by a pre-trained classification model that uses driver and order characteristics as features. Meanwhile, idle drivers may either relocate based on a random walk model trained using their historical movement data or follow the company's instructions to move to specific locations as determined by a pre-trained repositioning algorithm. Finally, idle drivers could go offline before the next order dispatching round, while new drivers may appear online, according to historical data. 
\end{enumerate}
Similar to Section~\ref{sec:exp_simulator_small}, the observation 
in this city-scale simulator 
is also three-dimensional, including the number of orders, the number of drivers, and the 
driver income, which is the outcome of interest. 
For each design, we conduct the online experiments over four days 
to estimate the ATE 
and replicate the experiment 30 times to calculate its root MSE. 

\begin{table}[b!]
    \centering
    \scalebox{0.9}{
    
        \begin{tabular}{c|ccccccc|cc}
        \toprule[1pt]
            \textbf{Designs} & AT & UR & Greedy & AD & Switch & NMDP  & TMDP & CO & RL    \\
            \hline
            \textbf{Average RMSE$(\times10^4)$} & 29.9 & 26.9 & 26.2 & 7.5 & 7.3 & 5.9 & 5.1 & \textbf{2.3} & 2.6  \\
            \hline
        \end{tabular}
    }
    \caption{\small Average RMSEs under different designs in environments created by the city-level real-data-based simulator. }
    \label{table:experiment_largesimulator}
\end{table}

\begin{figure*}[t!]
		\centering
		\begin{subfigure}[t]{0.44\textwidth}
			\centering
			\includegraphics[width=\textwidth,trim=10 38 10 10,clip]{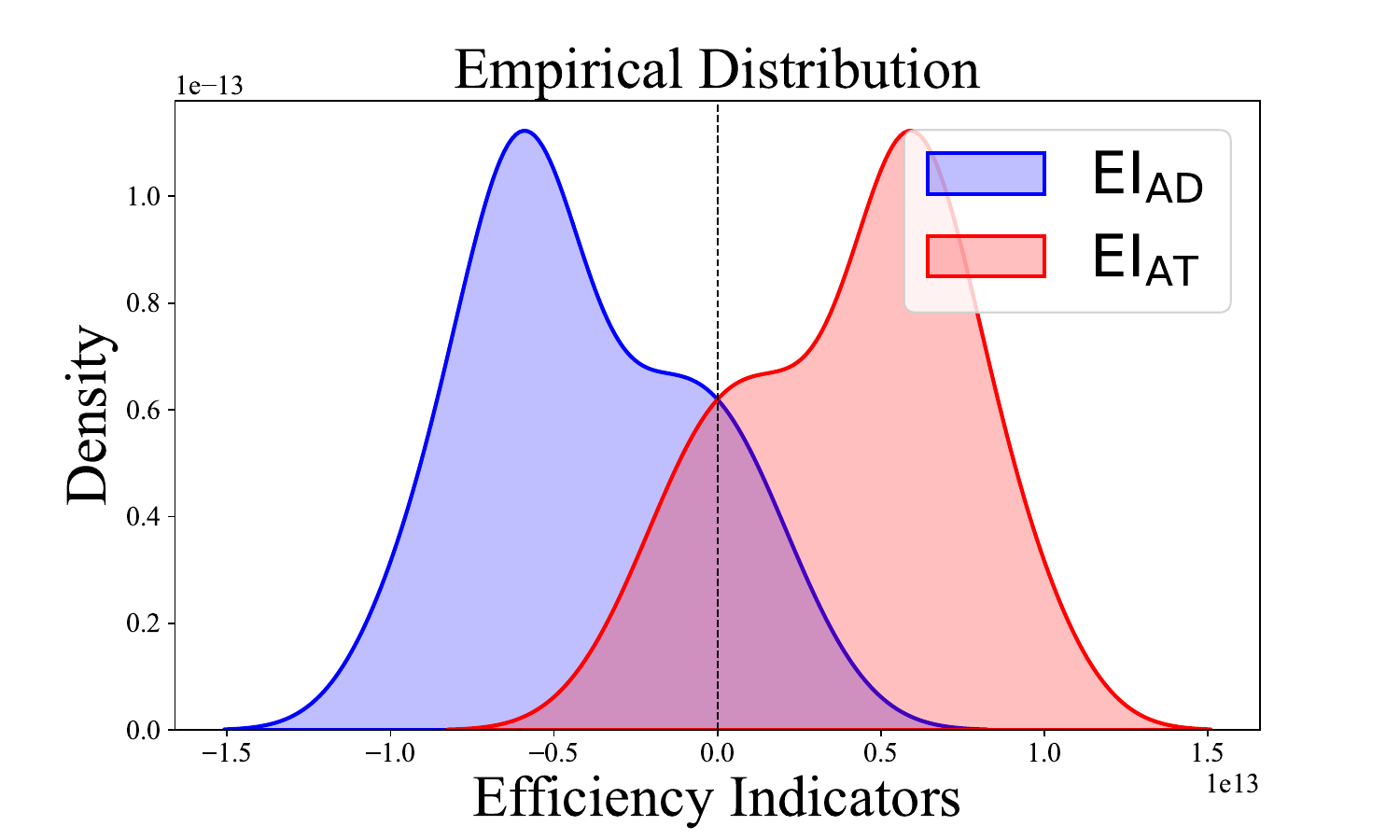}
			\caption{Efficiency indicators}
		\end{subfigure}
		\begin{subfigure}[t]{0.38\textwidth}
			\centering
			\includegraphics[width=\textwidth,trim=5 58 10 5,clip]{F4_Simulator_violin.pdf}
			\caption{RMSE under different designs}
		\end{subfigure}
		\caption{\small (a) The KDE plot for the empirical distribution of the two efficiency indicators $\text{EI}_{\text{AD}}$ and $\text{EI}_{\text{AT}}$.  (b) Violin plots of the MSEs of ATE estimators under different designs in environments created by the city-level real-data-based simulator.}
		\label{fig:experiment3_simulator}
\end{figure*}

\noindent \textit{\textbf{Results}}. The empirical distribution of efficiency indicators and the root MSE (RMSE) of ATE estimators under different 
designs are visualized in Figure~\ref{fig:experiment3_simulator}(a) and (b), respectively. Additionally, the values of these RMSEs are reported in Table~\ref{table:experiment_largesimulator} as well. We summarize the results as follows.

\begin{itemize}[leftmargin=*]
    \item \textbf{Efficiency Indicators.} Figure~\ref{fig:experiment3_simulator}(a) suggests that the estimated $\textbf{EI}_{\text{AD}}$ values are mostly negative across the 30 replications, whereas the estimated $\textbf{EI}_{\text{AT}}$ values are mostly positive. 
    This suggests that AD is more efficient than AT and UR in this environment, which aligns with the results reported in Figure~\ref{fig:experiment3_simulator}(b) and Table~\ref{table:experiment_largesimulator}. 

    \item \textbf{Designs.} Figure~\ref{fig:experiment3_simulator} and Table~\ref{table:experiment_largesimulator} demonstrate the superiority of 
    our proposed CO and RL designs, 
    which achieves the lowest MSE among all considered designs. As mentioned earlier, both the synthetic simulator in Section \ref{sec:exp_simulator_small} and the real-data-based simulator in this section are built based on physical models to simulate driver and customer behaviors. Even though the data from the two dispatch simulators might not follow the proposed model, our designs consistently deliver the best performance. This outperformance demonstrates the robustness of our designs against model misspecification, enabling more accurate A/B testing than existing state-of-the-art in real practice.

\end{itemize}

\subsection{Real Data-based Analyses}\label{sec:realdata}

\begin{figure*}[t!]
    \centering
    \begin{subfigure}[t]{0.75\textwidth}
        \centering
    \includegraphics[width=\textwidth,trim=0 0 0 0,clip]{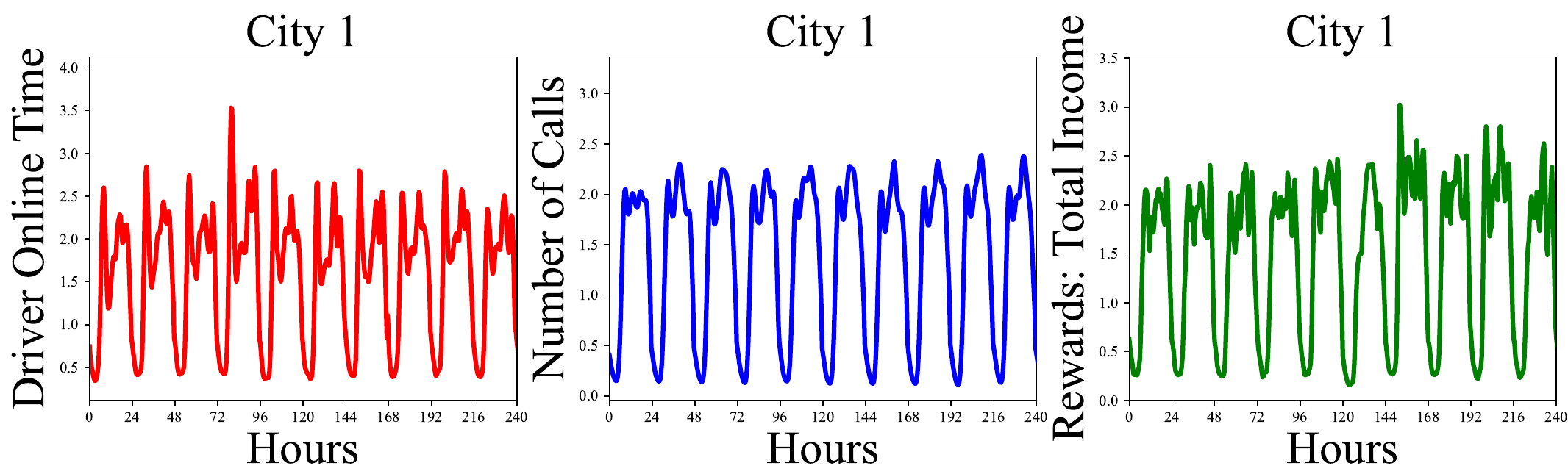}
    \end{subfigure}
    \begin{subfigure}[t]{0.75\textwidth}
        \centering
    \includegraphics[width=\textwidth,trim=0 0 0 0,clip]{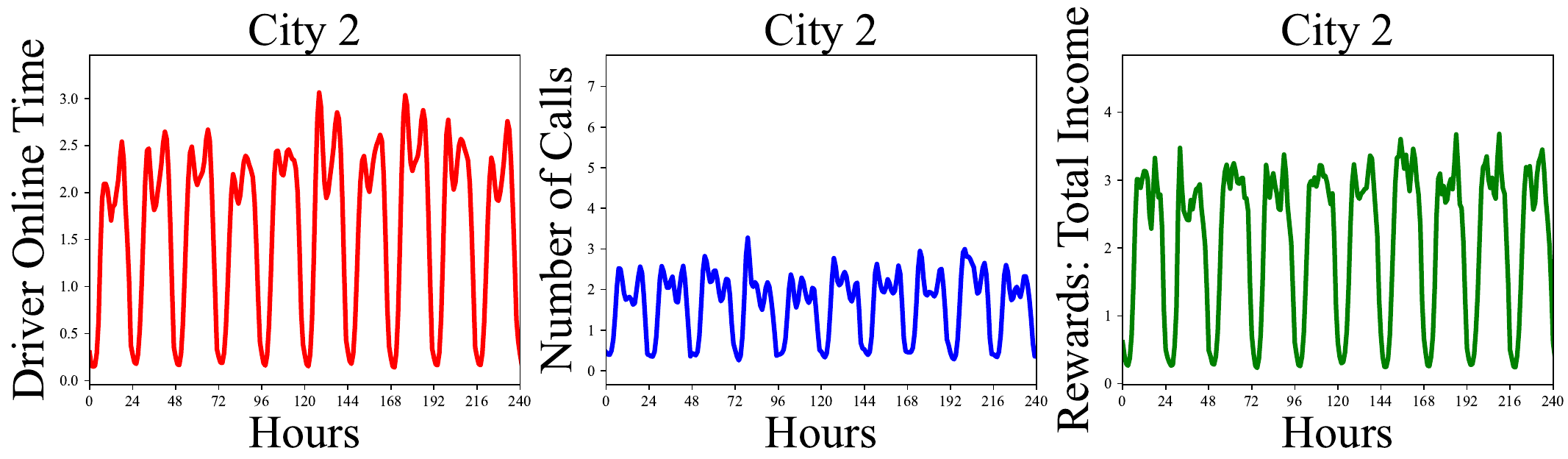}
    \end{subfigure}
    \caption{\small Trend of observations: driver online time, number of calls, and reward~(the drivers' income) on City 1~(the first row) and City 2~(the second row) across 240 hours~(10 days).}
    \label{fig:experiment2_city}
\end{figure*}

\noindent \textit{\textbf{Data.}} 
We use two real datasets from two different cities, provided by the ride-sharing company, to create simulation environments for investigating the finite sample performance of the proposed efficiency indicators and designs. Both datasets are generated under A/A experiments, where a single order dispatching strategy is consistently deployed over time. Each dataset contains 40 days of data and is summarized as a three-dimensional time series. The first dimension records the drivers' total income at each time interval, serving as the outcome. The last two elements are the number of order requests and drivers' online time at each time interval, respectively, measuring the demand and supply of the market. The time units in the datasets differ, with the first being 30 minutes and the second being one hour. See Figure \ref{fig:experiment2_city} for visualizations of these three-dimensional time series.

\noindent \textit{\textbf{
Bootstrap-based Simulation.}} 
Figure~\ref{fig:experiment2_city} reveals clear daily trends in both time series, with a significant rise and a subsequent decline in driver income and the number of call orders during the morning and evening peak hours. To effectively capture these seasonal patterns, we incorporate a dummy variable, $D_t$, as an exogenous variable in our controlled VARMA model to fit the three-dimensional observation. This variable is set to one during peak hours between 8 am to 8 pm and zero otherwise. 

Next, we employ the parametric bootstrap to create simulated data. Specifically, we first fit the following VARMA model 
    \begin{equation}
        \begin{aligned}\nonumber
            \mathbf{Y}_{t}=\bm{\mu}+\sum_{j=1}^{p} \mathbf{A}_j \mathbf{Y}_{t-j} + \boldsymbol{\eta} D_{t} + \mathbf{Z}_{t},
        \end{aligned}
    \end{equation}
and record all estimated parameters, i.e., $\widehat{\bm{\mu}}$, $\{\widehat{\mathbf{A}}_i\}_{i=1}^p$, $\widehat{\boldsymbol{\eta}}$, $\widehat{\bm{\Sigma}}$,  $\{\widehat{\mathbf{M}}_{i=1}^q \}$. 
Next, we simulate time series $\{	\widehat{\mathbf{Y}}_{t} \}$ according to the following equation:
    \begin{equation}
        \begin{aligned}\label{eqn:bootsimu}
            \widehat{\mathbf{Y}}_{t}=\widehat{\bm{\mu}} + \sum_{j=1}^{p} \widehat{\mathbf{A}}_j \widehat{\mathbf{Y}}_{t-j}+ b \mathbf{1}  U_{t} + \widehat{\boldsymbol{\eta}} D_{t} + \widehat{\mathbf{Z}}_{t},
        \end{aligned}
    \end{equation}
where $\mathbf{1}$ denotes a vector of ones, $b$ is some pre-specified parameter that determines the size of the ATE, $\{U_t\}_t$ are determined by different designs, and $\{\widehat{\mathbf{Z}}_{t}\}$ follow the estimated MA process and are generated prior to $\{\widehat{\mathbf{Y}}_{t}\}$.

\noindent \textit{\textbf{Evaluation and Results.}} For each design and each choice of $b$, we apply 
the bootstrap-based simulation to generate an experimental dataset. 
We next apply the controlled VARMA model to this experimental dataset to 
estimating the ATE and evaluating its MSE. We choose an appropriate range of $b$ for each city to ensure that the resulting ATE falls between 0.5$\%$ and 2$\%$, a range that aligns with our empirical observations \citep{tang2019deep}. 

\begin{figure*}[t!]
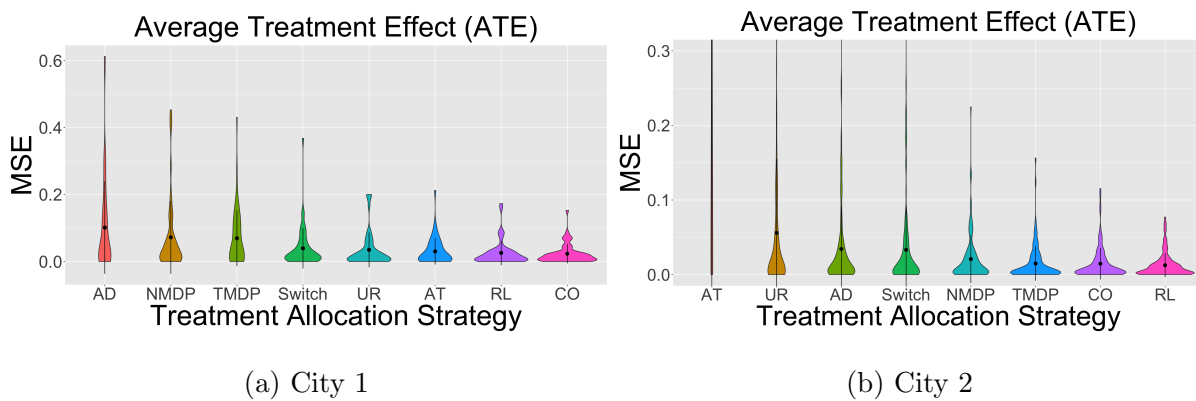

		\centering
		\begin{subfigure}[t]{0.41\textwidth}
			\centering
			\includegraphics[width=\textwidth,trim=0 145 10 10,clip]{F3_City_ATE1.pdf}
			\caption{City 1}
		\end{subfigure}
		\begin{subfigure}[t]{0.40\textwidth}
			\centering
			\includegraphics[width=\textwidth,trim=0 142 10 10,clip]{F3_City_ATE2.pdf}
			\caption{City 2}
		\end{subfigure}
		\caption{\small Violin plots of the MSEs of ATE estimators under different treatment allocation strategies in two cities. \textbf{The designs are ranked in descending order from left to right regarding average MSEs}. The positions of the middle points in each violin denote the mean, while the black solid lines indicate the standard deviation.}
    \label{fig:experiment2_ate}
\end{figure*}

Given that the magnitude of the estimated ATE and the associated MSE vary with $b$, averaging all MSEs across different values of $b$ may not accurately evaluate each design. To address this, we report a \textit{performance ranking} metric across the eight considered designs, which serves as a more robust measure alongside the average MSE. All results are summarized in Figure~\ref{fig:experiment2_ate} and Table~\ref{table:experiment2}.

\begin{table*}[b!]
    \centering
\scalebox{0.75}{
\begin{tabular}{l|cc|c|rrrrrr|rr}
            \toprule[1pt]
            \textbf{City} & $\textbf{EI}_{\text{AD}}$ &$\textbf{EI}_{\text{AT}}$ & \textbf{MSE} & \textbf{AD} & \textbf{UR} & \textbf{AT} &  \textbf{Switch} & \textbf{NMDP} & \textbf{TMDP} &\textbf{CO} & \textbf{RL} \\
            \hline
            \multirow{2}*{City 1} &\multirow{2}*{13.72}&\multirow{2}*{-15.78}& $\downarrow \text{Average} \ (10^{-2})$ &10.15&3.49 &3.02& 3.95& 7.22 & 6.94 & \bf{2.31}& \underline{2.58} \\
            ~&~&~& $\downarrow$ \text{Ranking} $\in$ [1, 8] &5.34&4.20&\underline{3.86}& 4.38&5.00&5.42 & 3.98 &  \bf{3.82}\\
            \hline
            \multirow{2}*{City 2} &\multirow{2}*{-16.83}&\multirow{2}*{4.87}& $\downarrow \text{Average} \ (10^{-2})$ &3.43&5.59 &59.79& 3.31 & 2.08 & 1.48 & \underline{1.46} & \bf{1.42} \\
            ~&~&~& $\downarrow$ \text{Ranking} $\in$ [1, 8] &4.46&4.70& 6.36& 4.20&4.28& \bf{3.87} & \underline{3.99} & 4.13 \\
            \bottomrule[1pt]
        \end{tabular}
}
    \caption{\small Comparison of different designs in estimating ATE. The bold number indicates the best result, while the underlined number denotes the second-best. The symbol $\downarrow$ represents an inverse indicator, meaning that a lower value denotes a more effective design for estimating the ATE.} 
    \label{table:experiment2}
\end{table*}
\begin{itemize}[leftmargin=*]
    \item \textbf{Efficiency indicators}. 
As evidenced by both the average MSE and the performance ranking metric, the AT design yields a more accurate ATE estimator for City 1 compared to AD and UR. 
These findings are consistent with a negative $\text{EI}_{\text{AT}}$ and a positive $\text{EI}_{\text{AD}}$. 
In contrast, the results in City 2 are reversed, where the AD design significantly outperforms AT with a considerably smaller average MSE and a higher ranking. Meanwhile, AD generally outperforms UR. These results are, again, consistent with a negative $\text{EI}_{\text{AD}}$ and a positive $\text{EI}_{\text{AT}}$. 
    \item \noindent \textbf{Designs}. The violin plots in Figure~\ref{fig:experiment2_ate} visualize the distribution of MSEs of ATE estimators under various designs, where the width of the violin indicates the density of data points at different MSE values. The designs are arranged in descending order from left to right according to the average MSE across the range of $b$. 
    In both cities, the distributions of MSEs under the proposed CO and RL designs are more 
    tightly centered around zero when compared to other designs. Table~\ref{table:experiment2} also suggests a consistent improvement of statistical efficiency for our proposed optimal designs over alternatives. 
    It is also worth mentioning that the AT design achieves a competitive second-best performance ranking in City 1. 
    In City 2, the TMDP design outperforms ours in terms of performance ranking, partly because it additionally leverages observational data to determine optimal treatments in the online experiment, whereas our designs are observation-agnostic. However, neither AT nor TMDP performs well in the other city. On the contrary, the performance of our designs is more consistent and robust across the two cities.
\end{itemize}

\section{Discussion}\label{sec:discussion}

In this paper, we study experimental designs for A/B testing in partially observable online experiments where the data does not satisfy the Markov assumption. Specifically, we propose the controlled (V)ARMA model---a rich subclass of POMDPs---for fitting experimental data, establish asymptotic MSEs of ATE estimators, derive two efficiency indicators to assess the statistical efficiency of three commonly used designs, and develop two data-driven algorithms to learn the optimal observation-agnostic design. Our work bridges several vital research areas, including time series analysis, experimental design, causal inference, RL, and A/B testing, opening numerous exciting avenues for future research across these fields, covering theory, methodology, and applications. 
{\singlespacing
\subsection{Applications: Extensions to Spatially Dependent Experiments}
}
In our ride-sharing example, we consider evaluating order dispatching policies, which are typically randomized over time and implemented in the whole city at each time, leading to temporally dependent experiments where the data generated is summarized into a single time series. However, the company is equally keen on applying different subsidizing policies in different spatial locations of the city to balance the driver supply and customer demand across the city \citep{shi2023multiagent}. These experiments are inherently spatially dependent. 

Spatially dependent experiments are common in many applications that involve a group of experimental units that receive (sequential) treatments across different locations \citep{ugander2013graph, baird2018optimal, johari2022experimental, leung2022rate,  jia2023clustered, viviano2023causal, liu2024cluster, zhan2024estimating}; see also \citet{bajari2023experimental} for a recent overview. In these experiments, in addition to the carryover effect over time, the spatial spillover effect also exists where the treatment of one experimental unit can affect the outcome of others. There is growing interest in developing causal inference methods that account for spatial interference \citep[see][for a recent overview]{reich2021review}. It would be practically interesting to integrate the proposed designs with these methodologies to adapt them to such settings.


\subsection{Methodology: Model-based v.s. Model-free Approaches}
Our proposal is model-based in that it employs classical time series models for estimating the ATE and designing the experiment. Alternatively, model-free methods that do not directly model the time series are equally applicable. Depending on how they estimate the ATE, these model-free methods can be roughly categorized into two types:
\begin{enumerate}[leftmargin=*]
    \item[(i)] The first type assumes that the experimental data follows an MDP to handle carryover effects \citep{farias2022markovian,shi2022dynamic,cao2024orthogonalized} and adapts existing model-free off-policy evaluation methods developed in the RL literature for MDPs \citep[see e.g.,][]{liao2022batch,kallus2022efficiently,uehara2022review} to evaluate the ATE.
    \item[(ii)] The second type is completely model-agnostic and employs generic importance sampling methods \citep[see e.g.,][]{zhang2013robust,bojinov2019time,hu2023off}. 
\end{enumerate}
Both model-based and model-free methods have their own merits. Model-based approaches are often more data efficient, leading to less variable ATE estimators. However, they can be vulnerable to model misspecification. The first type of model-free method does not rely on a specific model, but it may fail under partial observability. The second type of method allows partial observability, but it leads to more variable ATE estimators. This increased variability is undesirable for A/B testing, particularly in settings with small sample sizes and weak signals. 

Our choice of the model-based approach is guided by the principle that ``all models are wrong, but some are useful'' \citep{box1979all}. Unlike the aforementioned model-free methods, our approach not only addresses the four practical challenges mentioned in the introduction but also demonstrates its usefulness in our numerical experiments under model misspecification.  Additionally, our collaborators in the ride-sharing company prefer the model-based approach for its interpretability. 

Meanwhile, the proposed controlled (V)ARMA model serves as a stepping stone. It can be extended to a variety of models for future research. For instance, autoregressive fractionally integrated moving-average (ARFIMA) models could be explored to handle long-term dependencies, which are common in practice. Similarly, more general linear state space models such as those in \citet{liang2023randomization} could be considered. Despite these changes in modeling, our paper establishes a foundational framework for analysis and design. It covers both theoretical techniques, such as the small signal asymptotics, and methodological developments, including the RL algorithms, which can be adapted to these new models. 


\subsection{Theory: Small Signal Asymptotic Framework}

At the core of our asymptotic theories is the proposed small signal asymptotic framework, which substantially simplifies the asymptotic calculations in time dependent experiments. As mentioned earlier, it aligns with our empirical observations where most improvements from new strategies are not substantial. When this assumption is violated, our designs are not guaranteed to be optimal. However, in such cases, the treatment effect becomes non-negligible. Our approach, although potentially sub-optimal, remains consistent in detecting this effect. This ensures that our design remains safe to use, regardless of whether the signal is small or large.

Meanwhile, similar assumptions have been proposed in the literature on either A/B testing or other fields to simplify theoretical or methodological development. For instance, \citet{kuang2023weak} introduced a weak signal asymptotic framework in a different context for solving multi-armed bandit problems. The main differences include: (i) Our small signal condition requires the ATE to decay to zero at an arbitrary rate, whereas \citet{kuang2023weak} requires the difference in mean outcomes between different arms to decay to zero at a more restrictive parametric rate. (ii) Unlike our framework, which is designed to simplify the asymptotic analysis, their theoretical framework is developed to derive a diffusion convergence limit theorem for sequentially randomized Markov experiments.

Additionally, \citet{farias2022markovian} and \citet{wen2024analysis} imposed similar small signal conditions in the same context as ours for A/B testing in time dependent experiments, aiming either to derive more efficient ATE estimators methodologically or to analyze these estimators theoretically. However, their focus on Markovian environments is more restrictive than ours. They also required the difference of Markov state transition probabilities under the two treatments to be small, a condition less interpretable than our requirement for the ATE to approach zero. 
\cite{viviano2023causal} assumed a small peer effect condition (Assumption 3) to handle the spatial spillover effect. Such an assumption shares similar spirits to ours, but is designed to develop the optimal design in spatially dependent experiments.

\bibliographystyle{apalike}
\spacingset{1.4}
\bibliography{references}

\appendix


THIS SUPPLEMENT IS STRUCTURED as follows. Appendix \ref{appendix:optimal} outlines the proofs of  Theorems~\ref{them:asymptotic}, \ref{thm:ATEestV} and \ref{thm:optimal}. Appendix~\ref{appendix:equivalence} establishes the equivalence between the controlled ARMA model and POMDP in Section~\ref{subsec:ARMAmodel} of the main manuscript. Appendix~\ref{appendix:efficiency_ARMA} and \ref{appendix:efficiency_VARMA} provide detailed proof of the estimation, asymptotic MSEs, and efficiency indicators in the controlled ARMA and VARMA, respectively. Finally, the procedure to simplify asymptotic MSEs for the optimal Markov design can be found in Appendix~\ref{appendix:optimal_markov}.

\section{Proofs of Theorems~\ref{them:asymptotic}, \ref{thm:ATEestV} and \ref{thm:optimal}}\label{appendix:optimal}
As the proofs of Theorems~\ref{them:asymptotic}, \ref{thm:ATEestV} and \ref{thm:optimal} are closely related, we put them together in a section. 
The proofs in this section are organized as follows:
\begin{itemize}
    \item Appendix~\ref{appendix:thm_ARMA} presents the proof of Theorem~\ref{them:asymptotic} which establishes the asymptotic MSEs of ATE estimators under the controlled ARMA model.
    \item Appendix~\ref{appendix:thm_VARMA} presents the proof of Theorem~\ref{thm:ATEestV} which generalizes the proof of Theorem~\ref{them:asymptotic} to the controlled VARMA model. 
    \item Appendix~\ref{appendix:thm_optimal} presents the proof of Theorem~\ref{thm:optimal} which establishes the optimality conditions for the optimal design.
\end{itemize}

\subsection{Proof of Theorem~\ref{them:asymptotic} in Controlled ARMA}\label{appendix:thm_ARMA} 

For a given observation-agnostic treatment allocation strategy $\pi$, recall that $\xi_{\pi} = \lim_{t\rightarrow+\infty}\Mean(U_t)$. 
Notably, $\xi_\pi = 0$ under the balanced design, such as AT, UR, and AD. However, for 
other designs, $\xi_\pi$ may not be zero. Thus, unlike traditional ARMA models where responses are typically centered and the intercept term is zero, our controlled ARMA model requires the inclusion of an intercept term $\mu$, as in Equation
\eqref{eq:controlledARMA}:
		\begin{equation}
			\begin{aligned}\nonumber
				Y_{t}=\mu + \sum_{j=1}^{p} a_j Y_{t-j}+b U_{t}+Z_{t}.
			\end{aligned}
		\end{equation}
    According to Lemma~\ref{lemma:ATE}, $\text{ATE}=2b(1-a)^{-1}$ where $a=\sum_{j=1}^p a_j$ even with the intercept term. As analyzed in \eqref{eqn:ATE_delta}, 
    an application of the delta method yields
        \begin{equation}\nonumber
    \widehat{\textrm{ATE}}-\textrm{ATE}=\frac{2(\widehat{b}-b)}{1-a}+\frac{2b}{(1-a)^2}\sum_{j=1}^p(\widehat{a}_j-a_j)+o_p(T^{-1/2}), 
\end{equation}
    where the third term is a high-order reminder, which becomes negligible as $T\rightarrow+\infty$, and the
    second term is $O_p(T^{-1/2}\text{ATE})$, which becomes $o_p(T^{-1/2})$ as well under the small signal condition. Consequently, we obtain
    \begin{eqnarray}\label{eqn:ATEasymptotic}
        \widehat{\textrm{ATE}}-\textrm{ATE}=\frac{2(\widehat{b}-b)}{1-a}+o_p(T^{-1/2}),
    \end{eqnarray}
    and it suffices to compute the asymptotic variance of $\widehat{b}$ to calculate the asymptotic variance of the ATE estimator. 
    
    We first define $T_p=T-q-p$. Following the Yule-Walker estimation procedure presented in the main context of this paper (detailed in Appendix~\ref{appendix:efficiency_ARMA}), 
    we obtain that
		\begin{equation}
			\begin{aligned}\nonumber
				\left(\begin{array}{c}
					\widehat{b}-b \\
					\widehat{\mu}-\mu \\
					\widehat{a}_1-a_1  \\
					\vdots \\
					\widehat{a}_p-a_p  \\
				\end{array}\right)= 
				\left(\begin{array}{cc|ccc}
					1  & \xi_\pi & \xi_{uy^{-1}} & \cdots & \xi_{uy^{-p}} \\
					\xi_\pi & 1 & \xi_y & \cdots & \xi_y \\
     \hline
					\xi_{uy^{-q-1}} & \xi_y & \xi_{y^{-1} y^{-q-1}} & \cdots & \xi_{y^{-p} y^{-q-1}}\\
					\vdots & \vdots & \vdots & & \vdots \\
					\xi_{uy^{-q-p}} & \xi_y & \xi_{y^{-1}y^{-q-p}} & \cdots & \xi_{y^{-p}y^{-q-p}}
				\end{array}
				\right)^{-1} \frac{1}{T_p} \left(\begin{array}{c}
					\displaystyle \sum_{t} U_{t} Z_{t}  \\
					\displaystyle  \sum_{t} Z_{t} \\
					\vdots\\
					\displaystyle  \sum_{t} Y_{t-p-q}Z_{t}
				\end{array}
				\right) + o_p(1),
			\end{aligned}
		\end{equation}
	where we define $\xi_y= \frac{1}{T}\sum_{t} \Mean (Y_{t}), \xi_{y^{-i}y^{-q-j}} = \frac{1}{T_p}\sum_{t}\Mean (Y_{t-i} Y_{t-q-j})$ for $i,j=1,\ldots,p$, and $\xi_{uy^{-j}} = \frac{1}{T_p}\sum_{t}\Mean (Y_{t-j} U_{t})$ for $j=1,\ldots,p$ and $q+1,\ldots,q+p$. By \eqref{eqn:ATEasymptotic}, it suffices to compute the first row of the matrix inverse  in the above expression to obtain the asymptotic linear representation of $\widehat{\textrm{ATE}}-\textrm{ATE}$. Using the block matrix inverse formula, it can be shown that most entries in the first row are approximately zero. 
 In particular, 
 the first row of the matrix inverse is asymptotically equivalent to $(\frac{1}{1-\xi_\pi^2},-\frac{\xi_\pi}{1-\xi_\pi^2},0,\ldots,0),$ where the first two entries are derived by calculating the inverse matrix of the upper-left sub-matrix and the remaining terms are $\mathcal{O}(\text{ATE})$, which tends to 0 under the small signal condition. This calculation follows similar arguments to those presented, particularly for the UR and AT designs, in Appendix~\ref{appendix:sec_armapq} of the Supplementary Material. Therefore, we omit the details here to save space. Together with \eqref{eqn:ATEasymptotic}, we obtain the following  asymptotic linear representation of $\widehat{\textrm{ATE}}-\textrm{ATE}$:
		\begin{equation}            \begin{aligned}\nonumber
    \widehat{\textrm{ATE}}-\textrm{ATE} & = \frac{2}{1-a} \left( \frac{1}{1-\xi_\pi^2} \frac{1}{T} \sum_t U_t Z_t - \frac{\xi_\pi}{1-\xi_\pi^2} \frac{1}{T} \sum_t Z_t \right)+o_p(T^{-1/2}) \\
    & =  \frac{2}{(1-a)(1-\xi_\pi^2) T} \Big[\sum_{t=1}^T (U_{t}-\xi_\pi)Z_{t}\Big]+o_p(T^{-1/2}).
			\end{aligned}
		\end{equation}
This yields the following formula of the asymptotic MSE: 
		\begin{equation}            \begin{aligned}\nonumber
				\lim_{\substack{T \rightarrow +\infty\\ \text{ATE}\to 0}}    \text{MSE}( \sqrt{T} \widehat{\text{ATE}}(\pi))  = 	\lim_{T \rightarrow +\infty} \frac{4}{(1-a)^2(1-\xi_\pi^2)^2 T} \Var\Big[\sum_{t=1}^T (U_{t}-\xi_\pi)Z_{t}\Big].
			\end{aligned}
		\end{equation}

\subsection{Proof of Theorem~\ref{thm:ATEestV} in Controlled VARMA}\label{appendix:thm_VARMA} 
The proof 
extends the results from the controlled ARMA model outlined in Appendix \ref{appendix:thm_ARMA} to the controlled VARMA and relies largely on the arguments detailed in Appendix~\ref{appendix:efficiency_VARMA} of the Supplementary Material. 
Recall that our controlled VARMA model is given by 
$$\mathbf{Y}_{t}=\boldsymbol{\mu} + \sum_{j=1}^{p} \mathbf{A}_j \mathbf{Y}_{ t-j}+ \mathbf{b} U_{t} + \mathbf{Z}_{t}.$$ 
 We begin by introducing the following estimating equations for estimating $\{\mathbf{A}_j\}_j$ and $\mathbf{b}$: 
 \begin{equation}
			\begin{aligned}\nonumber
   	\underbrace{\frac{1}{T_p} \sum_{t} U_{t} \mathbf{Y}_{t}}_{\widehat{\xi}_{u y}}&= \boldsymbol{\mu} \underbrace{\frac{1}{T_p} \sum_t U_t}_{\widehat{\xi}_\pi} + \sum_{j=1}^p {\mathbf{A}_j}	\underbrace{\frac{1}{T_p} \sum_{t} U_{t} \mathbf{Y}_{t-j}}_{\widehat{\xi}_{u y^{-j} }}+ {\mathbf{b}} \\
    \underbrace{\frac{1}{T_p} \sum_{t} \mathbf{Y}_{t}}_{\widehat{\xi}_{y}}&= \boldsymbol{\mu}  + \sum_{j=1}^p {\mathbf{A}_j}	\underbrace{\frac{1}{T_p} \sum_{t}  \mathbf{Y}_{t-j}}_{\widehat{\xi}_{y }}+ {\mathbf{b}} \underbrace{\frac{1}{T_p} \sum_{t}  U_{t}}_{\widehat{\xi}_{u}} \\
				\underbrace{\frac{1}{T_p} \sum_{t} \mathbf{Y}_{t} \mathbf{Y}^\top_{t-q-1}}_{\widehat{\xi}_{y y^{-q-1}}}&=  \boldsymbol{\mu} \underbrace{\frac{1}{T_p} \sum_t \mathbf{Y}^\top_{t-q-1}}_{\widehat{\xi}_y} + \sum_{j=1}^p{\mathbf{A}_j}	\underbrace{\frac{1}{T_p} \sum_{t} \mathbf{Y}_{t-j} \mathbf{Y}^\top_{t-q-1}}_{\widehat{\xi}_{y^{-j} y^{-q-1}}}+ {\mathbf{b}} \underbrace{\frac{1}{T_p} \sum_{t} U_t \mathbf{Y}^\top_{t-q-1}}_{\widehat{\xi}_{u y^{-q-1}}^\top} \\
            & \cdots \\
			\underbrace{\frac{1}{T_p} \sum_{t} \mathbf{Y}_{t} \mathbf{Y}^\top_{t-q-p}}_{\widehat{\xi}_{y y^{-q-p}}}&=\boldsymbol{\mu} \underbrace{\frac{1}{T_p} \sum_t \mathbf{Y}^\top_{t-q-p}}_{\widehat{\xi}_y} +  \sum_{j=1}^p {\mathbf{A}}_j	\underbrace{\frac{1}{T_p} \sum_{t} \mathbf{Y}_{t-j} \mathbf{Y}^\top_{t-q-p}}_{\widehat{\xi}_{y^{-j} y^{-q-1}}}+ {\mathbf{b}} \underbrace{\frac{1}{T_p} \sum_{t} U_{t} \mathbf{Y}^\top_{t-q-p}}_{\widehat{\xi}_{u y^{-q-p}}^\top} .
			\end{aligned}
		\end{equation}
Following the proof technique in Appendix~\ref{appendix:efficiency_VARMA}, solving these estimating equations leads to:
	\begin{small}
		\begin{equation}
			\begin{aligned}\nonumber
				\left[\widehat{\mathbf{b}} - \mathbf{b}, \widehat{\boldsymbol{\mu}} - \boldsymbol{\mu}, \widehat{\mathcal{A}} - \mathcal{A} \right] 
				&=\frac{1}{T_p}  \left[\sum_{t} U_{t}\mathbf{Z}_{t}, \sum_t Z_t, \sum_{t} \mathbf{Z}_{t} \mathbf{Y}_{t-q-1}^\top, \ldots,\sum_{t} \mathbf{Z}_{t} \mathbf{Y}_{t-q-p}^\top   \right]
				\boldsymbol{\xi}_{p,q}^{-1} + o_p(1),
			\end{aligned}
		\end{equation}
	\end{small}
 where the matrix $\boldsymbol{\xi}_{p,q}$ is given by
 \begin{equation}
    \begin{aligned}\nonumber 
\boldsymbol{\xi}_{p,q} \equiv 
        \left(\begin{array}{cc|ccc}
            1  & \xi_\pi & \xi^\top_{u y^{-q-1}} & \cdots & \xi^\top_{u y^{-q-p}} \\
            \xi_\pi & 1 & \xi_y & \cdots & \xi_y \\
            \hline
            \xi_{uy^{-1}} & \xi_y & {\xi}_{y^{-1}y^{-q-1}}  & \cdots & {\xi}_{y^{-1}y^{-q-p}}  \\
            \cdots & \cdots  & \cdots & \cdots & \cdots \\
            \xi_{uy^{-p}} & \xi_y & {\xi}_{y^{-p} y^{-q-1}} & \cdots & {\xi}_{y^{-p}y^{-q-p}}      
        \end{array}
        \right),
    \end{aligned}
\end{equation}
where $\xi_y$, $\xi_{uy^{-j}}$ and $\xi_{y^{-i}y^{-q-j}}$ are population-level limits of $\widehat{\xi}_y$, $\widehat{\xi}_{uy^{-j}}$ and $\widehat{\xi}_{y^{-i}y^{-q-j}}$, defined similarly to those in Appendix \ref{appendix:thm_ARMA}.

Applying the vectorization to the above equation, we have the following equation:
		\begin{equation}
			\begin{aligned}\nonumber
				\left(\begin{array}{c}
            \widehat{\mathbf{b}} - \mathbf{b}   \\
            \widehat{\boldsymbol{\mu}} - \boldsymbol{\mu} \\
					\text{vec}(\widehat{\mathcal{A}} - \mathcal{A}) 
				\end{array}
				\right)
				=  \frac{1}{T_p}  \left( 	\boldsymbol{\xi}_{p,q}^{-1} \otimes \mathbb{I}_d \right) 	
				\left(\begin{array}{c}
					\displaystyle \sum_{t} U_{t}\mathbf{Z}_{t} \\
            \displaystyle \sum_{t} \mathbf{Z}_{t} \\
					\displaystyle \text{vec}(\sum_{t} \mathbf{Z}_{t} \mathbf{Y}_{t-q-1}^\top) \\
                    \vdots \\
                    \displaystyle \text{vec}(\sum_{t} \mathbf{Z}_{t} \mathbf{Y}_{t-q-p}^\top)
				\end{array}
				\right)
				+ o_p(1).
			\end{aligned}
		\end{equation}
Applying the Taylor expansion and using the small asymptotic conditions, the ATE estimator under the controlled VARMA model can be similarly shown to satisfy:
		\begin{equation}
			\begin{aligned}\nonumber
				\widehat{\textrm{ATE}} - \text{ATE}
    = 2\bm{e}^\top (\mathbb{I} - \mathbf{A})^{-1} (\widehat{\mathbf{b}} - \mathbf{b})+o_p(T^{-1/2}). 
			\end{aligned}
		\end{equation}
Similar to the proof in Appendix~\ref{appendix:thm_ARMA}, the first row of $\boldsymbol{\xi}_{p,q}^{-1}$ is asymptotically equivalent to $(\frac{1}{1-\xi_\pi^2},-\frac{\xi_\pi}{1-\xi_\pi^2},0,\ldots,0),$ by using the small signal conditions. Consequently, the resulting ATE estimator has the following form:
		\begin{equation}            \begin{aligned}\nonumber
				\lim_{\substack{T \rightarrow +\infty\\ \text{ATE}\to 0}}     \widehat{\textrm{ATE}}-\textrm{ATE} 
    & =  \frac{2}{(1-\xi_\pi^2) T} \bm{e}^\top (\mathbb{I} - \mathbf{A})^{-1} \Big[\sum_{t=1}^T (U_{t}-\xi_\pi)\mathbf{Z}_{t}\Big]
			\end{aligned}
		\end{equation}
 Therefore, the asymptotic MSE of the ATE estimator satisfies
		\begin{equation}            \begin{aligned}\nonumber
				& \lim_{T \rightarrow +\infty} \text{MSE}( \sqrt{T} \widehat{\text{ATE}}(\pi))  = 	\lim_{T \rightarrow +\infty} \frac{4}{(1-\xi_\pi^2)^2 T} \bm{e}^\top (\mathbb{I} - \mathbf{A})^{-1} \Var\Big[\sum_{t} (U_{t}-\xi_\pi)\mathbf{Z}_{t}\Big] (\mathbb{I} - \mathbf{A})^{-1} e.
			\end{aligned}
		\end{equation}
This completes the proof. 

\subsection{Proof of Theorem~\ref{thm:optimal}: Optimality Conditions for Optimal Design}\label{appendix:thm_optimal} 
We begin with the controlled ARMA model. 
Recall that the asymptotic MSE takes the following form:
		\begin{equation}            \begin{aligned}\nonumber
				\lim_{\substack{T \rightarrow +\infty\\ \text{ATE}\to 0}}    \text{MSE}( \sqrt{T} \widehat{\text{ATE}}(\pi))  = 	\lim_{T \rightarrow +\infty} \frac{4}{(1-a)^2(1-\xi_\pi^2)^2 T} \Var\Big[\sum_{t=1}^T (U_{t}-\xi_\pi)Z_{t}\Big].
			\end{aligned}
		\end{equation}
According to the following formula: 
$$\text{Cov}(AC,BD)=\text{Cov}(A, B) \text{Cov}(C, D)+\mathbb{E}(A) {E}(B) \text{Cov}(C, D)+\mathbb{E}(C) \mathbb{E}(D) \text{Cov}(A, B)$$
for random variables $A, B, C$ and $D$, where $A$is independent of $C$ and $D$, and $B$ is independent of $C$ and $D$, we have for any $k\le q$ and $t>k$ that
     \begin{eqnarray}\label{eqn:someeqn}
       \begin{aligned}\nonumber
            \Cov((U_{t} -\xi_\pi) Z_{t}, (U_{t-k} - \xi_\pi) Z_{t-k}) & = \Cov(U_{t}-\xi_{\pi}, U_{t-k}-\xi_{\pi}) \text{Cov}(Z_{t}, Z_{t-k}) \\
            &=\Cov(U_t,U_{t-k})\Cov(Z_{t}, Z_{t-k}),
       \end{aligned}
    \end{eqnarray}
as $t\to \infty$, provided the limit $\xi_{\pi}=\lim_t U_t$ exists. The above equation holds as we consider the observation-agnostic design, where $U_t$ is independent of $Z_t$.

Equation \eqref{eqn:someeqn} implies that $\Var\Big[\sum_{t} (U_{t}-\xi_\pi)Z_{t}\Big]$ is independent of $\xi_\pi$. Notice that for any treatment sequence $\{U_t\}_t$, we can define another sequence $\{U_t^*\}_t$ such that either $U_t^*=U_t$ for all $t$, or $U_t^*=-U_t$ for all $t$. Both events occur with a probability of 0.5. By definition, it is immediate to see that the treatment allocation strategy $\pi^*$ for generating $\{U_t^*\}_t$ is balanced. Meanwhile,  $\{U_t^*\}_t$ shares the common covariance matrix with $\{U_t\}_t$. Together with \eqref{eqn:someeqn}, it implies that for any $\pi$, there exists another $\pi^*$ such that $\xi_{\pi}^*=0$ and its generated treatments $\{U_t^*\}_t$ satisfy
\begin{equation*}
    \frac{1}{T}\Var\Big[\sum_{t=1}^T (U_{t}-\xi_\pi)Z_{t}\Big]=\frac{1}{T}\Var\Big[\sum_{t=1}^T U_{t}^*Z_{t}\Big].
\end{equation*}
This proves the balanced condition for the optimal design. Meanwhile, under any balanced design $\pi$, we have
$$\text{Cov}(U_{t}, U_{t-k}) \text{Cov}(Z_{t}, Z_{t-k}) = \mathbb{E} (U_{t} U_{t-k}) \sum_{j=k}^q \theta_j\theta_{j-k}.$$ 

The asymptotic MSE 
of the resulting ATE estimator can be simplified as:
			\begin{eqnarray}
				\begin{aligned}\label{eq:appendix_optimality_arma}
					& \lim_{\substack{T \rightarrow +\infty\\ \text{ATE}\to 0}}  \text{MSE}( \sqrt{T} \widehat{\text{ATE}}(\pi)) \\ & = \frac{4}{(1-a)^2} [\sum_{j=0}^q \theta_j^2+2\sum_{k=1}^q  \lim_{T\rightarrow +\infty} \frac{1}{T}\sum_{t=q+1}^T \mathbb{E} (U_{t} U_{t-k}) \sum_{j=k}^q \theta_j\theta_{j-k} ].
				\end{aligned}
			\end{eqnarray}
    The optimal treatment allocation strategy is thus achieved by minimizing
        \begin{equation}\label{eq:appendix_optimal}
            \lim_{T\rightarrow +\infty} \sum_{k=1}^q c_k \left[\frac{1}{T}\sum_{t=q+1}^T \mathbb{E} (U_{t} U_{t-k}) \right], \quad c_k = \sum_{j=k}^q \theta_j\theta_{j-k},
        \end{equation}
    subject to $\xi_{\pi}=0$. 
    
    Based on the discussions in Section \ref{sec:optimal_history}, we can cast of the problem of minimizing \eqref{eq:appendix_optimal} into estimating the optimal policy of an MDP with the past $q$-dependent treatments defined as the new state. Using the properties of the optimal policy in MDP \citep{puterman2014markov, ljungqvist2018recursive}, we can show that the optimal $\pi$ is $q$-dependent, stationary and deterministic.
    
    Under the controlled VARMA model, we can similarly show that the optimal treatment allocation strategy is balanced, $q$-dependent, deterministic, and stationary. The asymptotic MSE of its ATE estimator is given by:
			\begin{eqnarray}
				\begin{aligned}\nonumber
				 &	\lim_{\substack{T \rightarrow +\infty\\ \text{ATE}\to 0}} \text{MSE}( \sqrt{T} \widehat{\text{ATE}}(\pi^*))   \\ 
        & = 4 \bm{e}^\top (\mathbb{I} - \mathbf{A})^{-1} \left( \sum_{j=0}^{q} \mathbf{M}_j \Sigma \mathbf{M}_j + 2 \sum_{k=1}^{q}\lim_{T\rightarrow +\infty} \frac{1}{T}\sum_{t=q+1}^T  \mathbb{E} (U_{t} U_{t-k}) \sum_{j=k}^{q}  \mathbf{M}_j \Sigma \mathbf{M}_{j-k} \right)  (\mathbb{I} - \mathbf{A})^{-1} \bm{e}.
				\end{aligned}
			\end{eqnarray}

\section{Equivalence between Controlled ARMA and POMDP}\label{appendix:equivalence}
\begin{proof}
We show that the controlled ARMA($p, q$) without an intercept term can be written as a special form of POMDP with linear state transition and observation emission functions in \eqref{eq:statespace}. Recap the controlled ARMA($p, q$) model:
$Y_{t} = \sum_{i=1}^p a_i Y_{t-i} + b U_{t} + \sum_{i=0}^q \theta_i \epsilon_{t-i}$. We denote a new latent variable $X_t$ and let $d = \max\{p, q+1\}$, $a_0=1, \theta_0=1$ and $\theta_{-1}=1$. We start from the following special form of state space model:
		\begin{equation}
			\begin{aligned}\label{eq:proof_statespace}
				\underbrace{\left(\begin{array}{c}
					X_{t} \\
					X_{t-1} \\
					\vdots \\
					X_{t-(d-1)} \\
				\end{array}
				\right)}_{\boldsymbol{X}_{t+1}} & = 	\underbrace{\left(\begin{array}{cccc|c}
					a_1  & a_2 & \cdots & a_{d-1} & a_d \\
					1 & 0  &  \cdots & 0 & 0 \\
					& & \vdots & & \\
					0 & 0 & \cdots & 1 & 0\\
				\end{array}
				\right)}_{F}	\underbrace{\left(\begin{array}{c}
					X_{t-1} \\
					X_{t-2} \\
					\vdots \\
					X_{t-d} \\
				\end{array}
				\right)}_{\boldsymbol{X}_t}+ b a_1 
				\underbrace{\left(\begin{array}{c}
					U_{t-1} \\
					0 \\
					\vdots \\
					0 \\
				\end{array}
				\right)}_{\boldsymbol{U}_t} + \underbrace{\left(\begin{array}{c}
					\epsilon_{t} \\
					0 \\
					\vdots \\
					0 \\
				\end{array}
				\right)}_{\boldsymbol{V}_t}, \\
				Y_{t} & = \sum_{i=0}^{d-1} \theta_i X_{t-i} + b \theta_{-1} U_t,
			\end{aligned}
		\end{equation}
    where each vector or matrix in \eqref{eq:statespace} has a specific form in the above equations. Next, the state transition equation in the above model regarding the latent variable $X_t$ can be rewritten as:
		\begin{equation}
		\nonumber
				\begin{cases}
					\theta_0 X_{t} &= \theta_0 \sum_{i=1}^{d} a_i X_{t-i} + b a_1 \theta_0 U_{t-1} + \theta_0 \epsilon_t \\
					\theta_1 X_{t-1} &= \theta_1 \sum_{i=1}^{d} a_i X_{t-1-i} + b a_2 \theta_1 U_{t-2} + \theta_1 \epsilon_{t-1} \\
					&\cdots \\
					\theta_{d-1} X_{t-(d-1)} &= \theta_{d-1} \sum_{i=1}^{d} a_i X_{t-(d-1)-i} + b a_{d} \theta_{d-1} U_{t-1-(d-1)} + \theta_{d-1} \epsilon_{t-(d-1)}.
				\end{cases}    
		\end{equation}
	Summing over the LHS and RHS in the above equations and using the observation emission equation regarding $Y_t$ in \eqref{eq:proof_statespace}, we attain that:
		\begin{equation}
			\begin{aligned}\nonumber
				Y_{t}  - b U_t & = \sum_{i=1}^{d} a_i (\sum_{j=0}^{d-1} \theta_j X_{t-i -j}) + b \sum_{i=1}^{d} a_i \theta_{i-1} U_{t-i} + \sum_{i=0}^{d-1} \theta_i \epsilon_{t-i}  \\
				& =  \sum_{i=1}^{d} a_i (Y_{t-i} - b \theta_{i-1} U_{t-i}) + b \sum_{i=1}^{d} a_i \theta_{i-1} U_{t-i} + \sum_{i=0}^{d-1} \theta_i \epsilon_{t-i} \\
				& =  \sum_{i=1}^{p} a_i Y_{t-i} + \sum_{i=0}^{q} \theta_i \epsilon_{t-i}.
			\end{aligned}
		\end{equation}
	After rearranging the above equation, we have the controlled ARMA($p, q$) model as:
		\begin{equation}
			\begin{aligned}\nonumber
				Y_{t} =  \sum_{i=1}^{p} a_i Y_{t-i} +  b U_t  +  \sum_{i=0}^{q} \theta_i \epsilon_{t-i}.
			\end{aligned}
		\end{equation}
In summary, any controlled ARMA($p, q$) can be expressed as a special form of linear state space model with a controlled variable and noise-free observation equations. It is also noted that many other choices exist to transform a controlled ARMA to its state space form, including Hamilton, Harvey, and Akaike forms, while preserving the correlation structure. Similarly, it is still possible to cast a state space model with a control variable to a special case of Controlled ARMA if we suppress the noise in the observation equation.
\end{proof}

\section{Estimation, Asymptotic MSEs, and Efficiency Indicators in Controlled ARMA}\label{appendix:efficiency_ARMA}
The outline of our proof in this section is:
    \begin{itemize}
        \item \textbf{Appendix~\ref{appendix:sec_arma1q}} Controlled ARMA($1, q$) with the proof transition from AD, UR, to AT design.
        \item \textbf{Appendix~\ref{appendix:sec_armapq}}  Controlled ARMA($p, q$) with the proof transition from AD, UR, to AT design.
    \end{itemize} 
 Since AD, UR, and AT are all balanced designs, i.e., $\xi_\pi=0$, we present the proof in the controlled ARMA model without the intercept term $\mu$ in this section.

\subsection{Proof in Controlled  ARMA(\texorpdfstring{$1, q$}{})}\label{appendix:sec_arma1q}
We start from controlled ARMA$(1,q)$ model with the state transition as: $Y_{t}=a Y_{t-1}+ b U_{t} + Z_{t}$. In controlled ARMA, we assume $R_{t} = Y_{t}$, a function of the current state $Y_{t-1}$ and action $U_{t}$. As demonstrated in Lemma~\ref{lemma:ATE}, the true ATE is $2b/(1-a)$. We multiply $U_{t}$ and $Y_{t-q-1}$ on both sides to estimate $\widehat{a}$ and $\widehat{b}$ due to their independence of $Z_{t}$ and then take the expectation on both sides. This leads to the Yule-Walker equations as follows: 
\begin{equation}\label{eqn:YW_arma1q}
\begin{aligned}\nonumber
\underbrace{ \frac{1}{T-q} \sum_{t=q+1}^ T \Mean (Y_{t} U_{t} )}_{\xi_{uy}}&= a \underbrace{ \frac{1}{T-q} \sum_{t=q+1}^ T \Mean (\frac{1}{T-q} \sum_{t=q+1}^ T Y_{t-1} U_{t})}_{\xi_{uy^{-1} }}+b,\\
      \underbrace{\frac{1}{T-q} \sum_{t=q+1}^ T\Mean(Y_{t} Y_{t-q-1}  )}_{\xi_{y y^{-q-1}}}&= a \underbrace{\frac{1}{T-q} \sum_{t=q+1}^ T\Mean (Y_{t-1} Y_{t-q-1}  )}_{\xi_{y^{-1} y^{-q-1} }}+b \underbrace{\frac{1}{T-q} \sum_{t=q+1}^ T\Mean (U_{t} Y_{t-q-1})}_{\xi_{uy^{-q-1}}},\\
\end{aligned}
\end{equation}
where we use $\xi_{uy}, \xi_{uy^{-1}}, \xi_{y y^{-q-1}}, \xi_{y^{-1} y^{-q-1}}$ and $\xi_{uy^{-q-1}}$ to represent their corresponding expectation forms in the above equations. To apply the method of moment, we replace these expectation terms with their moments, denoted by $\widehat{\xi}_{uy}, \widehat{\xi}_{uy^{-1}}, \widehat{\xi}_{y y^{-q-1}}, \widehat{\xi}_{y^{-1}y^{-q-1}}$ and $\widehat{\xi}_{uy^{-q-1}}$, respectively. We then arrive at the following equations regarding $\widehat{a}$ and $\widehat{b}$:
		\begin{equation}
			\begin{aligned}\label{eq:estimation_arma1q}
   	\underbrace{\frac{1}{T-q}\sum_{t=q+1}^{T} Y_{t} U_t}_{\widehat{\xi}_{uy}}&= \widehat{a} \underbrace{\frac{1}{T-q}\sum_{t=q+1}^{T} Y_{t-1}U_{t}}_{\widehat{\xi}_{uy^{-1}}} +  \widehat{b}, \\
				\underbrace{\frac{1}{T-q}\sum_{t=q+1}^{T}  Y_{t} Y_{t-q-1}}_{\widehat{\xi}_{yy^{-q-1}}}&= \widehat{a} \underbrace{\frac{1}{T-q}\sum_{t=q+1}^{T} Y_{t-1} Y_{t-q-1} }_{\widehat{\xi}_{y^{-1}y^{-q-1}}}+ \widehat{b}\underbrace{\frac{1}{T-q}\sum_{t=q+1}^{T} U_{t} Y_{t-q-1}}_{\widehat{\xi}_{uy^{-q-1}}}.
			\end{aligned}
		\end{equation}
We next multiply $U_{t}$ and $Y_{t-q-1}$ on both sides of $Y_{t}=a Y_{t-1}+ b U_{t} + Z_{t}$ and obtain:
		\begin{equation}
			\begin{aligned}\nonumber
		&	\widehat{\xi}_{uy} = a \widehat{\xi}_{uy^{-1}} + b  + \frac{1}{T-q} \sum_{t=q+1}^{T} U_{t} Z_{t}, \\ 
            & \widehat{\xi}_{yy^{-q-1}} = a \widehat{\xi}_{y^{-1}y^{-q-1}} + b \widehat{\xi}_{uy^{-q-1}} + \frac{1}{T-q} \sum_{t=q+1}^{T} Y_{t-q-1} Z_{t}.
			\end{aligned}
		\end{equation}
	Solving the two groups of equations and using the $o_p$ notations, we have:
		\begin{equation}\label{eq:estimation_arma1q_2}
			\left(\begin{array}{c} 
				\widehat{a}-a \\
				\widehat{b}-b
			\end{array}
			\right)=\left(\begin{array}{cc}
				\xi_{uy^{-1}} & 1 \\
				\xi_{y^{-1}y^{-q-1}} & \xi_{uy^{-q-1}} \\
			\end{array}
			\right)^{-1}\left(
			\begin{array}{c}
				\displaystyle \frac{1}{T-q} \sum_{t=q+1}^{T} U_{t} Z_{t} \\
				\displaystyle \frac{1}{T-q} \sum_{t=q+1}^{T} Y_{t-q-1} Z_{t}
			\end{array}
			\right)+o_p(1),
		\end{equation}
	where $	\xi_{uy^{-1}}=\mathbb{E}( \widehat{\xi}_{uy^{-1}})$, $\xi_{y^{-1}y^{-q-1}}= \mathbb{E}( \widehat{\xi}_{y^{-1}y^{-q-1}})$, and $\xi_{uy^{-q-1}} = \mathbb{E}(\widehat{\xi}_{uy^{-q-1}})$, which also correspond to the expectation terms defined in \eqref{eq:estimation_arma1q}. Since ATE$=2b/(1-a)$, we apply the Delta method on \eqref{eq:estimation_arma1q_2} and have the following ATE estimator equation:
	\begin{equation}\begin{aligned}\nonumber
				\widehat{\text{ATE}} - \text{ATE} & = \left(\begin{array}{cc} 
					\frac{2b}{(1-a)^2},  & \frac{2}{1-a} 
				\end{array}
				\right) \left(\begin{array}{cc}
					\xi_{uy^{-1}} & 1 \\
					\xi_{y^{-1}y^{-q-1}} & \xi_{uy^{-q-1}} \\
				\end{array}
				\right)^{-1}\left(
				\begin{array}{c}
					\displaystyle \frac{1}{T-q} \sum_{t=q+1}^{T} U_{t} Z_{t} \\
					\displaystyle \frac{1}{T-q} \sum_{t=q+1}^{T} Y_{t-q-1} Z_{t}
				\end{array}
				\right) \\
                & +o_p(T^{-1/2}),
		\end{aligned} \end{equation}
 where $(\frac{2b}{(1-a)^2},  \frac{2}{1-a})$ is the Jacobian vector of $2b/(1-a)$ on $a$ and $b$. By applying the inverse matrix formula, we have the general form of the inverse matrix:
		\begin{equation} 	\begin{aligned} \nonumber
				& \left(\begin{array}{cc}
					\xi_{uy^{-1}} & 1 \\
					\xi_{y^{-1}y^{-q-1}} & \xi_{uy^{-q-1}} \\
				\end{array}
				\right)^{-1}  =  \\
				& \left(\begin{array}{cc}
					- (\xi_{y^{-1}y^{-q-1}} - \xi_{uy^{-1}} \xi_{uy^{-q-1}})^{-1}  \xi_{uy^{-1}} & (\xi_{y^{-1}y^{-q-1}} - \xi_{uy^{-1}} \xi_{uy^{-q-1}})^{-1} \\
					1+ \xi_{uy^{-1}} (\xi_{y^{-1}y^{-q-1}} - \xi_{uy^{-1}} \xi_{uy^{-q-1}})^{-1} \xi_{uy^{-q-1}} & -\xi_{uy^{-1}} (\xi_{y^{-1}y^{-q-1}} - \xi_{uy^{-1}} \xi_{uy^{-q-1}})^{-1} 
				\end{array}\right).
			\end{aligned} 
		\end{equation}
To evaluate the ATE estimator's asymptotic MSEs, we consider each treatment assignment strategy: AD, UR, and AT. 

\noindent \textbf{For the  AD design},  as $\tau\rightarrow +\infty$, we can show that $\xi_{uy^{-1}}=\xi_{uy^{-q-1}}=b/(1-a)$  regardless of $U_{t}=1$ or $-1$. Particularly, we find the following equation that holds strictly:
		\begin{equation} 	\begin{aligned} \nonumber
				\left(\begin{array}{cc} 
					\frac{2}{1-a},  & 0 
				\end{array}
				\right) \left(\begin{array}{cc}
					\xi_{uy^{-1}} & 1 \\
					\xi_{y^{-1}y^{-q-1}} & \xi_{uy^{-q-1}} \\
				\end{array}
				\right) =	\left(\begin{array}{cc} 
					\frac{2b}{(1-a)^2},  & \frac{2}{1-a}
				\end{array}\right).
			\end{aligned} 
		\end{equation}
    Immediately, it suggests a concise result regarding the ATE estimator:
		\begin{equation}\begin{aligned}\label{eq:proof_ate_arma_ad}
				\widehat{\text{ATE}} - \text{ATE} = \frac{2}{1-a}  \frac{1}{T-q} \sum_{t=q+1}^{T} U_{t} Z_{t} +o_p(T^{-1/2}).
		\end{aligned} \end{equation}
   We then simplify the asymptotic MSEs by using the concise form above: 
    \begin{equation}
        \begin{aligned}\label{eq:asymptoticMSE_proof}
 \lim_{T\rightarrow+\infty} \text{MSE}(\sqrt{T}\widehat{\text{ATE}}(\pi))&= \lim_{T\rightarrow +\infty}  \mathbb{E} \left[ \sqrt{T} (\widehat{\text{ATE}} - \text{ATE}) \right]^2 \\
  &= \lim_{T\rightarrow +\infty}  \mathbb{E} \left[ 
\frac{2}{1-a} \frac{1}{\sqrt{T}} \sum_{t=1}^T U_{t} Z_{t}  + o_p(1) \right]^2 \\ 
&= \frac{4}{(1-a)^2} \lim_{T\rightarrow+\infty} \frac{1}{T} \mathbb{E}\left[ \sum_{t=1}^T U_{t} Z_{t} \right]^2  \\
& = \frac{4}{(1-a)^2} \lim_{T\rightarrow+\infty} \frac{1}{T} \left( \text{Var} \left( \sum_{t=1}^T U_{t} Z_{t} \right) +  \left(\mathbb{E}\left[ \sum_{t=1}^T U_{t} Z_{t} \right] \right)^2 \right) \\
    & = \frac{4}{(1-a)^2} \lim_{T\rightarrow+\infty} \frac{1}{T} \text{Var} \left( \sum_{t=1}^T U_{t} Z_{t} \right),
        \end{aligned}
        \end{equation}
where we utilize $\mathbb{E}\left[ U_{t} Z_{t} \right]=0$ as $U_t$ is independent of the white noises. This implies that the asymptotic MSE only relies on the asymptotic variance of $\sum_{t=1}^T U_{t} Z_{t}$. Since $\mathbb{E}\left[U_{t}\right] = 0$ and $\mathbb{E}\left[Z_{t}\right] = 0$, we have $\text{Cov}(U_{j}Z_{j}, U_{k} Z_{k}) = \text{Cov}(U_{j}, U_{k}) \text{Cov}(Z_{j},  Z_{k})$ when $j-k<q$. As $\lim_{\tau \rightarrow +\infty} \text{Cov}(U_{j}, U_{k}) = 1$, we have $\lim_{\tau \rightarrow +\infty} \text{Cov}(U_{j}Z_{j}, U_{k} Z_{k}) = \text{Cov}(Z_{j},  Z_{k}) = \sum_{i=j-k}^{q} \theta_i \theta_{i-(j-k)} \sigma^2$. Consequently, within the small signal asymptotic framework, the asymptotic MSE under the AD treatment allocation strategy $\pi_{\text{AD}}$ has the following form:
		\begin{equation}
			\begin{aligned}\label{eq:asymptoticMSE_controlarma_1p_ad}
				\lim_{\substack{T \rightarrow +\infty\\ \tau\to +\infty}} \text{MSE}(\sqrt{T}\widehat{\text{ATE}}(\pi_{\text{AD}}))
    &  =
			\frac{4\sigma^2}{(1-a)^2}\Big[ \sum_{j=0}^q \theta_j^2+2\sum_{k=1}^q \sum_{j=k}^q \theta_j\theta_{j-k} \Big],
			\end{aligned}
		\end{equation}
 where the asymptotic MSE depends on the parameters in AR(1) and MA($q$), regardless of $b$.
 
\noindent \textbf{For the UR design,} since all treatments are i.i.d. generated from Bernouli(0.5), it is immediate to attain that $\xi_{uy^{-1}}=\xi_{uy^{-q-1}}=0$. The ATE estimator equation is simplified as:
		\begin{equation}\begin{aligned}\label{eq:proof_UR}
				& \widehat{\text{ATE}} - \text{ATE} \\
                    &= \left(\begin{array}{cc} 
					\frac{2b}{(1-a)^2},  & \frac{2}{1-a} 
				\end{array}
				\right) \left(\begin{array}{cc}
					0 &  \quad \xi_{y^{-1}y^{-q-1}}^{-1} \\
					1 & 0 \\
				\end{array}
				\right)\left(
				\begin{array}{c}
					\displaystyle \frac{1}{T-q} \sum_{t=q+1}^T U_{t} Z_{t} \\
					\displaystyle \frac{1}{T-q} \sum_{t=q+1}^T Y_{t-q-1} Z_{t}
				\end{array}
				\right)+o_p(T^{-1/2}) \\
				& = \frac{2}{1-a}  \frac{1}{T-q} \sum_{t=q+1}^T U_{t} Z_{ t} 
				+ \frac{2b}{(1-a)^2} \xi_{y^{-1}y^{-q-1}}^{-1}  \frac{1}{T-q} \sum_{t=q+1}^T Y_{t-q-1} Z_{t}
				+o_p(T^{-1/2}).
		\end{aligned} \end{equation}
	Now, we consider the complicated term $ \xi_{y^{-1}y^{-q-1}}$ to further simplify the ATE estimator above. We start our derivations from the AD design. In the AD design with $U_{t}=1$, we have:
		\begin{equation}\begin{aligned} \nonumber
				Y_{t-1}&=\frac{b}{1-a}+\theta_0 \epsilon_{t}+(\theta_1+a)\epsilon_{t-1}+  (\theta_2+a \theta_1+a^2)\epsilon_{t-2}+\cdots \\
				& +(\theta_q+a \theta_{q-1}+\cdots+a^q) \epsilon_{t-q}
				+\sum_{j=1}^{\infty} a^j (\theta_q+a \theta_{q-1}+\cdots+a^q) \epsilon_{t-q-j}. 
		\end{aligned} \end{equation}
	Consequently, as $T\to\infty$, we have 
		\begin{equation}\begin{aligned} \nonumber
				\xi_{y^{-1} y^{-q-1}} &= \Mean (Y_{t-1} Y_{t-q-1}) \to \frac{b^2}{(1-a)^2}+\sigma^2 (\theta_q+a \theta_{q-1}+\cdots+a^q) [1+a(\theta_1+a)+  \\ 
        & a^2(\theta_2+a\theta_1+a^2) +\cdots +\frac{a^q}{1-a^2} (\theta_q+a\theta_{q-1}+\cdots+a^q)] \equiv \frac{b^2}{(1-a)^2} + \sigma^2 d(a, \theta),
		\end{aligned} \end{equation}
	where the second term of RHS in the second last equation is denoted as $d(a, \theta)$, which is a function of $a$ and $\theta=[\theta_0, \theta_1, \ldots,  \theta_q]$. The same form of $\xi_{y^{-1} y^{-q-1}}$ holds when $U_{t}=-1$.  Therefore, for the AD design, we have 
$ 		\xi_{y^{-1} y^{-q-1}} - \xi_{u y^{-1}} \xi_{u y^{-q-1}} = \sigma^2 d(a, \theta).$ 	
	Similarly, for the UR design, 
	as $\mathbb{E}\left[U_{j} U_{k}\right] = 0$, it suggests that 
     \begin{equation}
     \begin{aligned}\nonumber
         \xi_{y^{-1} y^{-q-1}} = \Mean(Y_{t-1} Y_{t-q-1}) = \sigma^2 d(a, \theta), \ \text{and} \ \xi_{y^{-1} y^{-q-1}} - \xi_{u y^{-1}} \xi_{u y^{-q-1}}= \xi_{y^{-1} y^{-q-1}}  = \sigma^2 d(a, \theta),
     \end{aligned}
 \end{equation}
where $	\xi_{y^{-1} y^{-q-1}}$ is only a function of $a$ and $\theta$, which is independent of $b$ or the magnitude of the ATE. We then look at the second term in \eqref{eq:proof_UR}, i.e., $\frac{2b}{(1-a)^2} \xi_{y^{-1}y^{-q-1}}^{-1}  \frac{1}{T-q} \sum_{t=q+1}^T Y_{t-q-1} Z_{t}$. As $Y_{t-q}$ is uncorrelated with $Z_t$, $ \frac{1}{T-q} \sum_{t} Y_{t-q-1} Z_{t}$ converges to zero by weak law of large number with the convergence rate $T^{-1/2}$. Therefore, we have
$ \frac{1}{T-q} \sum_{t} Y_{t-q-1} Z_{t} = O_p(T^{-1/2})$. Meanwhile, $2b/(1-a)^2 \propto \text{ATE}$ (especially $b$ is small), which is denoted by $O(\text{ATE})$. Consequently, the second term in \eqref{eq:proof_UR} is $O(\text{ATE})O_p(T^{-1/2})$, which tends to 0 within the small signal asymptotics by letting $\text{ATE}\rightarrow0$ and $T\rightarrow+\infty$. This indicates that:
		\begin{equation}\begin{aligned}\nonumber
			\lim_{\substack{T \rightarrow +\infty\\ \text{ATE}\to 0}}	\widehat{\text{ATE}} - \text{ATE} = \frac{2}{1-a}  \frac{1}{T-q} \sum_{t=q+1}^T U_{t} Z_{t}  + o_p(T^{-1/2}),
		\end{aligned} \end{equation}
 where the limit regarding the ATE estimator under the UR design has the same form as the AD design in \eqref{eq:proof_ate_arma_ad}. Based on the general asymptotic MSE in \eqref{eq:asymptoticMSE_proof}, we have the simplified asymptotic MSEs under the UR treatment allocation strategy $\pi_{\text{UR}}$ as follows:
		\begin{equation}
			\begin{aligned}\label{eq:asymptoticMSE_controlarma_1p_ur}
				\lim_{\substack{T \rightarrow +\infty\\ \text{ATE}\to 0}} \text{MSE}(\sqrt{T} \widehat{\text{ATE}}(\pi_{\text{UR}})) &  =
				\frac{4}{(1-a)^2}  	\lim_{T \rightarrow +\infty} \frac{1}{T} \text{Var}(\sum_{t=1}^T U_{t} Z_{t}) =\frac{4\sigma^2}{(1-a)^2} \sum_{j=0}^q \theta_j^2,
			\end{aligned}
		\end{equation}
	where  $\text{Cov}(U_{j}Z_{j}, U_{k} Z_{k}) = \text{Cov}(U_{j}, U_{k}) \text{Cov}(Z_{j},  Z_{k}) = 0$ for the UR design when $j \neq k$ because $\text{Cov}(U_{j}, U_{k})=0$. 
	
	\noindent \textbf{For the AT design, } after calculations, we have $ \xi_{u y^{-1}} = -b + ab - a^2 b + ... = -b/(1+a)$, $ \xi_{u y^{-q-1}} = (-1)^{q+1} b / (1+a)$, and $\xi_{y^{-1} y^{-q-1}} = (-1)^{q+2} b / (1+a) + \sigma^2 d(a, \theta)$. Then, we have 
$            \xi_{y^{-1} y^{-q-1}} - \xi_{u y^{-1}} \xi_{u y^{-q-1}} = \sigma^2 d(a, \theta),$ 
which shares the same form as the AD and UR designs. As such, we have:
		\begin{equation}\begin{aligned}\nonumber
				& \widehat{\text{ATE}} - \text{ATE}  \\
				& \frac{\left(\begin{array}{cc} 
						\frac{2b}{(1-a)^2},  & \frac{2}{1-a} 
					\end{array}
					\right) }{ (\xi_{y^{-1}y^{-q-1}} - \xi_{uy^{-1}} \xi_{uy^{-q-1}})}
				\left(\begin{array}{cc}
					- \xi_{uy^{-q-1}} & 1\\
					\xi_{y^{-1}y^{-q-1}}  & -\xi_{uy^{-1}} 
				\end{array}\right)
				\left(
				\begin{array}{c}
					\displaystyle \frac{1}{T-q} \sum_{t} U_{t} Z_{t} \\
					\displaystyle \frac{1}{T-q} \sum_{t} Y_{t-q-1} Z_{t}
				\end{array}
				\right) +o_p(T^{-\frac{1}{2}}) \\
				& \overset{(a)}{\rightarrow} \left(\begin{array}{cc} 
					\frac{2}{1-a},  & \frac{\frac{2b}{(1-a)^2}   - \frac{2  \xi_{u y^{-1}}}{1-a} }{\xi_{y^{-1}y^{-q-1}} - \xi_{uy^{-1}} \xi_{uy^{-q-1}}} 
				\end{array} 	\right)
				\left(
				\begin{array}{c}
					\displaystyle \frac{1}{T-q} \sum_{t} U_{t} Z_{t} \\
					\displaystyle \frac{1}{T-q} \sum_{t} Y_{t-q-1} Z_{t}
				\end{array}
				\right)+o_p(T^{-\frac{1}{2}}) \\
				&  \overset{(b)}{\rightarrow}  \frac{2}{1-a}  \frac{1}{T-q} \sum_{t} U_{t} Z_{t} +o_p(T^{-1/2}),
		\end{aligned} \end{equation}
	where $(a)$ utilizes $\frac{2b}{(1-a)^2}  \xi_{u y^{-1}}  = O(\text{ATE}^2)$ and $\xi_{u y^{-q-1}} \xi_{u y^{-1}} = O(\text{ATE}^2)$, both of which tend to 0 as $\text{ATE} \rightarrow 0$. $(b)$ relies on $\frac{2b}{(1-a)^2}  \sum_{t} Y_{t-q-1} Z_{t} = O(\text{ATE})O_p(T^{-1/2})$ and $\xi_{u y^{-1}} \sum_{t} Y_{t-q-1} Z_{t} =  O(\text{ATE}) O_p(T^{-1/2})$, which tend to 0 as $\text{ATE} \rightarrow 0$. Consequently, under the small signal asymptotic, we significantly simplify the calculations, and it eventually leads to the following asymptotic MSEs form under the AT design $\pi_{\text{AT}}$:
		\begin{equation}
			\begin{aligned}\nonumber
				\lim_{\substack{T \rightarrow +\infty\\ \text{ATE}\to 0}} \text{MSE}(\sqrt{T} \widehat{\text{ATE}}(\pi_{\text{AT}})) &  =
				\frac{4\sigma^2}{(1-a)^2}\Big[ \sum_{j=0}^q \theta_j^2+2\sum_{k=1}^q \sum_{j=k}^q (-1)^k \theta_j\theta_{j-k} \Big],
			\end{aligned}
		\end{equation}
	where  $\text{Cov}(U_{j}Z_{j}, U_{k} Z_{k}) =  (-1)^{j-k} \sum_{i=j-k}^{q} \theta_i \theta_{i-(j-k)} \sigma^2$ when $j-k<q$. By comparing the asymptotic MSEs under the three designs, we define $\textrm{EI}_{\textrm{AD}}=\sum_{k=1}^q \sum_{j=k}^q \theta_j\theta_{j-k}$ and $\textrm{EI}_{\textrm{AT}}=\sum_{k=1}^q (-1)^k \sum_{j=k}^q \theta_j\theta_{j-k}$. Under the small signal conditions, these two efficiency indicators determine the statistical efficiency of the ATE estimator in the controlled ARMA($1, q$) among AT, UR, and AD. More explanation is provided in Section~\ref{subsec:smallsignal}.  One remark is that if the ATE signal is large, the asymptotic MSE will include one extra bias term for UR and two for AT, potentially leading to larger asymptotic MSEs than AD.
 

 \subsection{Proof in Controlled ARMA(\texorpdfstring{$p, q$}{})}\label{appendix:sec_armapq}
 
     We next extend our analysis to control ARMA($p, q$) model: $Y_{t} = \sum_{j=1}^{p} a_j Y_{t-j} + b U_{t} + Z_{t}$, where we additionally consider an AR($p$) part with coefficients $a_1, \ldots,  a_p$ and the true ATE is also $2b/(1-a)$ with $a=a_1 + \ldots+ a_p$.   Due to extra coefficients to estimate in AR($p$) part, we multiply $U_{t}, Y_{t-q-1}, Y_{t-q-2}, \ldots,  Y_{t-q-p}$ on the model equation:
\begin{equation}
\begin{cases}
    \underbrace{\frac{1}{T_p} \sum_t \Mean (Y_{t} U_{t})}_{\xi_{uy}}&=\sum_{j=1}^{p} a_j \underbrace{\frac{1}{T_p} \sum_t \Mean (Y_{t-j} U_{t})}_{\xi_{uy^{-j}}}+b, \\
     \underbrace{\frac{1}{T_p} \sum_t \Mean(Y_{t} Y_{t-q-1})}_{\xi_{y y^{-q-1}}}&= \sum_{j=1}^{p} a_j\underbrace{\frac{1}{T_p} \sum_t \Mean (Y_{t-j} Y_{t-q-1})}_{\xi_{y^{-j}y^{-q-1}}}+b \underbrace{\frac{1}{T_p} \sum_t \Mean (U_{t} Y_{t-q-1})}_{\xi_{u y^{-q-1}}},\\
      &\cdots \\
      \underbrace{\frac{1}{T_p} \sum_t \Mean (Y_{t} Y_{t-q-p})}_{\xi_{y y^{-q-p}}}&= \sum_{j=1}^{p} a_j \underbrace{\frac{1}{T_p} \sum_t \Mean (Y_{t-j} Y_{t-q-p})}_{\xi_{y^{-j}y^{-q-p}}}+b \underbrace{\frac{1}{T_p} \sum_t \Mean (U_{t} Y_{t-q-p})}_{\xi_{uy^{-q-p}}}.
\end{cases}
\end{equation}
We denote $T_p = T-q-p$ and replace the expectation terms above by their moments:
		\begin{equation}\nonumber
				\begin{cases}
					\underbrace{\frac{1}{T_p}\sum_{t} Y_{t} U_{t}}_{\widehat{\xi}_{uy}} &= \sum_{j=1}^{p} \widehat{a}_j \underbrace{\frac{1}{T_p}\sum_{t} Y_{t-j} U_{t}}_{\widehat{\xi}_{uy^{-j}}} + \widehat{b} \\
					\underbrace{\frac{1}{T_p}\sum_{t} Y_{t} Y_{t-q-1}}_{\widehat{\xi}_{y y^{-q-1}}}&= \sum_{j=1}^{p} \widehat{a}_j  \underbrace{\frac{1}{T_p}\sum_{t} Y_{t-j} Y_{t-q-1}}_{\widehat{\xi}_{y^{-j}y^{-q-1}}} + \widehat{b} \underbrace{\frac{1}{T_p}\sum_{t} Y_{t-q-1} U_{t}}_{\widehat{\xi}_{uy^{-q-1}}},\\
					&\cdots \\
					\underbrace{\frac{1}{T_p}\sum_{t} Y_{t} Y_{t-q-p}}_{\widehat{\xi}_{y y^{-q-p}}}  &= \sum_{j=1}^{p} \widehat{a}_j  \underbrace{\frac{1}{T_p} \sum_{t} Y_{t-j} Y_{t-q-p}}_{\widehat{\xi}_{y^{-j}y^{-q-p}}} + \widehat{b} \underbrace{\frac{1}{T_p}\sum_{t} Y_{t-q-p} U_{t}}_{\widehat{\xi}_{uy^{-q-p}}}.
				\end{cases}    
		\end{equation}
	    Similar to controlled ARMA$(1, q)$, we have:
		\begin{equation}
			\begin{aligned}\nonumber
				\left(\begin{array}{c} 
					\widehat{a}_1 - a_1 \\
					\widehat{a}_2 - a_2 \\
					\cdots \\
					\widehat{b} - b
				\end{array}
				\right) & = 
				\left(\begin{array}{ccc|c} 
					\xi_{uy^{-1}}  & \cdots & \xi_{uy^{-p}} & 1 \\
					\hline 
					{\xi}_{y^{-1}y^{-q-1}}  & \cdots & {\xi}_{y^{-p}y^{-q-1}} & {\xi}_{uy^{-q-1}}   \\
					\cdots  & \cdots & \cdots & \cdots \\
					{\xi}_{y^{-1} y^{-q-p}} & \cdots & {\xi}_{y^{-p}y^{-q-p}} & {\xi}_{uy^{-q-p}}   \\
				\end{array}
				\right)^{-1}  \frac{1}{T_p}	\left(\begin{array}{c} 
						\sum_{t} U_{t} Z_{t} \\
					\sum_{t} Y_{t-q-1} Z_{t}\\
					\cdots \\
					\sum_{t} Y_{t-q-p} Z_{t}
				\end{array}
				\right)  + o_p(1) \\
				& = 	\left(\begin{array}{c|c} 
					{\xi}_A	& 1 \\
					\hline 
					{\xi}_B  & {\xi}_C  \\
				\end{array}
				\right)^{-1} \frac{1}{T_p}	\left(\begin{array}{c} 
						\sum_{t} U_{t} Z_{t} \\
					\sum_{t} Y_{t-q-1} Z_{t}\\
					\cdots \\
					\sum_{t} Y_{t-q-p} Z_{t}
				\end{array}
				\right)  + o_p(1),
			\end{aligned}
		\end{equation}
	where $\xi_A\in \mathbb{R}^{1\times p}$, $\xi_B \in \mathbb{R}^{p\times p}$ and $\xi_C\in \mathbb{R}^{p\times 1}$ represent each block component. Next,
		\begin{equation}
			\begin{aligned}\nonumber
				\widehat{\text{ATE}} - \text{ATE} = \left[ \underbrace{\frac{2b}{(1-a)^2}, \ldots, \frac{2b}{(1-a)^2}}_{p}, \frac{2}{1-a} \right]	\left(\begin{array}{c|c} 
					{\xi}_A	& 1 \\
					\hline 
					{\xi}_B  & {\xi}_C  \\
				\end{array}
				\right)^{-1}	\frac{1}{T_p}	\left(\begin{array}{c} 
					\sum_{t} U_{t} Z_{t} \\
					\sum_{t} Y_{t-q-1} Z_{t}\\
					\cdots \\
					\sum_{t} Y_{t-q-p} Z_{t}
				\end{array}
				\right)  + o_p(T_p^{-\frac{1}{2}}),				
			\end{aligned}
		\end{equation}
 \textbf{For the AD design,} we have $\xi_{u y^{-t}}=b/(1-a), t\geq 1$ as $\tau \rightarrow +\infty$, which has the equation:
		\begin{equation}
			\begin{aligned}\nonumber
				\left[ \frac{2}{1-a}, \mathbf{0}_p^\top \right]	\left(\begin{array}{c|c} 
					{\xi}_A	& 1 \\
					\hline 
					{\xi}_B  & {\xi}_C  \\
				\end{array}
				\right) = \left[ \underbrace{\frac{2b}{(1-a)^2}, \ldots, \frac{2b}{(1-a)^2}}_{p}, \frac{2}{1-a} \right],
			\end{aligned}
		\end{equation}
 where $\mathbf{0}_p \in \mathbb{R}^p$ is a zero-vector. The same results as controlled ARMA($1, q$) are obtained:
		\begin{equation}\begin{aligned}\nonumber
				\widehat{\text{ATE}} - \text{ATE} = \frac{2}{1-a}  \frac{1}{T_p} \sum_{t} U_{t} Z_{t} +o_p(T_p^{-1/2}).
		\end{aligned} \end{equation}
\textbf{For the UR design,} it suggests that $\xi_{u y^{-t}}=0$ for $t\geq1$ and we have:
		\begin{equation}
			\begin{aligned}\nonumber
				\left(\begin{array}{c|c} 
					{\xi}_A	& 1 \\
					\hline 
					{\xi}_B  & {\xi}_C  \\
				\end{array}
				\right)^{-1}	= 	\left(\begin{array}{c|c} 
					\mathbf{0}_p^\top	& 1 \\
					\hline 
					{\xi}_B  & \mathbf{0}_p  \\
				\end{array}
				\right)^{-1} = 	\left(\begin{array}{c|c} 
					\mathbf{0}_p& {\xi}_B^{-1} \\
					\hline 
					1  & \mathbf{0}_p^\top  \\
				\end{array}
				\right).
			\end{aligned}
		\end{equation}
 We simplify the first-order Taylor expansion of ATE estimation result under the UR design:
		\begin{equation}
			\begin{aligned}\label{eq:proof_UR_p}
				\widehat{\text{ATE}} - \text{ATE} & = \left[ \left(\frac{2b}{(1-a)^2}\right)_p^\top,  \frac{2}{1-a} \right]	\left(\begin{array}{c|c} 
					\mathbf{0}_p& {\xi}_B^{-1} \\
					\hline 
					1  & \mathbf{0}_p^\top  \\
				\end{array}	\right)	\frac{1}{T_p}
				\left(\begin{array}{c} 
						\sum_{t} U_{t} Z_{t}\\
					\sum_{t} Y_{t-q-1} Z_{t}\\
					\cdots \\
					\sum_{t} Y_{t-q-p} Z_{t}
				\end{array}
				\right) + o_p(T_p^{-1/2}) \\
				& = \frac{1}{T_p} \left(\frac{2b}{(1-a)^2}\right)_p^\top \xi_B^{-1} 		\left(\begin{array}{c} 
					\sum_{t} Y_{t-q-1} Z_{t}\\
					\cdots \\
					\sum_{t} Y_{t-q-p} Z_{t}
				\end{array}
				\right)	 + \frac{2}{1-a} \frac{\sum_{t} U_{t} Z_{t}}{T_p}  +o_p(T_p^{-1/2})  \\
				& \overset{(a)}{\rightarrow} \frac{2}{1-a} \frac{1}{T_p} \sum_{t} U_{t} Z_{t}  +o_p(T_p^{-1/2}),
			\end{aligned}
		\end{equation}
 where $\left(\frac{2b}{(1-a)^2}\right)_p^\top$ is the $p$-dimension of $2b/(1-a)^2$ and $\xi_B$ is the $p$-dimensional extension of $\xi_{y^{-1} y^{-q-1}}=\sigma^2 d(a, \theta)$, independent of $b$. For $(a)$, the first term in the second line of \eqref{eq:proof_UR_p} is still $O(\text{ATE})O_p(T_p^{-1/2})$, which tends 0 under the small signal asymptotic by letting $\text{ATE} \rightarrow 0$. Eventually, the asymptotic MSE in controlled AMMA($p, q$) under the UR design has a similar form as the controlled ARMA($1, q$) in \eqref{eq:asymptoticMSE_controlarma_1p_ur}. \\
 \textbf{For the AT design,} by symmetry, $\xi_{u y^{-t}}$ is a periodic function with the period 2. If $p$ is even, the two equations when we take the limit regarding $\xi_{u y^{-t}}$ are given by:
		\begin{equation}
			\begin{aligned}\nonumber
				\xi_{u y} = a_1 \xi_{u y^{-1}} + a_2 \xi_{u y} +\ldots + a_p \xi_{u y} - b, \quad \xi_{u y^{-1}} = a_1 \xi_{u y} + a_2  \xi_{u y^{-1}} +\ldots+ a_p \xi_{u y^{-1}} + b,
			\end{aligned}
		\end{equation}
	where we denote $a_e = a_2 + a_4 + \ldots+ a_p$ and $a_o=a_1 + a_3 + \ldots+ a_{p-1}$ as the sum of even and odd coefficients in the AR(p) part, respectively. The solution is then given by $\xi_{u y} = -\frac{b}{1-a_e + a_o}$, $\xi_{u y^{-1}} = \frac{b}{1-a_e + a_o}$, and $\xi_{u y^{-t}} = (-1)^{t+1} \frac{b}{1-a_e + a_o}$. When $p$ is odd, the solution of $\xi_{u y^{-t}}$ is also the same. Notice that $\xi_{u y^{-t}} = O(\text{ATE})$ as the true ATE is typical $\propto b$. Based on the inverse matrix formula of the two-dimensional block matrix, we obtain that
		\begin{equation}
			\begin{aligned}\nonumber
				\left(\begin{array}{c|c} 
					{\xi}_A	& 1 \\
					\hline 
					{\xi}_B  & {\xi}_C  \\
				\end{array}
				\right)^{-1}	= 	\left(\begin{array}{c|c} 
					-(\xi_B - \xi_C \xi_A)^{-1} \xi_C &(\xi_B - \xi_C \xi_A)^{-1} \\
					1 + \xi_A (\xi_B - \xi_C \xi_A)^{-1} \xi_C  & -\xi_A (\xi_B - \xi_C \xi_A)^{-1}  \\
				\end{array}
				\right),
			\end{aligned}
		\end{equation}
	where we find
		\begin{equation}
			\begin{aligned}\nonumber
				\xi_B - \xi_C \xi_A	= 	\left(\begin{array}{ccc} 
					\xi_{y^{-1} y^{-q-1}} & \cdots &\xi_{y^{-p} y^{-q-1}}  \\
					\cdots & \cdots & \cdots \\
					\xi_{y^{-1} y^{-q-p}} & \cdots &\xi_{y^{-p} y^{-q-p}}\\
				\end{array}
				\right) - 
				\left(\begin{array}{c} 
					\xi_{u y^{-q-1}} \\
					\cdots \\
					\xi_{u y^{-q-p}} 
				\end{array}
				\right) 
				\left(\begin{array}{ccc} 
					\xi_{u y^{-1}}, & \cdots &,  \xi_{u y^{-p}}
				\end{array}
				\right)
			\end{aligned}
		\end{equation}
	is exactly the p-dimensional extension of $\xi_{y^{-1} y^{-q-1}} - \xi_{u y^{-1}} \xi_{u y^{-q-1}} = \sigma^2 d(a, \theta)$ in controlled ARMA($1, q$), which is independent of $b$. Since each element in $\xi_A$ and $\xi_C$ satisfies $\xi_{u y^{-t}} = O(\text{ATE})$ for $t\geq1$, $\xi_A (\xi_B - \xi_C \xi_A)^{-1} \xi_C$ is therefore a quadratic function of ATE, i.e., $O(\text{ATE}^2)$. Then, the ATE estimator under the AT design can be simplified as:
		\begin{equation}
			\begin{aligned}\nonumber
				\widehat{\text{ATE}} - \text{ATE} 
				& \overset{(a)}{\rightarrow} \frac{2}{1-a} (1 + \xi_A (\xi_B - \xi_C \xi_A)^{-1} \xi_C) \frac{1}{T_p} \sum_{t} U_{t} Z_{i, t+1}  +o_p(T_p^{-1/2}) \\
				& \overset{(b)}{\rightarrow} \frac{2}{1-a} \frac{1}{T_p} \sum_{t} U_{t} Z_{t+1}  +o_p(T_p^{-1/2}),
			\end{aligned}
		\end{equation}
	where $(a)$ relies on $(\frac{2b}{(1-a)^2})_p^\top (\xi_B - \xi_C \xi_A)^{-1} \xi_C = O(\text{ATE}^2)$, $(\frac{2b}{(1-a)^2})_p^\top (\xi_B - \xi_C \xi_A)^{-1} \sum_{j} Y_{j} Z_{j+1} = O(\text{ATE})O_p(T_p^{-1/2})$ and $\xi_A (\xi_B - \xi_C \xi_A)^{-1}  \sum_{j} Y_{j} Z_{j+1} = O(\text{ATE})O_p(T_p^{-1/2})$, all of which tend to zero as $\text{ATE} \rightarrow 0$. $(b)$ leverages $\xi_A (\xi_B - \xi_C \xi_A)^{-1} \xi_C = O(\text{ATE}^2)$ as analyzed earlier. Therefore, within the small signal asymptotic, the ATE estimators under AT, UR, and AD in the controlled ARMA($p, q$) all have a similar form as those in the controlled ARMA($1, q$). This also implies that the resulting asymptotic MSEs of them in the controlled ARMA($p, q$) also share the same form as those in the controlled ARMA($1, q$), which we derived in Appendix~\ref{appendix:sec_arma1q}. This same form also applies to the efficiency indicators.

\section{Estimation, Asymptotic MSEs, and Efficiency Indicators in Controlled VARMA}\label{appendix:efficiency_VARMA}

  The outline of our proof in this section is:
    \begin{itemize}
        \item \textbf{Appendix~\ref{appendix:sec_varma1q}} Controlled VARMA($1, q$) from AD, UR to AT design.
        \item \textbf{Appendix~\ref{appendix:sec_varmapq}}  Controlled VARMA($p, q$) from AD, UR to AT design.
    \end{itemize}
    As AD, UR, and AT are all balanced designs, i.e., $\xi_\pi=0$, we consider the proof in the controlled VARMA model without the intercept term $\boldsymbol{\mu}$ in this section. Additionally, we exclude the exogenous variable as well since it remains unaffected by the treatments. 
    
\subsection{Controlled VARMA(\texorpdfstring{$1, q$}{})}\label{appendix:sec_varma1q}
We start our proof from controlled VARMA($1, q$), which is formulated as:
		\begin{equation}
			\begin{aligned}\nonumber
				\mathbf{Y}_{t}=\mathbf{A} \mathbf{Y}_{t-1}+ \mathbf{b} U_{ t} + \mathbf{Z}_{t},
			\end{aligned}
		\end{equation}
	where the response vector $\{\mathbf{Y}_{t}\}_{t}\in \mathbb{R}^{d}$ has 1-order lagging term with the coefficient matrix $\mathbf{A} \in \mathbb{R}^{d \times d}$. Next, we estimate $\mathbf{A}$ and $\mathbf{b}$ by multiplying $\mathbf{Y}_{t-q-1}^\top$ and $U_{t}$, and then take the expectation on both sides, which results in the following equations:
		\begin{equation}
			\begin{aligned}\nonumber
				\underbrace{\frac{1}{T-q} \sum_{t=q+1}^{T} \Mean( \mathbf{Y}_{t} \mathbf{Y}_{t-q-1}^\top )}_{{\xi}_{y y^{-q-1}}}&=  {\mathbf{A}}	\underbrace{ \frac{1}{T-q} \sum_{t=q+1}^{T} \Mean( \mathbf{Y}_{t-1} \mathbf{Y}_{t-q-1}^\top)}_{{\xi}_{y^{-1} y^{-q-1}}}+ {\mathbf{b}} \underbrace{\frac{1}{T-q} \sum_{t=q+1}^{T} \Mean( U_{t} \mathbf{Y}_{t-q-1}^\top)}_{{\xi}_{u y^{-q-1}}^\top}\\
				\underbrace{\frac{1}{T-q} \sum_{t=q+1}^{T} \Mean( U_{t} \mathbf{Y}_{t})}_{{\xi}_{u y}}&=  {\mathbf{A}}	\underbrace{\frac{1}{T-q} \sum_{t=q+1}^{T} \Mean( U_{t} \mathbf{Y}_{t-1} )}_{{\xi}_{u y^{-1} }}+ {\mathbf{b}} .
			\end{aligned}
		\end{equation}
 We next replace the expectation terms with their sample moments to apply the method of moments estimation:
		\begin{equation}
			\begin{aligned}\nonumber
				\underbrace{\frac{1}{T-q} \sum_{t=q+1}^{T} \mathbf{Y}_{t} \mathbf{Y}_{t-q-1}^\top}_{\widehat{\xi}_{y y^{-q-1}}}&=  \widehat{\mathbf{A}}	\underbrace{\frac{1}{T-q} \sum_{t=q+1}^{T} \mathbf{Y}_{t-1} \mathbf{Y}_{t-q-1}^\top}_{\widehat{\xi}_{y^{-1} y^{-q-1}}}+ \widehat{\mathbf{b}} \underbrace{\frac{1}{T-q}  \sum_{t=q+1}^{T} U_{t} \mathbf{Y}_{t-q-1}^\top}_{\widehat{\xi}_{u y^{-q-1}}^\top}\\
				\underbrace{\frac{1}{T-q} \sum_{t=q+1}^{T} U_{t} \mathbf{Y}_{t}}_{\widehat{\xi}_{u y}}&=  \widehat{\mathbf{A}}	\underbrace{\frac{1}{T-q} \sum_{t=q+1}^{T} U_{t} \mathbf{Y}_{t-1} }_{\widehat{\xi}_{u y^{-1} }}+ \widehat{\mathbf{b}} .
			\end{aligned}
		\end{equation}
Therefore, we have $d \times (d+1)$ equations to estimate the $d \times (d+1)$ parameters in  $\widehat{\mathbf{A}}$ and $\widehat{\mathbf{b}}$. To construct the ATE estimator equations, recall that we have another group of equations based on the true parameters $\mathbf{A}$ and  $\mathbf{b}$ without taking the expectation:
		\begin{equation}
			\begin{aligned}\nonumber
				\widehat{\xi}_{y y^{-q-1}} &=\mathbf{A} \widehat{\xi}_{y^{-1} y^{-q-1}} + \mathbf{b} \widehat{\xi}_{u y^{-q-1}}^\top + \frac{1}{T-q}\sum_{t=q+1}^{T} \mathbf{Z}_{t} \mathbf{Y}_{t-q-1}^\top \\
				\widehat{\xi}_{u y} &= \mathbf{A} \widehat{\xi}_{u y^{-1}} + \mathbf{b} + \frac{1}{T-q} \sum_{t=q+1}^{T} U_{t}\mathbf{Z}_{t}.
			\end{aligned}
		\end{equation}
	Similar to the proof procedure in controlled ARMA($1, q$), we combine these two groups of estimation equations and the resulting estimation function is given by:
		\begin{equation}
			\begin{aligned}\nonumber
				\left[\widehat{\mathbf{A}} - \mathbf{A}, \widehat{\mathbf{b}} - \mathbf{b}\right] 
				&=\frac{1}{T-q}  \left[\sum_{t=q+1}^{T} U_{t}\mathbf{Z}_{t} , \sum_{t=q+1}^{T} \mathbf{Z}_{t} \mathbf{Y}_{t-q-1}^\top   \right]
				\left(\begin{array}{cc}
					\xi_{uy^{-1}} & \xi_{y^{-1} y^{-q-1}} \\
					1 & \xi_{u y^{-q-1} }^\top 
				\end{array}
				\right)^{-1} + o_p(1),
			\end{aligned}
		\end{equation}
        where $\xi_{uy^{-1}}$, $\xi_{y^{-1} y^{-q-1}}$ and $\xi_{u y^{-q-1} }^\top $ are the expectation of $\widehat{\xi}_{u y^{-1}}$, $\widehat{\xi}_{y^{-1} y^{-q-1}}$ and $\widehat{\xi}_{u y^{-q-1}}^\top$, respectively. Then, we vectorize all the parameters:
		\begin{equation}
			\begin{aligned}\nonumber
				\left(\begin{array}{c}
					\text{vec}(\widehat{\mathbf{A}} - \mathbf{A}) \\
					\text{vec}(\widehat{\mathbf{b}} - \mathbf{b})
				\end{array}
				\right)
				=  \frac{1}{T-q}  \left( 	\left(\begin{array}{cc}
					\xi_{uy^{-1}}^\top & 1 \\
					\xi_{y^{-1} y^{-q-1}}^\top & \xi_{u y^{-q-1} } 
				\end{array}
				\right)^{-1} \otimes \mathbb{I}_d \right) 	
				\left(\begin{array}{c}
					\displaystyle \sum_{t} U_{t}\mathbf{Z}_{t} \\
					\displaystyle \text{vec}(\sum_{t} \mathbf{Z}_{t} \mathbf{Y}_{t-q-1}^\top )
				\end{array}
				\right)
				+ o_p(1),
			\end{aligned}
		\end{equation}
	where $\otimes$ is the Kronecker product and we use the formula $\text{vec}(AB) = (B^\top \otimes \mathbb{I}_k) \text{vec}(A)$ for any matrix $A$ and $B$ with $A \in \mathbb{R}^{k\times I}$ and $B \in \mathbb{R}^{I \times m}$. By applying the Taylor expansion to the formula of the true ATE, we have the following ATE estimator equation:
		\begin{equation}
			\begin{aligned}\label{eq:proof_VARMA_ATE}
				\widehat{\textrm{ATE}} - \text{ATE} &= J_{\text{vec}(\mathbf{A}), \mathbf{b}} 	\left(\begin{array}{c}
					\text{vec}(\widehat{\mathbf{A}} - \mathbf{A}) \\
					\text{vec}(\widehat{\mathbf{b}} - \mathbf{b})
				\end{array}
				\right)  + o_p(T^{-1/2})\\
				& = \left[ 2 \mathbf{b}^\top (\mathbb{I} - \mathbf{A}^\top)^{-1} \otimes \bm{e}^\top (\mathbb{I} - \mathbf{A})^{-1}, 2\bm{e}^\top (\mathbb{I} - \mathbf{A})^{-1} \right] 	\left(\begin{array}{c}
					\text{vec}(\widehat{\mathbf{A}} - \mathbf{A}) \\
					\text{vec}(\widehat{\mathbf{b}} - \mathbf{b})
				\end{array}
				\right)  + o_p(T^{-1/2}) ,
			\end{aligned}
		\end{equation}
	where $J_{\text{vec}(\mathbf{A}), \mathbf{b}}$ is the Jacobian matrix of the true ATE $=2 \bm{e}^\top (\mathbb{I} - \mathbf{A})^{-1} \mathbf{b}$ in terms of the vectorized coefficients $\text{vec}(\mathbf{A})$ and $\mathbf{b}$. We define $f(\mathbf{A}, \mathbf{b}) = 2 \bm{e}^\top (\mathbb{I} - \mathbf{A})^{-1} \mathbf{b}$ and highlight that the evaluation of  $J_{\text{vec}(\mathbf{A}), \mathbf{b}}$ is based on the following two formulas about the derivative over matrix and vectorization method: 
 \begin{enumerate}
     \item $(\frac{\partial f}{\partial \mathbf{A}})^\top =  2 (\mathbb{I} - \mathbf{A}^\top)^{-1} \bm{e} \mathbf{b}^\top (\mathbb{I} - \mathbf{A}^\top)^{-1}$ and  $(\frac{\partial f}{\partial \mathbf{b}})^\top =  2(\mathbb{I} - \mathbf{A}^\top)^{-1} \bm{e}$
     \item $\text{vec}(AXB) = (B^\top \otimes A) \text{vec}(X)$ and $(A \otimes B) (C \otimes D) = AC \otimes BD$ for any matrix $A, B$ and $X$.
 \end{enumerate}
Based on the two formulas above, we take the differential on the true ATE formula over $\mathbf{A}$:
		\begin{equation}
			\begin{aligned}\nonumber
				d f &= 2\bm{e}^\top (d(\mathbb{I} - \mathbf{A})^{-1} \textbf{b})  + 2(d \bm{e}^\top) (\mathbb{I} - \mathbf{A})^{-1} \textbf{b} = 2\bm{e}^\top((d (\mathbb{I} - \mathbf{A})^{-1}) \textbf{b} + (\mathbb{I} - \mathbf{A})^{-1} (d \textbf{b})) + 0\\
				& = 2\bm{e}^\top((d (\mathbb{I} - \mathbf{A})^{-1}) \textbf{b}
			\end{aligned}
		\end{equation}
	Next, we take the differential on the equation $(\mathbb{I} - \mathbf{A}) (\mathbb{I} - \mathbf{A})^{-1} = \mathbb{I}$ over $\mathbf{A}$ and obtain $(d (\mathbb{I} - \mathbf{A}) ) (\mathbb{I} - \mathbf{A})^{-1} + (\mathbb{I} - \mathbf{A}) (d (\mathbb{I} - \mathbf{A})^{-1})= 0$, which immediately implies:
		\begin{equation}
			\begin{aligned}\nonumber
				d (\mathbb{I} - \mathbf{A})^{-1}  = - (\mathbb{I} - \mathbf{A})^{-1} (d (\mathbb{I} - \mathbf{A}) ) (\mathbb{I} - \mathbf{A})^{-1}  = (\mathbb{I} - \mathbf{A})^{-1}  d \textbf{A} (\mathbb{I} - \mathbf{A})^{-1}.
			\end{aligned}
		\end{equation}
According to the derivative formula over matrix $d f = \text{tr}((\frac{\partial f}{\partial \textbf{A}})^\top d \textbf{A})$, we then have:
		\begin{equation}
			\begin{aligned}\nonumber
				df = \text{tr}(df) = \text{tr}(2\bm{e}^\top  (\mathbb{I} - \mathbf{A})^{-1}  d \textbf{A} (\mathbb{I} - \mathbf{A})^{-1} \textbf{b}) = \text{tr} (2 ((\mathbb{I} - \mathbf{A}^\top)^{-1} \bm{e} \mathbf{b}^\top (\mathbb{I} - \mathbf{A}^\top)^{-1})^\top d \mathbf{A}),
			\end{aligned}
		\end{equation}
	where $\frac{\partial f}{\partial \mathbf{A}} = 2(\mathbb{I} - \mathbf{A}^\top)^{-1} (\bm{e} \mathbf{b}^\top) (\mathbb{I} - \mathbf{A}^\top)^{-1}$. We further vectorize it, which leads to:
		\begin{equation}
			\begin{aligned}\nonumber
				\text{vec}(\frac{\partial f}{\partial \mathbf{A}}) & = 2[(\mathbb{I} - \mathbf{A})^{-1} \otimes (\mathbb{I} - \mathbf{A}^\top)^{-1}] \cdot \text{vec}(\bm{e} \mathbf{b}^\top)  
				 = 2[(\mathbb{I} - \mathbf{A})^{-1} \otimes (\mathbb{I} - \mathbf{A}^\top)^{-1}] \cdot (\mathbf{b} \otimes \bm{e}) \\
				& = 2 ((\mathbb{I} - \mathbf{A})^{-1} \mathbf{b}) \otimes ((\mathbb{I} - \mathbf{A}^\top)^{-1}  \bm{e}).
			\end{aligned}
		\end{equation}
	According to the formula $(A \otimes B)^\top = A^\top \otimes B^\top$ for any matrix $A$ and $B$ and $J_{\text{vec}(\mathbf{A}), \mathbf{b}} = (\text{vec}(\frac{\partial f}{\partial \mathbf{A}})^\top, \frac{\partial f}{\partial \mathbf{b}})^\top)^\top$, we finally arrive at the ATE estimation equation in \eqref{eq:proof_VARMA_ATE}. To simplify the asymptotic MSE within the small signal asymptotics, we next consider scenarios of different designs, including AD, UR, and AT.  
 
    \noindent \textbf{For the AD design,} after some calculations, we have $\xi_{u y^{-t}} = (\mathbb{I}-\mathbf{A})^{-1} \mathbf{b}$ for $t\geq 1$. Similar to Controlled AMRA($1, q$), we can also have an exact equation:
		\begin{equation}
			\begin{aligned}\nonumber
				& \left[ 2 \bm{e}^\top  (\mathbb{I} - \mathbf{A})^{-1}, \mathbf{0}^\top_{d^2}  \right] 
				\left(	\left(\begin{array}{cc}
					\xi_{uy^{-1}}^\top & 1 \\
					\xi_{y^{-1} y^{-q-1}}^\top & \xi_{u y^{-q-1} } 
				\end{array}
				\right) \otimes \mathbb{I}_d \right) \\
    & = \left[ 2 \mathbf{b}^\top (\mathbb{I} - \mathbf{A}^\top)^{-1} \otimes \bm{e}^\top (\mathbb{I} - \mathbf{A})^{-1}, 2\bm{e}^\top (\mathbb{I} - \mathbf{A})^{-1} \right], 
			\end{aligned}
		\end{equation}
	where $\mathbf{0}_{d^2}$ is a zero-vector with length $d^2$ and  we apply $\mathbf{\mu}^\top (\mathbf{\nu}^\top \otimes \mathbb{I}_d) = \mathbf{\nu}^\top \otimes \mathbf{\mu}^\top$ for arbitrary vectors $\mathbf{\mu}$ and $\mathbf{\nu}$. We have $2 \bm{e}^\top (\mathbb{I} - \mathbf{A})^{-1} (\mathbf{b}^\top (\mathbb{I} - \mathbf{A}^\top)^{-1} \otimes \mathbb{I}_d) = 2 \mathbf{b}^\top (\mathbb{I} - \mathbf{A}^\top)^{-1} \otimes \bm{e}^\top (\mathbb{I} - \mathbf{A})^{-1}$. Hence, under the AD design, the ATE estimation can be precisely simplified as:
		\begin{equation}
			\begin{aligned}\nonumber
				\widehat{\textrm{ATE}} - \text{ATE} &= 2 \bm{e}^\top (\mathbb{I} - \mathbf{A})^{-1}  \frac{1}{T-q}\sum_{t} U_{t}\mathbf{Z}_{t} + o_p(T^{-1/2}).
			\end{aligned}
		\end{equation}
	Due to the fact that $U_t$ is uncorrelated with $Z_{t}$, the asymptotic MSE of the ATE estimator can be simplified as:
    \begin{equation}
        \begin{aligned}\nonumber
 & \lim_{T\rightarrow +\infty}  \text{MSE}(\sqrt{T}\widehat{\text{ATE}}(\pi))  \\
 & = \lim_{T\rightarrow +\infty}  \mathbb{E} \left[ 
2 \bm{e}^\top (\mathbb{I} - \mathbf{A})^{-1} \frac{1}{\sqrt{T}} \sum_{t=1}^T U_{t} \mathbf{Z}_{t}  + o_p(1) \right]^2 \\ 
& =  \lim_{T\rightarrow+\infty} \frac{1}{T} \left( 4 \bm{e}^\top (\mathbb{I} - \mathbf{A})^{-1} \text{Var} \left( \sum_{t=1}^T U_{t} \mathbf{Z}_{t}  \right) (\mathbb{I} - \mathbf{A})^{-1} \bm{e} +  \left( 2 \bm{e}^\top (\mathbb{I} - \mathbf{A})^{-1} \mathbb{E}\left[ \sum_{t=1}^T U_{t} \mathbf{Z}_{t} \right] \right)^2 \right) \\
& = 4 \bm{e}^\top (\mathbb{I} - \mathbf{A})^{-1} \left( \lim_{T\rightarrow+\infty} \frac{1}{T} \text{Var} \left( \sum_{t=1}^T U_{t} \mathbf{Z}_{t} \right) \right)  (\mathbb{I} - \mathbf{A})^{-1} \bm{e}.
        \end{aligned}
        \end{equation}
As a consequence, the asymptotic MSE under the AD design $\pi_{\text{AD}}$ has the following form:
		\begin{equation}
			\begin{aligned}\nonumber
				\lim_{\substack{T \rightarrow +\infty\\ \tau\to +\infty}} \text{MSE}(\sqrt{T}\widehat{\text{ATE}}(\pi_{\text{AD}}))   & = 4 \bm{e}^\top (\mathbb{I} - \mathbf{A})^{-1} \left( \sum_{j=0}^{q} \mathbf{M}_j \Sigma \mathbf{M}_j + 2 \sum_{k=1}^{q}\sum_{j=k}^{q} \mathbf{M}_j \Sigma \mathbf{M}_{j-k} \right)  (\mathbb{I} - \mathbf{A})^{-1} \bm{e},
			\end{aligned}
		\end{equation}
 which is the direct extension of the asymptotic MSE in Controlled ARMA($1, q$) from \eqref{eq:asymptoticMSE_controlarma_1p_ad}. In particular, since $\mathbb{E}\left[U_{t}\right] = 0$ and $\mathbb{E}\left[\mathbf{Z}_{t}\right] = \mathbf{0}_d$, we have $\text{Cov}(U_{j}\mathbf{Z}_{j}, U_{k} \mathbf{Z}_{k}) = \text{Cov}(U_{j}, U_{k}) \text{Cov}(\mathbf{Z}_{j},  \mathbf{Z}_{k})$ when $j-k<q$. As $\lim_{\tau \rightarrow +\infty} \text{Cov}(U_{j}, U_{k}) = 1$, we have 
 $$\lim_{\tau \rightarrow +\infty} \text{Cov}(U_{j} \mathbf{Z}_{j}, U_{k} \mathbf{Z}_{k}) = \text{Cov}(\mathbf{Z}_{j},  \mathbf{Z}_{k}) = \sum_{i=j-k}^{q} \mathbf{M}_i \Sigma \mathbf{M}_{i-(j-k)}.$$
 
 \noindent \textbf{For the UR design,} specially we have $\xi_{u y^{-1}} = \xi_{u y^{-q-1}} = \mathbf{0}_d$. Therefore, we have:
		\begin{equation}
			\begin{aligned}\nonumber
				& \widehat{\textrm{ATE}} - \text{ATE} \\
                    &= J_{\text{vec}(\mathbf{A}), \mathbf{b}} 
				\left( 	\left(\begin{array}{cc}
					\mathbf{0}_d^\top & 1 \\
					\xi_{y^{-1} y^{-q-1}}^\top & \mathbf{0}_d 
				\end{array}
				\right)^{-1} \otimes \mathbb{I}_d \right) 	 \frac{1}{T-q}
				\left(\begin{array}{c}
					\displaystyle \sum_{t} U_{t}\mathbf{Z}_{t} \\
					\displaystyle \text{vec}(\sum_{t} \mathbf{Z}_{t} \mathbf{Y}_{t-q-1}^\top )
				\end{array}
				\right)
				+ o_p(T^{-1/2}) \\
				& = J_{\text{vec}(\mathbf{A}), \mathbf{b}}  
				\left( 	\left(\begin{array}{cc}
					\mathbf{0}_d & \xi_{y^{-1} y^{-q-1}}^{-\top} \\
					1 & \mathbf{0}_d^\top 
				\end{array}
				\right) \otimes \mathbb{I}_d \right)_{d(d+1) \times d(d+1)} 	 \frac{1}{T-q}
				\left(\begin{array}{c}
					\displaystyle  \sum_{t} U_{t}\mathbf{Z}_{t} \\
					\displaystyle \text{vec}(\sum_{t} \mathbf{Z}_{t} \mathbf{Y}_{t-q-1}^\top )
				\end{array}
				\right)
				+ o_p(T^{-1/2}) \\
				& =  J_{\text{vec}(\mathbf{A}), \mathbf{b}}   \frac{1}{T-q} \left(\begin{array}{c}
					\displaystyle  \xi_{y^{-1} y^{-q-1}}^{-\top}   \otimes  \mathbb{I}_d   \text{vec}(\sum_{t} \mathbf{Z}_{t} \mathbf{Y}_{t-q-1}^\top ) \\
					\displaystyle \sum_{t} U_{t}\mathbf{Z}_{t} 
				\end{array} 
				\right)_{d(d+1)} + o_p(T^{-1/2}) \\
				& \overset{(a)}{=} 2 \bm{e}^\top (\mathbb{I} - \mathbf{A})^{-1}  \frac{1}{T-q} \sum_{t=q+1}^{T} U_{t}\mathbf{Z}_{t} + O(\text{ATE})O_p(T^{-1/2}) + o_p(T^{-1/2}) \\
                & \rightarrow 2 \bm{e}^\top (\mathbb{I} - \mathbf{A})^{-1}  \frac{1}{T-q} \sum_{t=q+1}^{T} U_{t}\mathbf{Z}_{t} + o_p(T^{-1/2}) ,
			\end{aligned}
		\end{equation}
	where $J_{\text{vec}(\mathbf{A}), \mathbf{b}} = \left[ 2 \mathbf{b}^\top (\mathbb{I} - \mathbf{A}^\top)^{-1} \otimes \bm{e}^\top (\mathbb{I} - \mathbf{A})^{-1}, 2\bm{e}^\top (\mathbb{I} - \mathbf{A})^{-1} \right] $. In $(a)$, similar to controlled ARMA($1, q$), we can verify that the first term in the resulting product is $O(\text{ATE})O_p(T^{-1/2})$, which tends to 0 within the small signal asymptotics. We also denote  $(\xi_{y^{-1} y^{-q-1}}^{\top})^{-1} = \xi_{y^{-1} y^{-q-1}}^{-\top}$. Under UR design, we have the asymptotic MSE form:
		\begin{equation}
			\begin{aligned}\nonumber
					\lim_{\substack{T \rightarrow +\infty\\ \text{ATE}\to 0}}   \text{MSE}(\sqrt{T}\widehat{\text{ATE}}(\pi_{\text{UR}}))  & = 4 \bm{e}^\top (\mathbb{I} - \mathbf{A})^{-1} \left( \sum_{j=0}^{q} \mathbf{M}_j \Sigma \mathbf{M}_j \right)  (\mathbb{I} - \mathbf{A})^{-1} \bm{e}.
			\end{aligned}
		\end{equation}
	\noindent \textbf{For the AT design,} by multiplying $U_{t}$ and $U_{t-1}$, we solve the following equations, $\xi_{u y} = \mathbf{A} \xi_{u y^{-1}} + \mathbf{b}$ and $\xi_{u y^{-1}} = \mathbf{A} \xi_{u y^{-2}} - \mathbf{b}$, for which we can derive $\xi_{u y} =  \xi_{u y^{-2}}  = (\mathbb{I} + \mathbf{A})^{-1} \mathbf{b}$ and $\xi_{u y^{-1}}  = - (\mathbb{I} + \mathbf{A})^{-1} \mathbf{b}$. According to the inverse matrix formula, we have:
		\begin{equation}
			\begin{aligned}\nonumber
				& \left(\begin{array}{c|c}
					\xi_{uy^{-1}}^\top & 1 \\
					\hline
					\xi_{y^{-1} y^{-q-1}}^\top & \xi_{u y^{-q-1} } 
				\end{array}
				\right)^{-1}   \\
                    & = \left(\begin{array}{c|c}
					-(\xi_{y^{-1} y^{-q-1}}^\top - \xi_{u y^{-q-1}} \xi_{u y^{-1}}^\top)^{-1} \xi_{u y^{-q-1} }  & (\xi_{y^{-1} y^{-q-1}}^\top - \xi_{u y^{-q-1}} \xi_{u y^{-1}}^\top)^{-1} \\
					\hline
					1 + \xi_{u y^{-1}}^\top (\xi_{y^{-1} y^{-q-1}}^\top - \xi_{u y^{-q-1}} \xi_{u y^{-1}}^\top)^{-1} \xi_{u y^{-q-1}} &  -\xi_{u y^{-1}}^\top (\xi_{y^{-1} y^{-q-1}}^\top - \xi_{u y^{-q-1}} \xi_{u y^{-1}}^\top)^{-1}
				\end{array}
				\right) \\
                    & \equiv \boldsymbol{\xi}_{1, q},
			\end{aligned}
		\end{equation}
	where we use $\boldsymbol{\xi}_{1, q}$ to denote the matrix of interest for convenience. By applying the Taylor expansion, we have:
		\begin{equation}
			\begin{aligned}\nonumber
				& \widehat{\textrm{ATE}} - \text{ATE}  \\
                    & =  \left[ 2 \mathbf{b}^\top (\mathbb{I} - \mathbf{A}^\top)^{-1} \otimes \bm{e}^\top (\mathbb{I} - \mathbf{A})^{-1}, 2\bm{e}^\top (\mathbb{I} - \mathbf{A})^{-1} \right]  \frac{1}{T-q}  \left( \boldsymbol{\xi}_{1, q}  \otimes \mathbb{I}_d \right)  \left(\begin{array}{c}
					\displaystyle\sum_{t} U_{t}\mathbf{Z}_{t} \\
					\displaystyle \text{vec}(\sum_{t} \mathbf{Z}_{t} \mathbf{Y}_{t-q-1}^\top )
				\end{array}
				\right) \\
				& + o_p(T^{-1/2}) \\
				& \overset{(c)}{\rightarrow} 2 \bm{e}^\top (\mathbb{I} - \mathbf{A})^{-1} (1 + \xi_{u y^{-1}}^\top (\xi_{y^{-1} y^{-q-1}}^\top - \xi_{u y^{-q-1}} \xi_{u y^{-1}}^\top)^{-1} \xi_{u y^{-q-1}} ) \otimes \mathbb{I}_d \frac{1}{T-q} \sum_{t=q+1}^{T} U_{t}\mathbf{Z}_{t}  \\ 
				& - 2 (\mathbf{b}^\top (\mathbb{I} - \mathbf{A}^\top)^{-1} \otimes \bm{e}^\top (\mathbb{I} - \mathbf{A})^{-1}) ((\xi_{y^{-1} y^{-q-1}}^\top - \xi_{u y^{-q-1}} \xi_{u y^{-1}}^\top)^{-1} \xi_{u y^{-q-1} }  \otimes \mathbb{I}_d ) \frac{\sum_{t} U_{t}\mathbf{Z}_{t}}{T-q} 
				+ o_p(T^{-1/2}) \\
				& = 2 (1 + \xi_{u y^{-1}}^\top (\xi_{y^{-1} y^{-q-1}}^\top - \xi_{u y^{-q-1}} \xi_{u y^{-1}}^\top)^{-1} \xi_{u y^{-q-1}} ) \otimes \bm{e}^\top (\mathbb{I} - \mathbf{A})^{-1}   \frac{1}{T-q}\sum_{t=q+1}^{T} U_{t}\mathbf{Z}_{t} \\
				& - 2 \mathbf{b}^\top (\mathbb{I} - \mathbf{A}^\top)^{-1}   (\xi_{y^{-1} y^{-q-1}}^\top - \xi_{u y^{-q-1}} \xi_{u y^{-1}}^\top)^{-1} \xi_{u y^{-q-1} }  \otimes \bm{e}^\top (\mathbb{I} - \mathbf{A})^{-1} \frac{1}{T-q}\sum_{t=q+1}^{T} U_{t}\mathbf{Z}_{t}
				+ o_p(T^{-1/2}) \\
				& \overset{(d)}{\rightarrow} 2 \bm{e}^\top (\mathbb{I} - \mathbf{A})^{-1}  \frac{1}{T-q} \sum_{t=q+1}^{T} U_{t}\mathbf{Z}_{t} + o_p(T^{-1/2}),
			\end{aligned}
		\end{equation}
	where the limit in $(c)$ gets rid of terms involved with $\text{vec}(\sum_{t} \mathbf{Z}_{t+1} \mathbf{Y}_{t-q}^\top )$ based on $O(\text{ATE}) O_p(T^{-1/2}) \rightarrow 0$ within the small signal asymptotic framework. The limit in $(d)$ gets rid of $ \xi_{u y^{-1}}^\top (\xi_{y^{-1} y^{-q-1}}^\top - \xi_{u y^{-q-1}} \xi_{u y^{-1}}^\top)^{-1} \xi_{u y^{-q-1}}$ and $\mathbf{b}^\top (\mathbb{I} - \mathbf{A}^\top)^{-1}   (\xi_{y^{-1} y^{-q-1}}^\top - \xi_{u y^{-q-1}} \xi_{u y^{-1}}^\top)^{-1} \xi_{u y^{-q-1} }$ by relying on $O(\text{ATE}^2)\to 0$ as $\text{ATE} \rightarrow 0$. Finally, within the small signal asymptotic by letting $T\rightarrow +\infty$ and $\text{ATE} \rightarrow 0$, the simplified asymptotic MSE of the ATE estimator under the AT design $\pi_{\text{AT}}$ can be expressed as:
		\begin{equation}
			\begin{aligned}\nonumber
				& \lim_{\substack{T \rightarrow +\infty\\ \text{ATE}\to 0}}   
 \text{MSE} (\sqrt{n} \widehat{\textrm{ATE}} (\pi_{\text{AT}})) \\
    & = 4 \bm{e}^\top (\mathbb{I} - \mathbf{A})^{-1} \left( \sum_{j=0}^{q} \mathbf{M}_j \Sigma \mathbf{M}_j + 2 \sum_{k=1}^{q}\sum_{j=k}^{q} (-1)^k \mathbf{M}_j \Sigma \mathbf{M}_{j-k} \right)  (\mathbb{I} - \mathbf{A})^{-1} \bm{e} \\
 & = 4 \bm{e}^\top (\mathbb{I} - \mathbf{A})^{-1} \left( \sum_{j_1=0}^{q} \sum_{j_2=0}^{q}(-1)^{|j_2-j_1|}\mathbf{M}_{j_1} \bm{\Sigma} \mathbf{M}_{j_2} \right)  (\mathbb{I} - \mathbf{A})^{-1} \bm{e},
			\end{aligned}
		\end{equation}
where  $\text{Cov}(U_{j}\mathbf{Z}_{j}, U_{k} \mathbf{Z}_{k}) =  (-1)^{j-k} \sum_{i=j-k}^{q} \mathbf{M}_i \Sigma \mathbf{M}_{i-(j-k)} $ when $j-k<q$. Correspondingly, we define the multivariate efficiency indicators as 
    \begin{eqnarray}
\begin{aligned}
     \textrm{EI}_{\text{AD}}&=\bm{e}^\top (\mathbb{I} - \mathbf{A})^{-1}  \sum_{k=1}^{q}\sum_{j=k}^{q} \mathbf{M}_j \bm{\Sigma} \mathbf{M}_{j-k} (\mathbb{I} - \mathbf{A})^{-1}\bm{e},\\
    \textrm{EI}_{\text{AT}} &= \bm{e}^\top (\mathbb{I} - \mathbf{A})^{-1}  \sum_{k=1}^{q}\sum_{j=k}^{q} (-1)^k \mathbf{M}_j \bm{\Sigma} \mathbf{M}_{j-k} (\mathbb{I} - \mathbf{A})^{-1} \bm{e},
\end{aligned}
\end{eqnarray}
which is a direct multivariate version of those defined in Controlled ARMA($p, q$).

\subsection{Controlled VARMA(\texorpdfstring{$p, q$}{})}\label{appendix:sec_varmapq}
    Recap the controlled VARMA($p, q$) is formulated as $\mathbf{Y}_{t}=\sum_{j=1}^{p} \mathbf{A}_j \mathbf{Y}_{ t-j}+ \mathbf{b} U_{t} + \mathbf{Z}_{t}$, where $\{\mathbf{Y}_{t}\}_{t}\in \mathbb{R}^{d}$ has $p$-order lagging terms with the coefficient matrix $\mathcal{A} = [\mathbf{A}_1, \ldots,  \mathbf{A}_p]$. We denote $T_p = T - q - (p-1)$. By multiplying $\mathbf{Y}_{t-q-1}^\top,  \ldots,  \mathbf{Y}_{ t-q-p}^\top,$ and $U_{t}$ and then taking the expectation on both sides, we attain the following equations:
		\begin{equation}
			\begin{aligned}\nonumber
   	\underbrace{\frac{1}{T_p} \sum_t \Mean( U_{t} \mathbf{Y}_{t})}_{{\xi}_{u y}}&= \sum_{j=1}^q {\mathbf{A}_j}	\underbrace{\frac{1}{T_p} \sum_t \Mean( U_{t} \mathbf{Y}_{t-j} )}_{{\xi}_{u y^{-j} }}+ {\mathbf{b}} \\
				\underbrace{\frac{1}{T_p} \sum_t \Mean( \mathbf{Y}_{t} \mathbf{Y}_{t-q-1}^\top )}_{{\xi}_{y y^{-q-1}}}&=  \sum_{j=1}^q{\mathbf{A}_j}	\underbrace{\frac{1}{T_p} \sum_t \Mean( \mathbf{Y}_{t-j} \mathbf{Y}_{t-q-1}^\top)}_{{\xi}_{y^{-j} y^{-q-1}}}+ {\mathbf{b}} \underbrace{\frac{1}{T_p} \sum_t \Mean( U_{t} \mathbf{Y}_{t-q-1}^\top)}_{{\xi}_{u y^{-q-1}}^\top} \\
            & \cdots \\
			\underbrace{\frac{1}{T_p} \sum_t \Mean( \mathbf{Y}_{t} \mathbf{Y}_{t-q-p}^\top )}_{{\xi}_{y y^{-q-p}}}&=  \sum_{j=1}^q {\mathbf{A}}_j	\underbrace{\frac{1}{T_p} \sum_t \Mean( \mathbf{Y}_{t-j} \mathbf{Y}_{t-q-p}^\top)}_{{\xi}_{y^{-j} y^{-q-1}}}+ {\mathbf{b}} \underbrace{\frac{1}{T_p} \sum_t \Mean( U_{t} \mathbf{Y}_{t-q-p}^\top)}_{{\xi}_{u y^{-q-p}}^\top} .
			\end{aligned}
		\end{equation}
 We replace the expectation terms with sample moments, resulting in the equations:
 \begin{equation}
			\begin{aligned}\nonumber
   	\underbrace{\frac{1}{T_p} \sum_{t} U_{t} \mathbf{Y}_{t}}_{\widehat{\xi}_{u y}}&= \sum_{j=1}^p {\mathbf{A}_j}	\underbrace{\frac{1}{T_p} \sum_{t} U_{t} \mathbf{Y}_{t-j}}_{\widehat{\xi}_{u y^{-j} }}+ {\mathbf{b}} \\
				\underbrace{\frac{1}{T_p} \sum_{t} \mathbf{Y}_{t} \mathbf{Y}^\top_{t-q-1}}_{\widehat{\xi}_{y y^{-q-1}}}&=  \sum_{j=1}^p{\mathbf{A}_j}	\underbrace{\frac{1}{T_p} \sum_{t} \mathbf{Y}_{t-j} \mathbf{Y}^\top_{t-q-1}}_{\widehat{\xi}_{y^{-j} y^{-q-1}}}+ {\mathbf{b}} \underbrace{\frac{1}{T_p} \sum_{t} U_t \mathbf{Y}^\top_{t-q-1}}_{\widehat{\xi}_{u y^{-q-1}}^\top} \\
            & \cdots \\
			\underbrace{\frac{1}{T_p} \sum_{t} \mathbf{Y}_{t} \mathbf{Y}^\top_{t-q-p}}_{\widehat{\xi}_{y y^{-q-p}}}&=  \sum_{j=1}^p {\mathbf{A}}_j	\underbrace{\frac{1}{T_p} \sum_{t} \mathbf{Y}_{t-j} \mathbf{Y}^\top_{t-q-p}}_{\widehat{\xi}_{y^{-j} y^{-q-1}}}+ {\mathbf{b}} \underbrace{\frac{1}{T_p} \sum_{t} U_{t} \mathbf{Y}^\top_{t-q-p}}_{\widehat{\xi}_{u y^{-q-p}}^\top} .
			\end{aligned}
		\end{equation}
 Correspondingly, we have another group of equations without taking the expectation:
		\begin{equation}
			\begin{aligned}\nonumber
				\widehat{\xi}_{u y} &=\sum_{j=1}^{p}\mathbf{A}_j\widehat{\xi}_{u y^{-j}} + \mathbf{b} + \frac{1}{T_p}\sum_{t} U_{t}\mathbf{Z}_{t}\\
				\widehat{\xi}_{y y^{-q-1}} &=\sum_{j=1}^{p}\mathbf{A}_j \widehat{\xi}_{y^{-j} y^{-q-1}} + \mathbf{b} \widehat{\xi}_{u y^{-q-1}}^\top + \frac{1}{T_p} \sum_{t} \mathbf{Z}_{t} \mathbf{Y}_{t-q-1}^\top \\
				& \cdots \\
				\widehat{\xi}_{y y^{-q-p}} &=\sum_{j=1}^{p}\mathbf{A}_j \widehat{\xi}_{y^{-j} y^{-q-p}} + \mathbf{b} \widehat{\xi}_{u y^{-q-p}}^\top + \frac{1}{T_p} \sum_{t} \mathbf{Z}_{t} \mathbf{Y}_{t-q-p}^\top.
			\end{aligned}
		\end{equation}
	We next define:
		\begin{equation}
			\begin{aligned}\nonumber
				\left(\begin{array}{c|c} 
					\boldsymbol{\xi_{uy^{-p}}}	& \boldsymbol{\xi_{y^{-p} y^{-q-p}}} \\
					\hline 
					1  & {\boldsymbol{\xi^\top_{uy^{-q-p}}}}  \\
				\end{array}
				\right) = 
				\left(\begin{array}{c|ccc} 
                    \xi_{uy^{-1}} & {\xi}_{y^{-1}y^{-q-1}}  & \cdots & {\xi}_{y^{-1}y^{-q-p}}  \\
                    \cdots  & \cdots & \cdots & \cdots \\
                    \xi_{uy^{-p}} & {\xi}_{y^{-p} y^{-q-1}} & \cdots & {\xi}_{y^{-p}y^{-q-p}}    \\
                    \hline
					1 & {\xi}_{uy^{-q-1}}^\top  & \cdots & {\xi}_{uy^{-q-p}}^\top \\
				\end{array}
				\right),
			\end{aligned}
		\end{equation}
    where we pre-define this matrix of interest for proof of convenience. $\boldsymbol{\xi_{uy^{-p}}}$, $\boldsymbol{\xi_{y^{-p} y^{-q-p}}}$, and $\boldsymbol{\xi^\top_{uy^{-q-p}}}$ represent each block matrix component, respectively.
    
    Next, using the bold version  Controlled VARMA($p, q$) and replacing $\mathbf{A}$ with $\mathcal{A}$, we have:
		\begin{equation}
			\begin{aligned}\nonumber
				\left[\widehat{\mathcal{A}} - \mathcal{A}, \widehat{\mathbf{b}} - \mathbf{b}\right] 
				&=\frac{1}{T_p}  \left[\sum_{t} U_{t}\mathbf{Z}_{t} , \sum_{t} \mathbf{Z}_{t} \mathbf{Y}_{t-q-1}^\top, \ldots,\sum_{t} \mathbf{Z}_{t} \mathbf{Y}_{t-q-p}^\top   \right]
				\left(\begin{array}{cc} 
					{\boldsymbol{\xi_{uy^{-p}}}}	& \boldsymbol{\xi_{y^{-p} y^{-q-p}}} \\
					1  & {\boldsymbol{\xi^\top_{uy^{-q-p}}}}  \\
				\end{array}
				\right)^{-1} + o_p(1).
			\end{aligned}
		\end{equation}
	Applying the vectorization to the above equation, we have the following equation:
		\begin{equation}
			\begin{aligned}\nonumber
				\left(\begin{array}{c}
					\text{vec}(\widehat{\mathcal{A}} - \mathcal{A}) \\
					\text{vec}(\widehat{\mathbf{b}} - \mathbf{b})
				\end{array}
				\right)
				=  \frac{1}{T_p}  \left( 	\left(\begin{array}{cc}
					\boldsymbol{\xi_{uy^{-p}}^\top} & 1 \\
					\boldsymbol{\xi_{y^{-p} y^{-q-p}}^\top} & \boldsymbol{\xi_{u y^{-q-p} }}
				\end{array}
				\right)^{-1} \otimes \mathbb{I}_d \right) 	
				\left(\begin{array}{c}
					\displaystyle \sum_{t} U_{t}\mathbf{Z}_{t} \\
					\displaystyle \text{vec}(\sum_{t} \mathbf{Z}_{t} \mathbf{Y}_{t-q-1}^\top) \\
                    \vdots \\
                    \displaystyle \text{vec}(\sum_{t} \mathbf{Z}_{t} \mathbf{Y}_{t-q-p}^\top)
				\end{array}
				\right)
				+ o_p(1).
			\end{aligned}
		\end{equation}
Applying the Delta method, the ATE estimator in Controlled VARMA($p, q$) formulates:
		\begin{equation}
			\begin{aligned}\nonumber
				\widehat{\textrm{ATE}} - \text{ATE} &= J_{\text{vec}(\mathcal{A}), \mathbf{b}} 	\left(\begin{array}{c}
					\text{vec}(\widehat{\mathcal{A}} - \mathcal{A}) \\
					\text{vec}(\widehat{\mathbf{b}} - \mathbf{b})
				\end{array}
				\right)  + o_p(T_p^{-1/2})\\
				& = \left[ (2 \mathbf{b}^\top (\mathbb{I} - \mathbf{A}^\top)^{-1} \otimes \bm{e}^\top (\mathbb{I} - \mathbf{A})^{-1})_p, 2\bm{e}^\top (\mathbb{I} - \mathbf{A})^{-1} \right] 	\left(\begin{array}{c}
					\text{vec}(\widehat{\mathcal{A}} - \mathcal{A}) \\
					\text{vec}(\widehat{\mathbf{b}} - \mathbf{b})
				\end{array}
				\right)  + o_p(T_p^{-1/2}) ,
			\end{aligned}
		\end{equation}
	where $(2 \mathbf{b}^\top (\mathbb{I} - \mathbf{A}^\top)^{-1} \otimes \bm{e}^\top (\mathbb{I} - \mathbf{A})^{-1})_p$ represents the repeating along the row and we denote $\mathbf{A} = \sum_{j=1}^{p} \mathbf{A}_j$ with slight abuse of notation. The remaining parts to derive the asymptotic MSE of ATE estimators under the AD, UR, and AT design are the same as controlled VARMA($1, q$). In particular, $\xi_{u y^{-t}}=(\mathbb{I} - \mathbf{A})^{-1} \mathbf{b}$ for the AD design, $\xi_{u y^{-t}}=\mathbf{0}_d$ for the UR design,  and $\xi_{u y^{-t}}=(-1)^{t+1}(\mathbb{I} + \mathbf{A})^{-1} \mathbf{b}$ for the AT design for $t \geq 1$. In particular, within the small signal asymptotic by letting $T\rightarrow+\infty$, $\tau\to +\infty$ and $\text{ATE}\rightarrow 0$, the simplified ATE estimators of the three designs have the same limit:
		\begin{equation}
			\begin{aligned}\nonumber
				\widehat{\textrm{ATE}} - \text{ATE} &\rightarrow 2 \bm{e}^\top (\mathbb{I} - \mathbf{A})^{-1}  \frac{1}{T_q}\sum_{t} U_{t}\mathbf{Z}_{t} + o_p(T_q^{-1/2}).
			\end{aligned}
		\end{equation}
 The asymptotic MSEs under the three treatment allocation strategies retain the same forms as those in Controlled VARMA($1, q$) by simply replacing the 1-order coefficient matrix $\mathbf{A}$ by the compound one $\mathbf{A} = \sum_{j=1}^{p} \mathbf{A}_j$. A similar form applies to the efficiency indicators.

\section{Derivation of Optimal Markov Design}\label{appendix:optimal_markov}

\begin{proof}


We study the correlation structure of the stationary treatments, i.e., $\text{Cov}(U_{t}, U_{t-k})$, which determines the asymptotic MSE. By mathematical induction, we find
		\begin{equation}
			\begin{aligned}\label{eq:appendix_stationary}
				P(U_{t} U_{t-k} = 1) = \sum_{i=0}^{\lfloor \frac{k}{2} \rfloor }\binom{k}{2j}  \alpha^{k-2j} (1-\alpha)^{2j}.
			\end{aligned}
		\end{equation}
 Assume $X$ is a ($n, p$)-binomial random variable, we have:
		\begin{equation}
			\begin{aligned}\nonumber
				((1-p)+p)^n & =\sum_{k=0}^n\left(\begin{array}{l}
					n \\
					k
				\end{array}\right) p^k(1-p)^{n-k} \\
				& =\sum_{k=0}^{\lfloor n / 2\rfloor}\left(\begin{array}{c}
					n \\
					2 k
				\end{array}\right) p^{2 k}(1-p)^{n-2 k}+\sum_{k=0}^{\lfloor n / 2\rfloor}\left(\begin{array}{c}
					n \\
					2 k+1
				\end{array}\right) p^{2 k+1}(1-p)^{n-(2 k+1)}, 
			\end{aligned}
		\end{equation}
 which equals $P\{X \text { even }\}+P\{X \text { odd }\}$. Therefore, the probability in \eqref{eq:appendix_stationary} can be interpreted as the sum of probabilities over events that occur an even number of times with probability $1-\alpha$. A similar result is also given by:
		\begin{equation}
			\begin{aligned}\nonumber
				((1-p)-p)^n & =\sum_{k=0}^n\left(\begin{array}{l}
					n \\
					k
				\end{array}\right)(-p)^k(1-p)^{n-k} \\
				& =\sum_{k=0}^{\lfloor n / 2\rfloor}\left(\begin{array}{c}
					n \\
					2 k
				\end{array}\right) p^{2 k}(1-p)^{n-2 k}-\sum_{k=0}^{\lfloor n / 2\rfloor}\left(\begin{array}{c}
					n \\
					2 k+1
				\end{array}\right) p^{2 k+1}(1-p)^{n-(2 k+1)}
			\end{aligned}
		\end{equation}
	which equals $P\{X \text { even }\}-P\{X \text { odd }\}$. This leads to $P\{X \text { even }\}=\frac{1}{2}\left(1+(1-2 p)^n\right)=\frac{1}{2}+\frac{1}{2}(1-2 p)^n$. In our case, we have $p = 1-\alpha$, and consequently, 
		\begin{equation}
			\begin{aligned}\nonumber
				P(U_{t} U_{t-k} = 1) = \sum_{i=0}^{[\frac{k}{2}]}\binom{k}{2j}  \alpha^{k-2j} (1-\alpha)^{2j} = \frac{1}{2} (1 + (1 - 2(1-\alpha))^k) = \frac{1}{2} (1 + (2\alpha - 1)^k).
			\end{aligned}
		\end{equation}
	Therefore, we have $\text{Cov}(U_{t}, U_{t-k}) = \mathbb{E}\left[U_{t} U_{t-k}\right] -0  = (2\alpha - 1)^k$. Another direct conclusion is $\xi_{u y^{-1}} = \frac{b(2\alpha - 1)}{ 1 - a(2\alpha - 1)}$ for Controlled ARMA($1, q$) and $\xi_{u y^{-1}} = \frac{b(2\alpha - 1)}{ 1 - \sum_{i=1}^{q} a_i (2\alpha - 1)^i}$ for Controlled ARMA($p, q$), which also unifies the three design policies. For example, $\xi_{u y^{-1}} = 0$ if $\alpha = \frac{1}{2}$ for the UR design. Following the proof in Appendix~\ref{appendix:thm_ARMA} and \ref{appendix:thm_optimal}, we have the asymptotic MSE form under the Markov policy $\pi_{\text{Mar}}$:
		\begin{equation}
			\begin{aligned}\nonumber
				\lim_{\substack{T \rightarrow +\infty\\ \text{ATE}\to 0}}  \text{MSE}( \sqrt{T} \widehat{\text{ATE}}(\pi_{\text{Mar}})) & =\frac{4\sigma^2}{(1-a)^2}\Big[ \sum_{j=0}^q \theta_j^2+2\sum_{k=1}^q (2\alpha - 1)^k \sum_{j=k}^q \theta_j\theta_{j-k}  \Big].
			\end{aligned}
		\end{equation}
 where $\lim_{T\rightarrow +\infty} \frac{1}{T}\sum_{t=q+1}^T \mathbb{E} (U_{t} U_{t-k}) = (2\alpha-1)^k$ in \eqref{eq:appendix_optimality_arma}. The proof of extension to Controlled VARMA($p, q$) is also similar. Below, we present the asymptotic MSE result in Controlled VARMA($p, q$) under the Markov design as
			\begin{eqnarray}
				\begin{aligned}\nonumber
			&	\lim_{\substack{T \rightarrow +\infty\\ \text{ATE}\to 0}}  \text{MSE}( \sqrt{T} \widehat{\text{ATE}}(\pi_{\text{Mar}}))  \\
   & = 4 \bm{e}^\top (\mathbb{I} - \mathbf{A})^{-1} \left( \sum_{j=0}^{q} \mathbf{M}_j \Sigma \mathbf{M}_j + 2 \sum_{k=1}^{q}\sum_{j=k}^{q} (2\alpha-1)^k  \mathbf{M}_j \Sigma \mathbf{M}_{j-k} \right)  (\mathbb{I} - \mathbf{A})^{-1} \bm{e}.
				\end{aligned}
			\end{eqnarray}
\end{proof}


\end{document}